\documentclass[]{aa}  

\usepackage{graphicx}
\usepackage{xcolor}
\bibpunct{(}{)}{;}{a}{}{,}
\usepackage[T1]{fontenc}
\usepackage{ae,aecompl}
\usepackage{adjustbox}

\usepackage{multicol}        
\usepackage{multirow}        
\usepackage{csquotes}
\usepackage{color}
\usepackage{url}
\usepackage{hyperref}
\usepackage{floatfig}
\usepackage{wrapfig}
\usepackage{txfonts}
\definecolor{purple}{RGB}{76, 0,153}
\definecolor{black}{RGB}{0, 0,0}
\definecolor{darkred}{RGB}{128, 0,0}
\definecolor{darkgreen}{RGB}{0,128,0}
\definecolor{ballblue}{rgb}{0.13, 0.67, 0.8}
\definecolor{darkblue}{rgb}{0.2, 0.2, 0.6}

\def \Eqt{Eq.\;}
\def \sect{Sect.\thinspace}
\def \fig{Fig.\thinspace}
\def \figs{Figs.\thinspace}
\def \tab{Table\thinspace}
\def \app{Appendix\thinspace}

\def\epsao{\epsilon_{a,1}}
\def\epsat{\epsilon_{a,2}}
\def\epsbo{\epsilon_{b,1}}
\def\epsbt{\epsilon_{b,2}}

\def\Om{{\Omega_{\rm m}}}
\def\C{{C_{\epsilon\epsilon}}}

\def\d{{\rm d}}

\begin{document} 



\title{KiDS-1000 Cosmology: Cosmic shear constraints and comparison between two point statistics}

\author{Marika Asgari\inst{1}\fnmsep\thanks{E-mail: ma@roe.ac.uk}
	    \and	Chieh-An Lin\inst{1}
		\and	Benjamin Joachimi\inst{2}
     	\and	Benjamin Giblin\inst{1}
     	\and	Catherine Heymans\inst{1,3}
     	\and	Hendrik Hildebrandt\inst{3}
        \and  Arun Kannawadi\inst{4}
        \and  Benjamin St\"olzner\inst{2}
		\and	Tilman Tr\"oster\inst{1}
		\and	Jan Luca van den Busch\inst{3}
		\and	Angus H. Wright\inst{3}
		\and	Maciej Bilicki\inst{5}
		\and Chris Blake\inst{6}
		\and Jelte de Jong\inst{7}
		\and Andrej Dvornik\inst{3}
		\and Thomas Erben\inst{8}
		\and Fedor Getman\inst{9}
		\and Henk Hoekstra\inst{10}
		\and Fabian K\"ohlinger\inst{3}
		\and Konrad Kuijken\inst{10}
		\and Lance Miller\inst{11}
		\and Mario Radovich\inst{12}
		\and Peter Schneider\inst{8}
		\and HuanYuan Shan\inst{13,14}
		\and Edwin Valentijn\inst{15}
		}
		
   \institute{Institute for Astronomy, University of Edinburgh, Royal Observatory,
Blackford Hill, Edinburgh, \; EH9 3HJ, U.K.
		\and
         Department of Physics and Astronomy, University College London, Gower Street, London WC1E 6BT, UK
         \and
        Ruhr-University Bochum, Astronomical Institute, German Centre for Cosmological Lensing, Universit\"atsstr. 150, 44801 Bochum, Germany
         \and
         Department of Astrophysical Sciences, Princeton University, 4 Ivy Lane, Princeton, NJ 08544, USA
         \and
         Center for Theoretical Physics, Polish Academy of Sciences, al. Lotników 32/46, 02-668 Warsaw, Poland
         \and
         Centre for Astrophysics \& Supercomputing, Swinburne University of Technology, P.O. Box 218, Hawthorn, VIC 3122, Australia
         \and
         Kapteyn Astronomical Institute, University of Groningen, PO Box 800, 9700 AV Groningen, the Netherlands
         \and
         Argelander-Institut für Astronomie, Auf dem H\"ugel 71, 53121 Bonn, Germany
         \and 
         INAF - Astronomical Observatory of Capodimonte, Via Moiariello 16, 80131 Napoli, Italy
         \and
         Leiden Observatory, Leiden University, Niels Bohrweg 2, 2333 CA Leiden, The Netherlands
	    \and
Department of Physics, University of Oxford, Denys Wilkinson Building, Keble Road, Oxford OX1 3RH, UK
		\and
INAF - Osservatorio Astronomico di Padova, via dell'Osservatorio 5, 35122 Padova, Italy
		\and
		Shanghai Astronomical Observatory (SHAO), Nandan Road 80, Shanghai 200030 China 
		\and 
		University of Chinese Academy of Sciences, Beijing 100049, China
		\and
		Kapteyn Institute, University of Groningen, PO Box 800, NL 9700 AV Groningen
             }

\date{Received XXX; accepted YYY}

\abstract{We present cosmological constraints from a cosmic shear analysis of the fourth data release of the Kilo-Degree Survey (KiDS-1000), doubling the survey area with nine-band optical and near-infrared photometry with respect to previous KiDS analyses.  
Adopting a spatially flat $\Lambda$CDM model, we find $S_8 = \sigma_8 (\Omega_{\rm m}/0.3)^{0.5} = 0.759^{+0.024}_{-0.021}$ for our fiducial analysis, which is in $3\sigma$ tension with the prediction of the \textit{Planck} Legacy analysis of the cosmic microwave background.
We compare our fiducial COSEBIs (Complete Orthogonal Sets of E/B-Integrals) analysis with complementary analyses of the two-point shear correlation function and band power spectra, finding results to be in excellent agreement.    
We investigate the sensitivity of all three statistics to a number of measurement, astrophysical, and modelling systematics, finding our $S_8$ constraints to be robust and dominated by statistical errors.  
Our cosmological analysis of different divisions of the data pass the Bayesian internal consistency tests, with the exception of the second tomographic bin.  As this bin encompasses low redshift galaxies, carrying insignificant levels of cosmological information, we find that our results are unchanged by the inclusion or exclusion of this sample.}


\titlerunning{KiDS1000 cosmic shear}
\authorrunning{Asgari, Lin, Joachimi et al.}
\maketitle
\tableofcontents


\section{Introduction}

\label{sec:introduction}

In this new era of precision cosmology reliable probes of the key parameters of the standard model, $\Lambda$CDM, are indispensable. The weak gravitational lensing effect that coherently distorts the shapes of galaxy images, commonly referred to as cosmic shear, was hailed as such a tool \citep{Albrecht06,Peacock06}, directly mapping the spatial distribution of all gravitating matter along the line of sight and therefore sensitive to the amplitude and shape of the matter power spectrum \citep[see][for a review]{Kilbinger_review}. This makes cosmic shear highly complementary to galaxy clustering, which as a spatially localised probe can trace line-of-sight modes of the matter distribution and localised features like baryon acoustic oscillations, but which suffers from the poorly known connection between the galaxy and matter distribution, known as galaxy bias.

First detected two decades ago \citep{Bacon00,Kaiser00,VanWaerbeke00,Wittman00}, cosmic shear has since matured into a primary probe in the golden era of galaxy surveys, featuring prominently alongside galaxy clustering in forthcoming experiments like the ESA \textit{Euclid} mission\footnote{\texttt{https://sci.esa.int/euclid}} \citep{Euclid11}, the Vera C. Rubin Observatory LSST\footnote{Legacy Survey of Space and Time; \texttt{https://www.lsst.org}} \citep{lsst12}, and NASA's {\it Nancy Grace Roman} Space Telescope\footnote{formerly Wide Field Infrared Survey Telescope; \texttt{https://nasa.gov/wfirst}} \citep{spergel15}. To meet the stringent accuracy requirements of these new surveys, all aspects of a cosmic shear analysis have to undergo critical revision and, in many cases, radical improvements. Vital lessons are being learnt by three concurrent surveys, whose analyses are on-going: the ESO Kilo-Degree Survey\footnote{\texttt{http://kids.strw.leidenuniv.nl}} (KiDS; \citealp{kuijken/etal:2015,hildebrandt/etal:2019}), the Dark Energy Survey\footnote{\texttt{https://www.darkenergysurvey.org}} (DES; \citealp{DrlicaWagner/etal:2018, Zuntz/etal:2018}), and the Hyper Suprime-Cam Subaru Strategic Program\footnote{\texttt{https://hsc.mtk.nao.ac.jp/ssp}} (HSC; \citealp{aihara18,Hikage18}). The current surveys already have the statistical power to independently test our cosmological standard model, in particular the amplitude of matter density fluctuations, by convention measured via the parameter $S_8=\sigma_8\,(\Omega_{\rm m}/0.3)^{0.5}$, where $\Om$ is the matter density parameter and $\sigma_8$ is the linear-theory standard deviation of matter density fluctuations in spheres of radius $8\,h^{-1}\,{\rm Mpc}$.

In this work we present the cosmic shear analysis of the fourth KiDS Data Release \citep{kuijken/etal:2019}, hereafter referred to as KiDS-1000. This more than doubles the survey area with respect to the previous KiDS cosmological analyses \citep{hildebrandt/etal:2017,hildebrandt/etal:2019}. While neither as deep as HSC, once it is completed, nor as wide as the final DES area, KiDS has unique properties that make it competitive in terms of controlling the two major measurement challenges for cosmic shear analyses -- the accurate measurement of gravitational shear, i.e. the image distortions imposed by the lensing effect, and the accurate determination of the redshift distribution of the galaxies used in the cosmic shear analysis. 
As all current cosmic shear analyses can already be considered systematics-limited to some degree (see \citealp{mandelbaum18} for a recent review of the major challenges), such benefits are likely to directly impact on the final cosmological constraints and could potentially outweigh a larger raw statistical power.

A robust and accurate analysis of cosmic shear data is of paramount importance for testing the concordance of the current standard cosmological model, flat $\Lambda$CDM. Currently the tightest constraints on the parameters of this model come from studies of full-sky CMB temperature and polarisation maps. Although these data are primarily sensitive to the physics of the early Universe, given a model they can make predictions for statistical properties of the structures formed in the late Universe as well as the current expansion rate. Since the first cosmological analysis of the \textit{Planck} data \citep{Planck14}, there have been indications of tension between the CMB and cosmic shear results \citep{heymans/etal:2013} as well as with the Hubble parameter estimated through the distance ladder \citep{riess/etal:2011}. 

Recently, there has been a high level of attention towards the ever-growing tension between the estimates of the Hubble parameter from early and late Universe probes \citep[see][for a recent summary]{verde/etal:2019}. 
Although not currently as significant, the level of tension in $S_8$ between the probes of the large-scale structures and the \textit{Planck} results has also been increasing. In particular, the cosmic shear analysis of the first-year data release of DES \citep[DES-Y1,][]{troxel/etal:2018a}, HSC \citep{Hikage18} and the KiDS results of \citet[KV450]{hildebrandt/etal:2019} all found values of $S_8$ that are lower than the Planck predictions \citep{Planck2018} by around $2\sigma$. 
Interestingly, these results are largely independent, as the images are taken over mostly different patches of the sky and the teams and pipelines analysing them were largely separate\footnote{We note that the \cite{hamana/etal:2020} re-analysis of HSC with 2PCFs, find an $S_8$ value that is closer to {\it Planck}, albeit still lower by $\sim 1\sigma$.}. 
Therefore, we can assume that the combined analysis of these data sets would result in deviations larger than $2\sigma$. For instance, \cite{joudaki/etal:2020} analysed the combination of DES-Y1 and KV450 data using the KV450 setup and redshift calibrations to find a tension of $2.5\sigma$, and the re-analysis of \cite{asgari/etal:2020a} increased the constraining power of DES by including smaller angular scales to find a DES-Y1 and KV450 joint result that is in $3.2\sigma$ tension with Planck. 

Aside from the importance of the quality of the data, we need to improve the model for a robust analysis.
Modelling challenges in cosmic shear prevail especially on small scales where the signal-to-noise ratio is highest and where non-linear structure growth \citep[e.g.][]{knabenhans19}, baryon feedback on the matter distribution \citep[e.g.][]{semboloni/etal:2011}, 
and complex matter-galaxy interactions affecting the intrinsic alignment of galaxies \citep[e.g.][]{fortuna20} all combine to lead to an uncertainty that is difficult to calibrate and quantify. 

It is standard to employ two-point statistics of the gravitational shear estimates as summary statistics, but which choice strikes a balance between the optimal extraction of information and the suppression of observational or modelling systematics?
While the KiDS-1000 approach to modelling and inference methodology is discussed in detail in \citet{joachimi/etal:2020}, here we focus on the choice of summary statistics and their sensitivity to different systematic and modelling effects. 

Two-point statistics of the shear field can be measured in configuration, Fourier or other spaces.  
In this analysis we consider Complete Orthogonal Sets of E/B-Integrals (COSEBIs; \citealp{SEK10}), band power estimates derived from the correlation functions \citep{schneider/etal:2002a,becker16b,vanuitert/etal:2017} and the shear two-point correlation functions (2PCFs).  As we discuss in \sect\ref{sec:method}, there are considerable advantages to the former two statistics, since they allow us to avoid scales that are affected by modelling uncertainties, although the latter method has been used in the clear majority of recent cosmic shear analyses \citep[see for example][]{heymans/etal:2013,jee/etal:2016,hildebrandt/etal:2017,Joudaki_KiDS450,troxel/etal:2018b,troxel/etal:2018a,hildebrandt/etal:2019,wright/etal:2020b,hamana/etal:2020}. With these statistics we connect previous work with this new analysis.

Consistent parameter constraints from a diverse set of summary statistics can add valuable corroboration to cosmological inference. However, care must be taken to accurately quantify the correlation between the different two-point statistics, which is strong, as they are calculated from the same catalogue, but not perfect, as scales are incorporated and weighted differently. 
 In this work we will apply all summary statistics to the same suite of realistic mock KiDS-1000 data, enabling us to map the expected differences in cosmological constraints. In addition we triplicate all of our cosmological analyses, including results for 2PCFs, COSEBIs and band powers for all cases.

The KiDS-1000 analysis methodology is discussed in \citet[J20]{joachimi/etal:2020}, while \citet[G20]{giblin/etal:2020} and \citet[H20b]{hildebrandt/etal:2020} detail the construction and calibration of the gravitational shear catalogues and the galaxy redshift distributions used in this analysis, respectively. Further KiDS-1000 companion papers include \citet{heymans/etal:2020} who present cosmological constraints from a combined-probe analysis of cosmic shear, galaxy-galaxy lensing and galaxy clustering. \citet{troster/etal:2020b} extend the cosmological inference from the combined weak lensing and clustering data beyond the spatially flat $\Lambda$CDM model considered in the remainder of the KiDS-1000 analyses.

This paper is structured as follows: in \sect\ref{sec:method} the modelling of the three two-point statistics employed in KiDS-1000 is described. \sect\ref{sec:data} provides an overview of the data set and the analysis pipeline. In \sect\ref{sec:results} the cosmological constraints are presented, including a range of validation tests as well as an assessment of consistency internal to the KiDS data vector and with \textit{Planck} CMB results, before concluding in \sect\ref{sec:conclusions}. More technical details of  the analysis are provided in the appendices. In particular we point the reader to \app\ref{app:extra} where we present constraints on all parameters, \app\ref{app:consistency} which details our internal and external consistency tests and \app\ref{app:cterm} where we model the impact of the residual constant additive shear biases on our two-point statistics. 

\section{Methods}

\label{sec:method}
We analyse the KiDS-1000 data with three sets of statistics: real-space shear two-point correlation functions (2PCFs), complete orthogonal sets of E/B-integrals (COSEBIs) and band power spectra estimated from 2PCFs (band powers). These statistics are all linear transformations of the observed cosmic shear angular power spectrum, $\C(\ell)$,
\begin{align}
\label{eq:stats_form}
 S_x = \int_0^{\infty}\d\ell\,\ell\, C_{\epsilon\epsilon}(\ell)\,W_x(\ell)\;,
\end{align} 
where $W_x(\ell)$ is a weight function that depends on the angular Fourier scale, $\ell$, as well as the argument of the statistics, $x$. 
The  $\C(\ell)$ in turn can be written as a sum of gravitational lensing (G) and intrinsic (I) alignments of galaxies, 
\begin{equation}
\label{eq:c_ee}
\C(\ell) = C_{\rm GG}(\ell) + C_{\rm GI}(\ell) + C_{\rm II}(\ell)\;.
\end{equation}
The observed cosmic shear signal can in principle consist of E and B-modes. 
Under the standard cosmological model, however, we do not expect to measure any significant B-modes for surveys such as KiDS\footnote{Effects such as  contributions beyond the Born approximation \citep{Schneider98}, source clustering \citep{schneider/etal:2002b}, intrinsic alignment models with tidal effects \citep[e.g.][]{blazek/etal:2015} and certain alternative cosmological models \citep[see for example][]{thomas/etal:2017} are able to produce B-modes. For current surveys, however, these effects are negligible.}. In this case we can substitute $\C(\ell)$ with $C_{\rm EE}$, the E-mode angular power spectrum and derive the three terms on the right hand side of \Eqt\eqref{eq:c_ee} from the matter power spectrum, using a modified Limber approximation \citep{LoverdeAfshordi08,kilbinger/etal:2017},
\begin{align}
\label{eq:limber}
 C_{\rm XY}^{(ij)}(\ell) = \int_0^{\chi_\mathrm{hor}}\mathrm{d}\chi\:\frac{W_{\rm X}^{(i)}(\chi)W_{\rm Y}^{(j)}(\chi)}{f_{\rm K}^2(\chi)} 
   P_{\rm m} \left(\frac{\ell+1/2}{f_\mathrm{K}(\chi)},\chi\right)\;,
\end{align} 
where X and Y stand for G or I, $i$ and $j$ denote two populations of galaxies,
$\chi$ is the radial comoving distance and $f_{\rm K}(\chi)$ is the comoving angular diameter distance which simplifies to $\chi$ for a spatially flat universe. The integral is taken from the observer, $\chi=0$ to the horizon, $\chi_{\rm hor}$. The kernels, $W_{\rm X/Y}$ depend on the redshift distribution of the two populations and their mathematical form can be found in equations 15 and 16 of J20.

It is common practice to divide galaxies based on their estimated photometric redshifts into tomographic bins, which has the advantage of improving the constraining power and reducing the degeneracy between redshift-dependent parameters in a cosmic shear analysis \citep{hu99}. In this case $i$ and $j$ in \Eqt\eqref{eq:limber} are the labels for the tomographic bins.  
 
From a theoretical point of view, spherical harmonic measures estimated from a pixelated sky may seem to be the most natural choice. Such direct power spectrum statistics have seen widespread application in other cosmological probes, most prominently in temperature and polarisation measurements of the cosmic microwave background (CMB: \citealp[e.g.][]{planck19}).
Analogous statistics, like pixel-based maximum-likelihood quadratic estimators \citep{Brown03,heymans05,Lin12,koehlinger16,koehlinger17} or pseudo-$C_\ell$ techniques \citep{Hikage11,becker/etal:2015,asgari/etal:2018,Hikage18,alonso19}, have also been developed for cosmic shear. These measurements are, however, affected directly by masking and finite field effects. Moreover, the significant noise component due to the random intrinsic orientations of galaxy shapes is spread out over all multipoles in harmonic space. For such analyses, these effects have to be either modelled or corrected for. 2PCFs, on the other hand, do not suffer from these limitations, as masking and noise effects do not bias their expectation value, although they should be included in their covariance estimation (see section 5 of J20 for a discussion on the importance of each effect). An additional motivation for employing 2PCFs is that measurement systematics are better traced in configuration space. 

This makes the 2PCFs the current method of choice to be applied to a catalogue of shear estimates.  However, considerable disadvantages are revealed in the further stages of the cosmological inference. Due to the very broad kernels linking the 2PCFs to the underlying power spectrum, the analyst  has little control over the physical scales entering the likelihood analysis, with undesirable consequences (see \fig\ref{fig:compare}). 
For instance, sensitivity to low multipoles where only few independent modes contribute leads to significant deviations from a Gaussian likelihood for 2PCFs measured on large-separations \citep{schneider_hartlap:2009,sellentin/etal:2018}, while a fairly wide range of small-scale 2PCF measurements are affected by non-linear modelling uncertainties such as baryon feedback \citep{asgari/etal:2020a}. In addition, the 2PCFs mix E-modes, which are expected to carry the cosmological signal, and B-modes, which are influenced by cosmological signals only at a very low level and hence provide a valuable null test for a range of systematics. There are also modes that cannot be uniquely identified as either E or B-modes. The 2PCFs are also impacted by these ambiguous modes.  

To remedy these shortcomings, we consider two promising alternatives: COSEBIs and band powers. 
COSEBIs offer a clean separation of E and B-modes over a finite range of available angular scales, with nearly lossless data compression and discrete abscissae as a bonus. 
Band powers allow for approximate E-/B-mode separation and closely follow the underlying angular power spectra, facilitating intuitive interpretation of the signals. 
These statistics are, in addition, insensitive to the ambiguous E and B-modes. 
We will demonstrate that both derived statistics avoid the modelling deficiencies of 2PCFs because of their more compact kernels. 
We note that direct power spectrum estimators will be applied to KiDS data in forthcoming work (Loureiro et al., in prep.).

In the following subsections we first introduce the 2PCFs and briefly review their measurement method (\sect\ref{sec:2pcfs}). We then introduce COSEBIs in \sect\ref{sec:cosebis} and summarise the main equations for band power spectra in \sect\ref{sec:bp}. Finally we compare the scale sensitivity of these statistics in \sect\ref{sec:compare}.

\subsection{Shear two-point correlation functions}
\label{sec:2pcfs}

The shear two-point correlation functions, $\xi_\pm$ \citep{Kaiser92}, are formally defined as
\begin{align}
\label{eq:xi-pm-def}
\xi_\pm(\theta) & =\langle\gamma_\mathrm{t} \gamma_\mathrm{t} \rangle (\theta)\pm 
\langle\gamma_\mathrm{\times} \gamma_\mathrm{\times} \rangle (\theta) \ ,
\end{align} 
where $\gamma_\mathrm{t}$ is the tangential shear and $\gamma_\times$ is the cross component of the shear defined with respect to the line connecting the pair of galaxies \citep[see][for details]{BartelmannSchneider01}.
2PCFs are functions of the angular separation, $\theta$, between pairs of galaxies whose ellipticities are used to estimate shear. In practice we bin the data into several $\theta$-bins and measure the signal using
\begin{equation}
\label{eq:xipm_meaure}
\hat\xi^{(ij)}_\pm(\bar{\theta})=\frac{\sum_{ab} w_a w_b
 \left[\epsilon^{\rm obs}_{{\rm t},a}\epsilon^{\rm obs}_{{\rm t},b}
 \pm\epsilon^{\rm obs}_{\times,a}\epsilon^{\rm obs}_{\times,b}\right] \Delta^{(ij)}_{ab}(\bar{\theta}) } 
 {\sum_{ab} w_a w_b (1+m_a)(1+m_b)  \Delta^{(ij)}_{ab}(\bar{\theta}) }\ ,
\end{equation}
where $\Delta^{(ij)}_{ab}(\bar{\theta})$ is a function that limits the sums to galaxy pairs of separation within the angular bin labelled by $\bar{\theta}$ and the tomographic bins $i$ and $j$. A galaxy indexed by $a$ is assigned a weight, $w_a$, based on the precision of its shear estimate. These weights are applied to the observed tangential and cross components of the ellipticity, $\epsilon^{\rm obs}_{\rm t}$ and $\epsilon^{\rm obs}_{\rm \times}$. Finally, the signal is normalised using the denominator, which takes the measurement biases into account, through an averaged multiplicative bias correction, $m_a$. As the value of $m$ is noisy for a single galaxy, we apply its corresponding correction, averaged over all the galaxies in the (tomographic) sample as shown in \Eqt\eqref{eq:xipm_meaure}.
This calibration is needed to correct for residual biases such as the effect of noise on the shear estimates \citep{melchior_viola:2012}, detection biases \citep{fenech-conti/etal:2017, kannawadi/etal:2019} as well as blending of the images of galaxies \citep{hoekstra/etal:2015}.

The 2PCFs are linear combinations of the E and B-mode angular power spectra, $C_{\rm EE/BB}(\ell)$,
\begin{align}
\label{eq:xipmPower}
 \xi_\pm(\theta) &=\int_0^{\infty} \frac{\d \ell\, \ell}{2\pi} {\rm J}_{0/4}(\ell\theta) \left[C_{\rm EE}(\ell)\pm C_{\rm BB}(\ell)\right]\;,
\end{align}
with Bessel functions of the first kind, $\mathrm{J}_{0/4}$, as their weights\footnote{From here on we drop the redshift dependence of $\C(\ell)$ as that has no effect on the linear transformations that produce our two-point statistics.}. Since we do not expect a significant B-mode signal of cosmological origin,
we can use the significance of the B-modes measured in the data as a null test of residual systematics \citep[see for example][]{hoekstra:2004, kilbinger/etal:2013, asgari/etal:2017,Hikage18, asgari/etal:2019a,asgari_heymans:2019}. As a result, this mixing of modes makes $\xi_\pm$ unsuitable for systematic tests that utilise B-modes. 

The measured 2PCFs are binned in $\theta$, and we match the binning procedure in their theoretical predictions. The theoretical value of $\xi_\pm$ has been estimated using an effective $\theta$ in previous cosmic shear analyses \citep[][]{hildebrandt/etal:2017, troxel/etal:2018b,troxel/etal:2018a}, although this approximation can result in biases \citep[see appendix A of][]{asgari/etal:2019a}. As the number of pairs of galaxies contributing to $\xi_\pm$ increases with angular separation, the correct method to bin the theory vector is to perform a weighted integral over $\xi_\pm(\theta)$ and include the effective number of pairs of galaxies, $N_{\rm pair}$, as the weight.
We employ $N_{\rm pair}$ as measured from the data, which includes all survey effects (see appendix C.3 of J20). The method used to measure the covariance matrix of $\xi_\pm$ is described in appendix E of J20.

\subsection{COSEBIs}
\label{sec:cosebis}

The complete orthogonal sets of E/B-integrals \citep[][]{SEK10} are two-point statistics defined on a finite angular range that cleanly separate all well-defined E and B-modes within that range, emoving any ambiguous modes that cannot be uniquely identified as E or B. COSEBIs form discrete values and can be measured through 2PCFs,
\begin{align}
\label{eq:COSEBIsReal}
 E_n &= \frac{1}{2} \int_{\theta_{\rm min}}^{\theta_{\rm max}}
 \d\theta\,\theta\: 
 [T_{+n}(\theta)\,\xi_+(\theta) +
 T_{-n}(\theta)\,\xi_-(\theta)]\;, \\ \nonumber
 B_n &= \frac{1}{2} \int_{\theta_{\rm min}}^{\theta_{\rm
     max}}\d\theta\,\theta\: 
 [T_{+n}(\theta)\,\xi_+(\theta) -
 T_{-n}(\theta)\,\xi_-(\theta)]\;,
\end{align} 
where $T_{\pm n}(\theta)$ are filter functions defined for a given angular range, i.e. between $\theta_{\rm min}$ and $\theta_{\rm max}$. 
\cite{SEK10} introduced two families of COSEBIs, linear-COSEBIs for which $T_\pm(\theta)$ have nearly linearly spaced oscillations, and also log-COSEBIs with nearly logarithmically spaced oscillations. These COSEBI $n$-modes are numbered with natural numbers, $n$, starting from 1, and their filters have $n+1$ roots in their range of support (see figure 1 of \citealt{asgari/etal:2019a}). 
Log-COSEBIs provide a more efficient data compression in that the first few $n$-modes are sufficient to essentially capture the full cosmological information \citep{asgari/etal:2012}. Therefore, we employ log-COSEBIs, which were also used for previous data analyses \citep[see for example][]{kilbinger/etal:2013,huff/etal:2014,asgari/etal:2020a}.

In practice, to measure COSEBIs accurately, we bin the 2PCFs into fine $\theta$-bins before applying the linear transformation in \Eqt\eqref{eq:COSEBIsReal}. The accuracy of the measured COSEBIs depends on the binning of the 2PCFs as well as the $n$-mode considered. For higher $n$-modes we need a larger number of bins. As our analysis employs log-COSEBIs, we adopt logarithmic binning of the 2PCFs, which results in a lower number of bins to reach the same accuracy requirement than for a linear binning approach. Previously we used linear binning with a million $\theta$-bins \citep{asgari/etal:2020a}. With log-binning we can reduce this number to $4000$ $\theta$-bins to reach the same level of accuracy (better than $0.03\%$), resulting in a speed gain in the measurement \citep[see appendix A of][for accuracy tests]{asgari/etal:2017}.

The theoretical prediction for COSEBIs can be found through
\begin{align}
\label{eq:EnBnFourier}
E_n &= \int_0^{\infty}
\frac{\d\ell\,\ell}{2\pi}C_{\mathrm{EE}}(\ell)\,W_n(\ell)\;,\\ \nonumber
B_n &= \int_0^{\infty}
\frac{\d\ell\,\ell}{2\pi}C_{\mathrm{BB}}(\ell)\,W_n(\ell)\;,
\end{align} 
where the weight functions, $W_n(\ell)$, are Hankel transforms of $T_\pm(\theta)$  \citep[see figure 2 in][]{asgari/etal:2012},
\begin{align}
\label{eq:Wn}
W_n(\ell) & =  \int_{\theta_{\rm{min}}}^{\theta_{\rm{max}}}\d\theta\:
\theta\:T_{+n}(\theta) \rm{J}_0(\ell\theta)\;, \nonumber \\ 
& = \int_{\theta_{\rm{min}}}^{\theta_{\rm{max}}}\d\theta\:
\theta\:T_{-n} (\theta) \rm{J}_4(\ell\theta)\;.
\end{align} 
These weight functions are highly oscillatory, but as we will see in \sect\ref{sec:compare}, they limit the effective range of support of COSEBIs in $\ell$, and as a result they allow for more control over which scales enter the analysis. To measure the covariance matrix of COSEBIs, we follow the formalism in appendix A of \cite{asgari/etal:2020a}, but with the updated $N_{\rm pair}$ and ellipticity dispersion, $\sigma_{\epsilon}$, definitions that are given in appendix C of J20. We also include the in-survey non-Gaussian term that was neglected in \cite{asgari/etal:2020a}, although that term has a negligible effect on the analysis \citep[][]{barreira/etal:2018}.

\subsection{Band powers}
\label{sec:bp}

The formalism for band power spectra is described in detail in J20 \citep[see also][]{schneider/etal:2002a,vanuitert/etal:2017}. Band powers are essentially binned angular power spectra, but estimated through 2PCFs. We can measure band powers, ${\mathcal C}_{{\rm E/B},l}$, via
\begin{equation}
\label{eq:bp_xipm}
{\mathcal C}_{{\rm E/B},l} = \frac{\pi}{{\mathcal N}_l}\; \int_0^\infty \d \theta\, \theta\; T(\theta) \left[\xi_+(\theta)\; g_+^l(\theta) \pm \xi_-(\theta)\; g_-^l(\theta)\right]\;, 
\end{equation}
where the normalisation, ${\mathcal N}_l$, is defined such that the band powers trace $\ell^2 C(\ell)$ at the logarithmic centre of the bin,
\begin{equation}
\label{eq:normalisation}
{\mathcal N}_l = \ln(\ell_{{\rm up},l})-\ln(\ell_{{\rm lo},l})\;,
\end{equation}
with $\ell_{{\rm up},l}$ and $\ell_{{\rm lo},l}$ defining the edges of the desired top-hat function for the bin indexed by $l$. The filter functions, $g_\pm^l(\theta)$, are given in equation 23 of J20. 
We note that the integral in \Eqt\eqref{eq:bp_xipm} is defined over an infinite range of $\theta$. In practice we cannot measure the 2PCFs over all angular distances, therefore, we need to truncate the integral at both ends. As a result it is impossible to produce perfect top-hat functions in Fourier space \citep{asgari_schneider:2015}. 
To reduce the ringing effect caused by the limited range of the 2PCFs we introduced apodisation in the selection function, $T(\theta)$, that softens the edges of the top hat (see equation 22 of J20). We note that $T(\theta)$ in \Eqt\eqref{eq:bp_xipm} and $T_{\pm n}(\theta)$ in \Eqt\eqref{eq:COSEBIsReal} are unrelated.

The relation between the band powers and the underlying angular power spectra is given by,
\begin{align}
\label{eq:bp}
{\mathcal C}_{{\rm E},l} &=  \frac{1}{2 {\mathcal N}_l} \int_0^\infty \d
\ell\, \ell \left[ W^l_{\rm EE}(\ell)\; C_{\rm
    EE}(\ell) + W^l_{\rm EB}(\ell)\; C_{\rm BB}(\ell) \right]\;, \\ \nonumber
{\mathcal C}_{{\rm B},l} &=  \frac{1}{2 {\mathcal N}_l} \int_0^\infty \d
\ell\, \ell \left[ W^l_{\rm BE}(\ell)\; C_{\rm
    EE}(\ell) + W^l_{\rm BB}(\ell)\; C_{\rm BB}(\ell)\right]\;,
\end{align}
where
\begin{align}
\label{eq:bpw}
W^l_{\rm EE}(\ell) &= W^l_{\rm BB}(\ell)  \\ \nonumber
&=\!\! \int_{0}^{\infty} \!\! \d \theta\, \theta\; T(\theta) \left[{ {\rm J}_0(\ell \theta)\; g_+^l(\theta) + {\rm J}_4(\ell \theta)\; g_-^l(\theta) }\right]\;, \\ \nonumber
W^l_{\rm EB}(\ell)& = W^l_{\rm BE}(\ell)    \\ \nonumber
&=\!\! \int_{0}^{\infty} \!\! \d \theta\, \theta\; T(\theta) \left[{ {\rm J}_0(\ell \theta)\; g_+^l(\theta) - {\rm J}_4(\ell \theta)\; g_-^l(\theta) }\right]\;.
\end{align}
These weight functions are no longer top hat functions (see \fig\ref{fig:compare}), however they allow for the correct transformation of the angular power spectra to band powers that can be compared to the measured values from \Eqt\eqref{eq:bp_xipm}. Similar to COSEBIs, we need to bin the 2PCFs before measuring the band powers. In this case we find that with 300 logarithmic $\theta$-bins in $[0\farcm5, 300']$ (with the binning extended on either side to allow for the apodisation) we can reach better than percent level accuracy, which is sufficient for the analysis of KiDS-1000 data. We define 8 logarithmically-spaced band power filters within the $\ell$-range of $100$ to $1500$. 
The covariance matrix of band powers is estimated by integrating over the covariance matrix of 2PCFs as described in appendix E.3 of J20. 

\subsection{Scale sensitivity of the two-point statistics}
\label{sec:compare}

\begin{figure}
   \begin{center}
     \begin{tabular}{c}
      \includegraphics[width=\hsize]{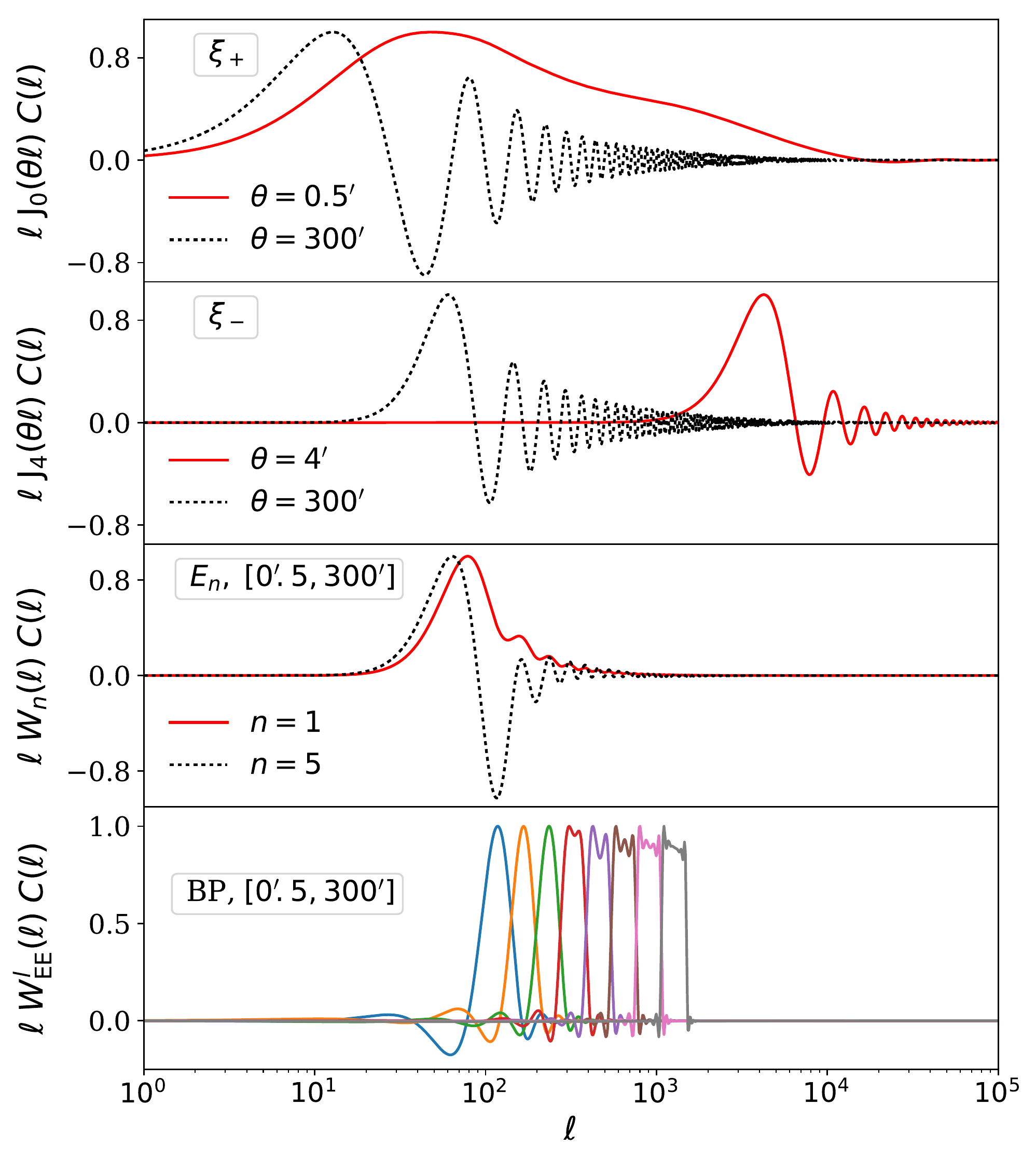}
     \end{tabular}
   \end{center}
     \caption{Integrands of the transformation between the angular power spectrum and 2PCFs (\Eqt.\ref{eq:xipmPower}), COSEBIs (\Eqt\ref{eq:EnBnFourier}) and band powers (\Eqt\ref{eq:bp}). All integrands are normalised by their maximum value.  $\xi_\pm$ results are shown for the maximum and minimum angular separations that are used in our analysis. For COSEBIs we chose $n=1$ and $n=5$, showing the range of $n$-modes that we consider. For band powers we show all 8 bins. COSEBIs are defined on the angular range of $[0\farcm5'300']$, while the band powers go beyond the indicated range to account for apodisation in their selection function, $T(\theta)$. We define 8 band power filters logarithmically spaced between $\ell=100$ and $\ell=1500$. }
     \label{fig:compare}
 \end{figure} 

 All two-point statistics considered here can be measured using linear combinations of finely binned 2PCFs. We set the full angular range for the measured 2PCFs to $\theta\in[0\farcm5,300']$ following the previous analysis of KiDS data, based on the extent of the survey and its resolution \citep[][]{hildebrandt/etal:2017}. \cite{hildebrandt/etal:2019}  applied extra $\theta$ cuts to their data vector. We apply their lower scale cut on $\xi_-$ to remove all $\theta<4'$, since $\xi_-$ for these scales are very sensitive to small physical scales where modelling becomes challenging. For COSEBIs and band powers, however, we use the full range of $\theta$-scales available. 

Our three sets of summary statistics place varying weights on different scales. Thus we do not expect them to have the same response to scale-dependent effects. Figure\;\ref{fig:compare} compares the integrands of these statistics, over the range that is used in the analysis. All integrands are normalised by their maximum value. The top two panels show results for $\xi_+$ and $\xi_-$, for the smallest and largest $\theta$ values that we consider in the analysis. The third panel demonstrates the integrands for the first and the fifth COSEBIs modes, since we only use the first 5 $n$-modes in our cosmological analysis defined on an angular range of $[0\farcm5,300']$. The bottom panel belongs to band powers and shows all of the bands that we use.

The first feature that we can immediately see from \fig\ref{fig:compare}, is that both correlation functions show substantial sensitivity to $\ell>1500$. In contrast both COSEBIs and band powers are essentially insensitive to these scales. As a result we expect the 2PCFs to be more sensitive to baryon feedback which becomes more important at smaller physical scales. In addition, $\xi_+$ is sensitive to scales below $\ell$ of about 10. Contributions from these scales can produce non-Gaussian distributions due to the small number of large-scale modes that enter the survey. Figure 17 of J20 compares the distributions of $\xi_+$ and band powers in the \textsc{Salmo}\footnote{Speedy Acquisition of Lensing and Matter Observables} simulations, which contain all KiDS-1000 survey effects. We show results for COSEBIs using the same suite of simulations in \fig\ref{fig:dist}. A comparison of these figures shows that
 the probability distribution of $\xi_\pm(\theta)$ for the largest values of $\theta$ deviates from a Gaussian, while this is not the case for band powers and COSEBIs.  \cite{louca_sellentin:2020} also showed that the COSEBI likelihood is well approximated by a Gaussian for a survey such as KiDS.
For our fiducial analysis we employ the angular ranges shown in \fig\ref{fig:compare}. We test the $\xi_\pm$ results for a reduced angular range in \sect\ref{sec:nuisance} and find that with our setup the non-Gaussian $\theta$-bins have a negligible effect on the cosmological results. 
In \app\ref{app:stagetests} we compare these statistics and their impact on parameter estimation, the results of which are summarised in \sect\ref{sec:consist}.

\section{Data and analysis pipeline}

\label{sec:data}

We measure the three summary statistics described in \sect\ref{sec:method} using the KiDS-1000 data and analyse them with the KiDS Cosmology Analysis Pipeline, {\sc KCAP}\footnote{{\sc KCAP} will become public once the KiDS-1000 analysis papers are accepted. Early access can be granted to interested parties on request.}.  This pipeline is built on {\sc CosmoSIS} \citep{cosmosis}, a modular cosmological parameter estimation code.  The measurements of the 2PCFs are performed with {\sc TreeCorr} \citep{jarvis/etal:2004,treecorr}. We applied our main analysis on blinded data (see G20 for details) and chose one of the blinds to test the effect of systematics prior to unblinding. More details on the small number of additional analyses done after unblinding can be found in \app\ref{app:unblinding}.

\subsection{KiDS-1000 data}

\begin{figure}
   \begin{center}
     \begin{tabular}{c}
      \includegraphics[width=\hsize]{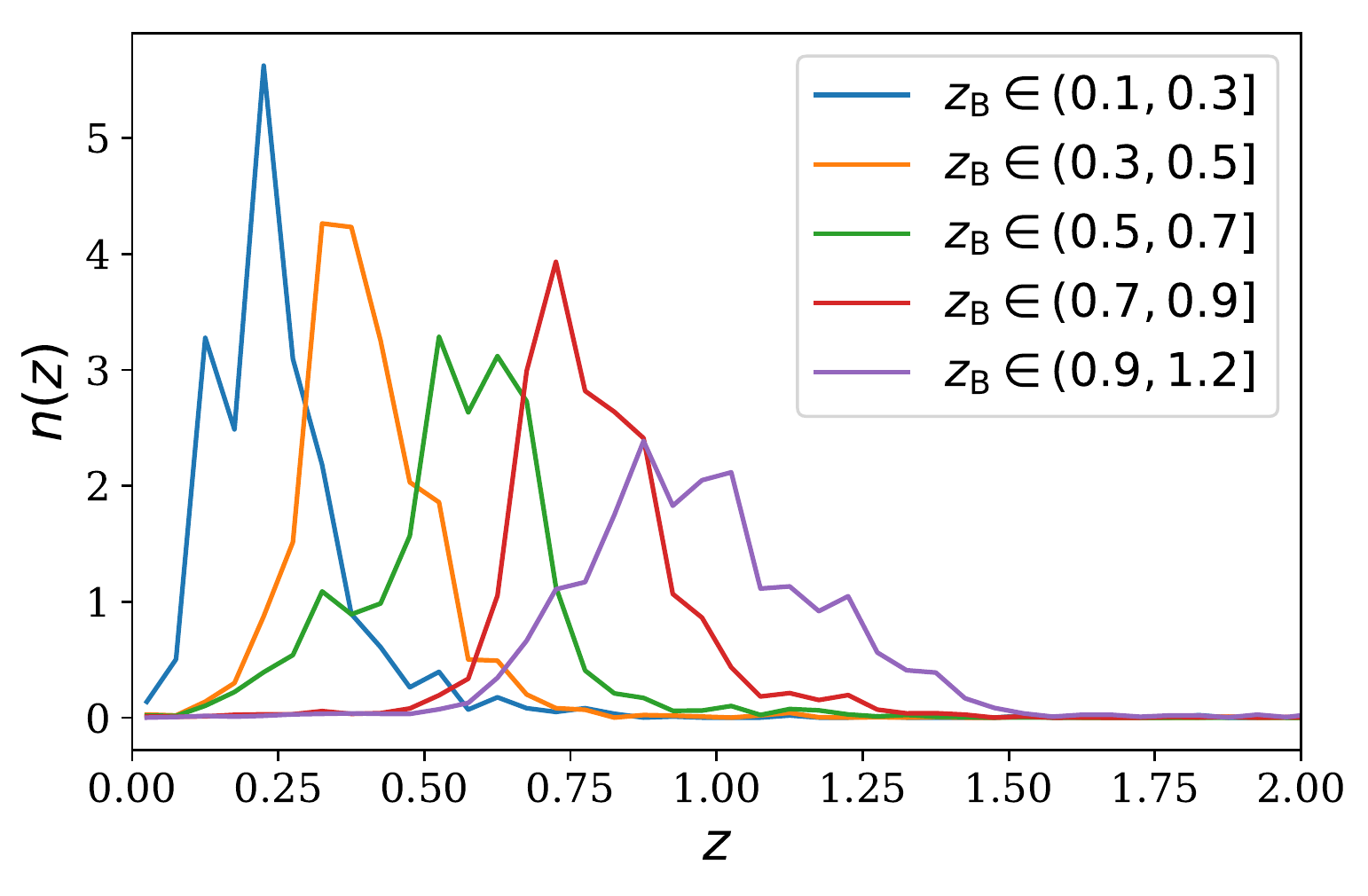}
     \end{tabular}
   \end{center}
     \caption{The redshift distribution of galaxies in five tomographic bins. The galaxies in each bin are selected based on their best-fitting photometric redshift, $z_{\rm B}$,  the range of which is shown in the legend. }
     \label{fig:photoz}
 \end{figure}

\begin{table*}
\centering
\caption{Data properties per tomographic redshift bin.}
\label{tab:dataprops}
\begin{tabular}{ c  c c c r r }
\hline
\hline \\[-2.2ex]
 Bin & $z_{\rm B}$ range & $n_{\rm eff}[{\rm arcmin}^{-2}]$ & $\sigma_{\epsilon,i}$ & $\Delta z=z_{\rm est}-z_{\rm true}$ & $m$ \quad\quad\quad \\ \\[-2.2ex] \hline  \\[-2.2ex]
1 & $0.1 < z_{\rm B} \leq 0.3$ &  $0.62$ & $0.27$ & $  0.000  \pm   0.0106 $ & $  -0.009 \pm    0.019 $ \\ \\[-2.2ex]
2 & $0.3 < z_{\rm B} \leq 0.5$ &  $1.18$ & $0.26$ & $  0.002  \pm   0.0113  $ & $  -0.011 \pm    0.020 $ \\ \\[-2.2ex]
3 & $0.5 < z_{\rm B} \leq 0.7$ &  $1.85$ & $0.27$ & $  0.013  \pm   0.0118  $ & $  -0.015 \pm    0.017 $ \\ \\[-2.2ex]
4 & $0.7 < z_{\rm B} \leq 0.9$ &  $1.26$ & $0.25$ & $  0.011  \pm   0.0087  $ & $   0.002 \pm    0.012 $ \\ \\[-2.2ex]
5 & $0.9 < z_{\rm B} \leq 1.2$ &  $1.31$  & $0.27$ & $ -0.006 \pm   0.0097 $ & $   0.007 \pm    0.010 $ \\ \\[-2.2ex]
\hline
\end{tabular}
\tablefoot{We list the index of the redshift bin, followed by the range of best-fitting photometric redshifts, $z_{\rm B}$, that divide galaxies into redshift bins. In the third column we show the effective number density, $n_{\rm eff}$, calculated with $A_{\rm eff}=777.4\;{\rm deg}^2$.  The values for the ellipticity dispersion per ellipticity component, $\sigma_{\epsilon,i}$, are presented in the fourth column. For explicit definitions of $n_{\rm eff}$ and $\sigma_{\epsilon,i}$ see appendix C of J20. The last two columns show the central values of calibration parameters, i.e. the shift in the mean of the redshift distributions, $\Delta z$, and the multiplicative shear bias, $m$, as well as their associated uncertainties. In the analysis the uncertainty on each of these calibration parameters is accounted for through their covariance matrices, since the values for the different tomographic bins are correlated.}
\end{table*}

The Kilo-Degree Survey (KiDS, \citealt{kuijken/etal:2015}, \citealt{dejong/etal:2015,dejong/etal:2017} and \citealt{kuijken/etal:2019})  is a public survey by the European Southern Observatory\footnote{Data products are made freely accessible through: \href{http://kids.strw.leidenuniv.nl/DR4}{{http://kids.strw.leidenuniv.nl/DR4} }}.
KiDS is a survey designed with weak lensing applications in mind, producing high-quality images with  VST-OmegaCAM. The primary images were taken in the $r$-band with a mean seeing of $0\farcs7$. In combination with infrared data from its partner survey, VIKING \citep[VISTA Kilo-degree INfrared Galaxy survey,][]{edge/etal:2013}, the observed galaxies have photometry in nine optical and near-infrared bands, $ugriZYJHK_{\rm s}$ \citep{wright/etal:2019}, allowing us to have a better estimate of their photometric redshifts compared to the four optical bands that KiDS observes \citep{hildebrandt/etal:2019}. 
We analyse the fourth KiDS data release \citep{kuijken/etal:2019}, named KiDS-1000 since it contains $1006\,{\rm deg}^{2}$ of images. After masking, the effective area of KiDS-1000 in the OmegaCAM pixel frame is $777.4\,{\rm deg}^{2}$. 

The KiDS data are processed with the {\sc theli} \citep{Erben13} and {\sc Astro-WISE} \citep{begeman/etal:2013} pipelines, and galaxy shear estimates are produced by \emph{lens}fit \citep{Miller13,fenech-conti/etal:2017}; for details see \cite{giblin/etal:2020} which also includes a series of null tests, showing that the impact we expect from known shear-related systematics detected in the data does not cause more than a $0.1\sigma$ shift in $S_8=\sigma_8(\Om/0.3)^{0.5}$ after calibration of multiplicative and global additive shear biases (see \app\ref{app:cterm} for the effect of this term on the two-point statistics). 

We perform a tomographic analysis of our cosmic shear data by dividing the galaxies based on their best-fitting photometric redshift, $z_{\rm B}$, into five tomographic bins. The $z_{\rm B}$ of each galaxy is estimated using the {\sc bpz} code \citep{bpz2000,bpz2004}. The redshift distribution of each tomographic bin is then calibrated using the self-organising map (SOM) method of \cite{wright/etal:2020a}. The SOM method organises galaxies into groups based on their nine-band photometry and finds matches within spectroscopic samples. Galaxies for which no matches are found are removed from the catalogue.
Following \cite{wright/etal:2020b}, we impose an extra quality requirement on our selection which removes galaxies with a $z_{\rm B}$ that is catastrophically different from the redshift of their matched spectroscopic sample (see equation 1 in H20b). 

The resulting catalogue  forms our \enquote{gold} sample for which redshift distributions with reliable mean redshifts can be obtained (see H20b for details of the selection criteria and accuracy tests of the redshift distributions). We note that a primary reason for the high accuracy of our redshift calibration is the nine-band photometry of our galaxy images. With those we can avoid
degeneracies of galaxy spectral energy distributions present in lower-dimensional colour spaces when calibrating the data with spectroscopic samples \citep{wright/etal:2020a}. Our calibration additionally benefits from dedicated KiDS-like observations of spectroscopic galaxy surveys beyond the KiDS footprint \citep{hildebrandt/etal:2019}. 

The means of the SOM redshift distributions are calibrated using KiDS-like mocks from the MICE2 simulations \citep{vdbusch20,fosalba/etal:2015,crocce/etal:2015,fosalba/etal:2015b,carretero/etal:2015,hoffmann/etal:2015}, these mocks are also used to determine the expected uncertainties on these means, which we incorporate into the inference via shift parameters for each redshift distribution. The redshift distributions of galaxies in each tomographic bin are shown in \fig\ref{fig:photoz} up to $z=2$. The full redshift distributions used in this analysis cover a range of $0\leq z\leq 6$ (see \fig\ref{fig:pofz_log} and \tab\ref{tab:nofz_stats}).
We validate our fiducial redshift distributions estimated with the SOM method in H20b using an alternative method that employs clustering cross-correlations with spectroscopic reference samples. 

The gold sample selection is repeated for all galaxies simulated in the image simulations of \cite{kannawadi/etal:2019}, which are then used to calibrate the shear estimates and estimate the uncertainty on the calibration parameters. This is done through an averaged multiplicative bias per redshift bin using \Eqt\eqref{eq:xipm_meaure}. The low-level contribution from the constant additive ellipticity bias is corrected in the catalogues as a global constant per tomographic bin and ellipticity component (see section 3.5.1 of G20 for details).  

In \tab\ref{tab:dataprops} we show the data properties that are relevant for covariance estimation, as well as the values of the calibration parameters. The $\Delta z$ parameters are defined as the difference between the mean of the estimated SOM distribution, $z_{\rm est}$, and the true redshift distribution of galaxies in the MICE2 mocks, $z_{\rm true}$, for a given redshift bin. We note that the effective area of the survey is relevant for the calculation of all the terms in the covariance matrix, except for the shape-noise only term. J20 found that for the cosmic variance (sample variance) term a larger effective area based on a {\sc Healpix} map with $N_{\rm side}=4096$ \citep{gorski/etal:2005}, provides a better match between the mock and theoretical covariances (see section 5.2 and appendix E of J20). Here we use this area for calculating the covariances matrices, although in \app\ref{app:pixelsize} we show that this choice has an insignificant effect on our analysis.

\subsection{Cosmological analysis pipeline}

\begin{table}
\centering
\caption{Fiducial sampling parameters and their priors.}
\label{tab:priors}
\begin{tabular}{ c  c  }
\hline
\hline\\[-2.2ex]
         Parameter                  &   prior                 \\ \hline\\[-2.2ex]
$S_8=\sigma_8(\Omega_{\rm m}/0.3)^{0.5}$                                &   $[0.1,1.3]$       \\ \\[-2.2ex]
$\omega_{\rm c}=\Omega_{\rm c} h^2$      &   $ [0.051, 0.255 ]$        \\ \\[-2.2ex]
$\omega_{\rm b}=\Omega_{\rm b} h^2$      &   $[0.019, 0.026]$         \\ \\[-2.2ex]
$h$                                    &   $ [0.64, 0.82]$       \\ \\[-2.2ex]
$n_{\rm s}$                       &   $[0.84,1.1]$        \\  \hline \\[-2.2ex]
$A_{\rm IA}$                     &   $ [-6,6]$            \\ \\[-2.2ex]
$A_{\rm bary}$                 &    $[2, 3.13]$                \\  \hline \\[-2.2ex]
$\delta_z$                      &    $\mathcal{N}(\vec{\mu},\tens{C})$       \\ \\[-2.2ex]
$\delta_{\rm c}$               &    $0\pm 2.3\times10^{-4}$              \\\\[-2.2ex]
\hline
\end{tabular}
\tablefoot{We vary five cosmological parameters assuming flat priors, with the ranges indicated in the second column. $\Omega_{\rm c}$ and $\Omega_{\rm b}$ are the density parameters for cold dark matter and baryonic matter, respectively. The dimensionless Hubble parameter is represented by $h$ and $n_{\rm s}$ is the spectral index of the primordial power spectrum. Two astrophysical nuisance parameters, $A_{\rm IA}$ and $A_{\rm bary}$, are allowed to vary over flat prior distributions. We also allow for freedom in the mean of the redshift distributions using five shift parameters, $\delta_z$, one per redshift bin. These parameters are correlated through their covariance matrix, $\tens{C}$. Their means $\vec{\mu}$ are fixed to the mean values for $\Delta z$ in \tab\ref{tab:dataprops}, where we also show the square root of the diagonals of $\tens{C}$. The $\delta_{\rm c}$ parameter is only applied to the $\xi_\pm$ chains, which mitigates the combined effect of constant additive ellipticity bias through a Gaussian prior centred on zero. For justification of prior ranges we refer to J20, G20 and H20b. }
\end{table}

For our cosmological analysis we assume a spatially flat $\Lambda$CDM model and infer the values of cosmological parameters through sampling of the likelihood with the {\sc MultiNest} sampler \citep{multinest}. We find the best-fitting values for each chain using the Nelder-Mead minimisation method \citep{neldmead:1965} implemented in {\sc SciPy}\footnote{We run the Nelder-Mead minimiser with the adaptive option, which is more reliable for higher dimensional and multi-modal problems. See \href{https://docs.scipy.org/doc/scipy/reference/optimize.minimize-neldermead.html}{docs.scipy.org/doc/scipy/reference/optimize.minimize-neldermead.html}.}, with the starting points taken from the  {\sc MultiNest} chains. We use this separate minimiser since the {\sc MultiNest} sampler is not optimised to find the best fitting point in the likelihood surface.

We calculate the linear matter power spectrum with {\sc camb} \citep{camb2000,camb12} and its non-linear evolution with {\sc HMCode} \citep{mead/etal:2015}. We also include the effect of the intrinsic alignment of galaxies through the non-linear alignment model \citep[NLA]{Bridle07}, before using the Limber approximation of \Eqt\eqref{eq:limber} to project the matter power spectrum along the line of sight and  obtain $\C(\ell)$. The $\C(\ell)$ are then transformed into $\xi_\pm$ (\Eqt\ref{eq:xipmPower}), COSEBIs (\Eqt\ref{eq:EnBnFourier}) and band powers (\Eqt\ref{eq:bp}), which are compared to their measured values, assuming Gaussian likelihoods with the analytic covariance model described in detail in J20. 

\tab\ref{tab:priors} lists the prior distributions of our sampled parameters. The cosmological model that we assume here contains five free parameters. We set the sum of the neutrino masses to a fixed value of $0.06\;$eV (\citealt{hildebrandt/etal:2019} showed that neutrinos have a negligible effect on cosmic shear analysis). 
In contrast to previous analyses of cosmic shear data, we sample over $S_8=\sigma_8(\Om/0.3)^{0.5}$.
Our primary results include constraints on $S_8$ and therefore we aim for an uninformative prior on this parameter. This choice is further justified in J20, by demonstrating that a flat prior over the amplitude of the primordial power spectrum $A_{\rm s}$ or its logarithm $\ln(10^{10} A_{\rm s})$ as employed in the previous analysis of KiDS and DES data produces informative priors for $S_8$.
Our constraints on the other cosmological parameters are mostly dominated by the prior, and we therefore set their prior range based on either the limitations in the theoretical modelling or previous observations (see section 6.1 of J20 for more details). Additionally, we allow for two astrophysical nuisance parameters, $A_{\rm IA}$ denoting the amplitude of the intrinsic alignment of galaxies and $A_{\rm bary}$, the baryon feedback parameter (by definition $A_{\rm bary}=3.13$ corresponds to a dark matter only case). 

We let the mean of the redshift distributions vary via a multivariate Gaussian prior for the five shift parameters shown in \tab\ref{tab:dataprops} (see figure 2 of H20b). For the analyses with $\xi_+$ we also allow for a $\delta_{\rm c}=\pm\sqrt{c_1^2+c_2^2}$ parameter which mitigates the uncertainty on the two additive ellipticity bias terms, $c_1$ and $c_2$, assuming that they are constants. The uncertainty on these parameters has a larger impact on $\xi_+$, while their effect on the other statistics is currently negligible (see \app\ref{app:cterm} for details on how to model this for the other statistics). We place a Gaussian prior on $\delta_{\rm c}$ centred at zero, since the catalogues have  already been corrected for a constant $c_i$. The width of the Gaussian is estimated using bootstrap samples of the data (see section 3.5.1 of G20 for details\footnote{G20 show that the 2D $c$-term is negligible. Therefore, we omit this term in our modelling.}). 

\section{Results}

\label{sec:results}

\begin{figure*}
   \begin{center}
     \begin{tabular}{c}
      \includegraphics[width=\hsize]{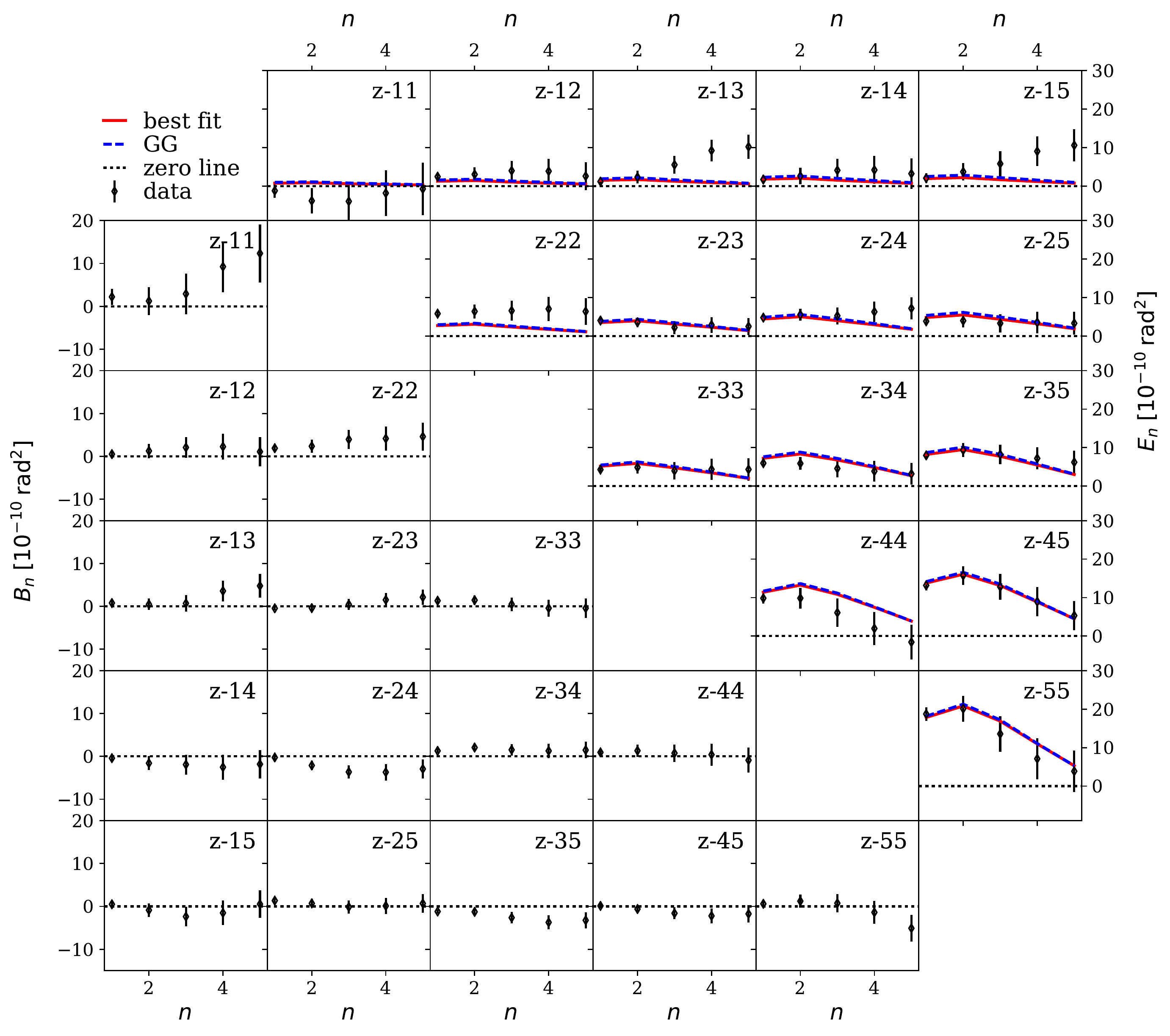}
     \end{tabular}
   \end{center}
     \caption{COSEBI measurements and their best fitting model (see \tab\ref{tab:bestfitall}). We show the best-fitting theoretical prediction with a red curve ($\chi^2_{\rm reduced} = 1.2$) and the gravitational lensing (GG) contribution with a blue dashed curve. A zero line is shown for reference (black dotted). The E-modes are shown in the top triangle , while the B-modes are shown in the bottom one. The predicted B-mode signal is zero.
We use the first five COSEBI E-modes in this analysis, as shown here. With the labels $z$-${ij}$ we show that redshift bins $i$ and $j$ are used for the corresponding panel. The COSEBIs modes are significantly correlated (see \fig\ref{fig:crosscov}), such that their goodness-of-fit cannot be established by eye. }
           \label{fig:COSEBIs}
 \end{figure*}

\begin{figure*}
   \begin{center}
     \begin{tabular}{c}
      \includegraphics[width=\hsize]{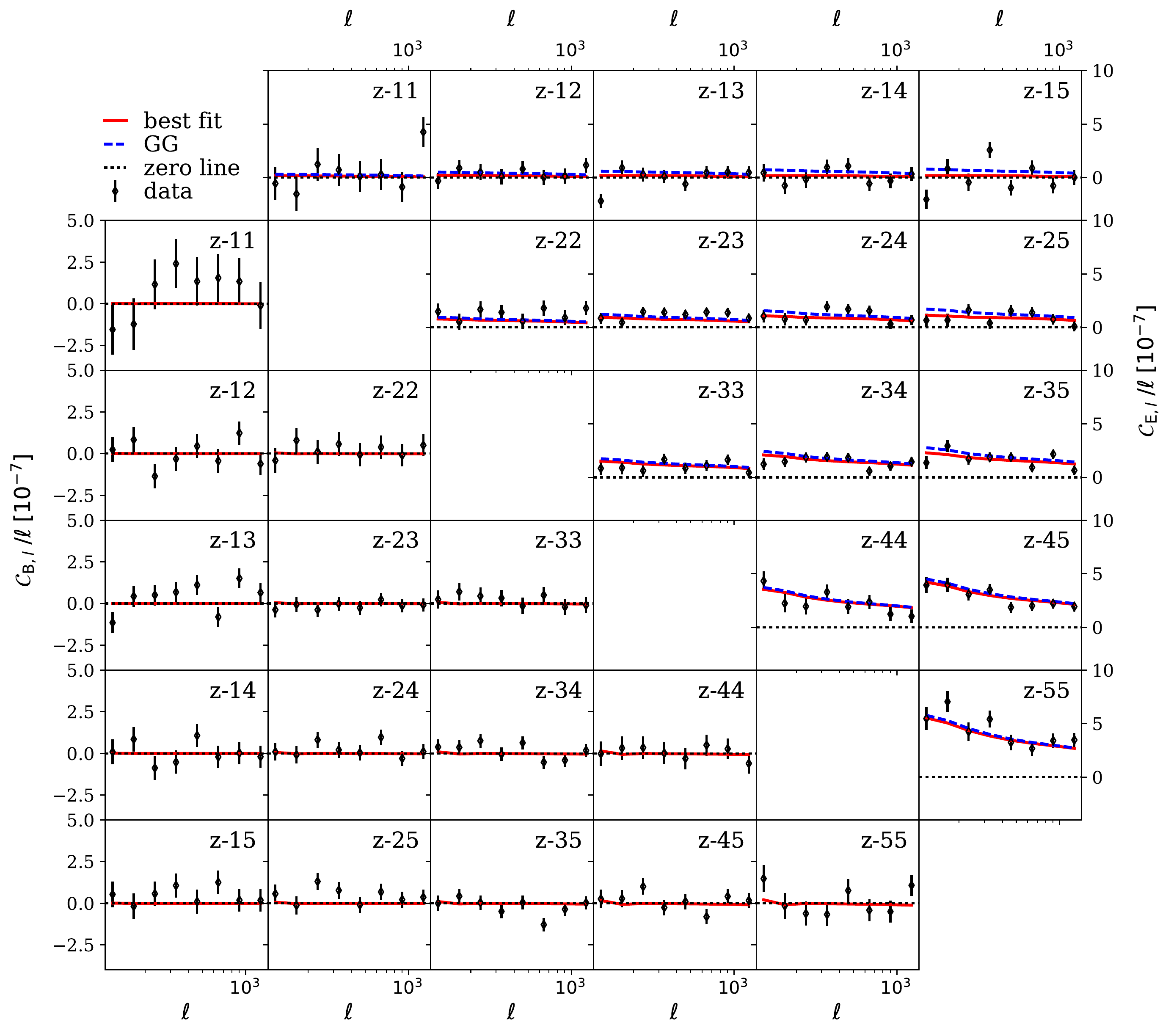}
     \end{tabular}
   \end{center}
     \caption{Band power measurements and their best fitting model (see \tab\ref{tab:bestfitall}). The red curves show the best fitting model fitted to the E-modes (top triangle, $\chi^2_{\rm reduced} = 1.3$) and the blue dashed curves show the  intrinsic alignment subtracted signal (GG). We also predict the B-modes (bottom triangle) using the same model, which results in small deviations from the zero line (black dotted, see \Eqt\ref{eq:bp}). We label the panels based on the pair of redshift bins used to measure the data. }
     \label{fig:bp}
 \end{figure*} 

\begin{figure*}
   \begin{center}
     \begin{tabular}{c}
      \includegraphics[width=\hsize]{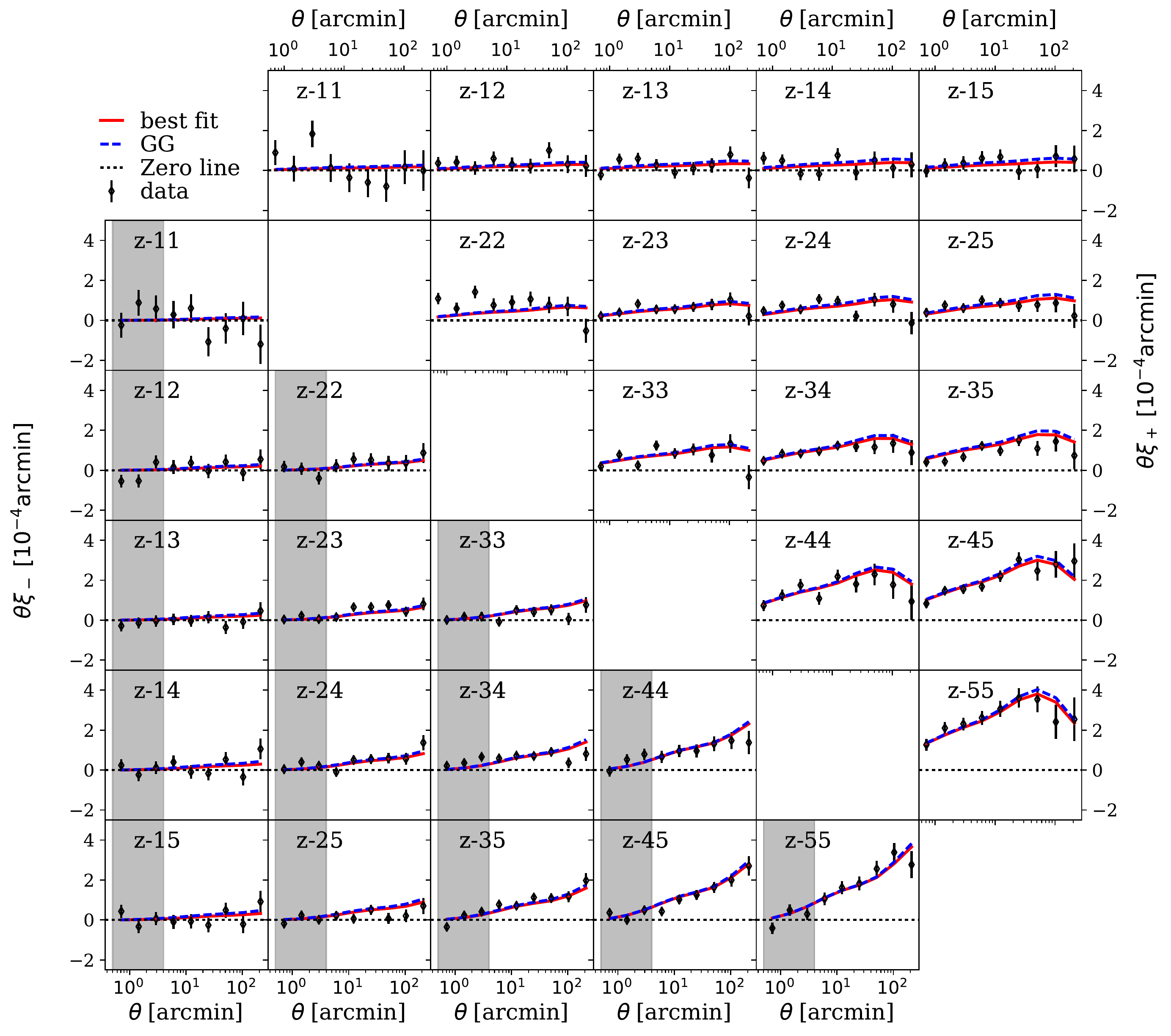}
     \end{tabular}
   \end{center}
     \caption{Measurements of the shear correlation functions. The best fitting curves are shown in red (see \tab\ref{tab:bestfitall}, $\chi^2_{\rm reduced} = 1.2$) and the gravitational-only (GG) signal is shown in blue (dashed). The top and bottom triangles show $\xi_+$ and $\xi_-$, respectively. The gray shaded region is excluded from the analysis, due to its sensitivity to small physical scale. Each panel is labelled based on the redshift bin pair that it represents.}
     \label{fig:xipm}
 \end{figure*}

In this section we present our cosmological results. We first report our headline constraints in \sect\ref{sec:fid}, and then we assess the sensitivity of our results to a range of systematic effects and the impact of omitting different tomographic bins in \sect\ref{sec:nuisance}. In \sect\ref{sec:consist} we summarise our internal consistency checks and in \sect\ref{sec:constst_external} compare our results with other cosmic shear surveys, and report the discrepancy between our results and the cosmic microwave background (CMB) results of the \textit{Planck} satellite. Throughout, we will use constraints from the \citet{Planck2018} TT, TE, EE + lowE temperature and polarisation power spectra, which extract cosmological information solely from the primary CMB anisotropies and are therefore independent of large-scale structure surveys\footnote{Except for the integrated Sachs–Wolfe effect which has a low-level impact on the CMB constraints.}.  

Before unblinding our data, we carried out a likelihood analysis on all blinds using a covariance matrix calculated from the sample properties of each blinded catalogue, which was generated assuming a fiducial cosmological model based on the parameter constraints from \cite{troster/etal:2020a} who analysed the third KiDS data release (KV450) in combination with Baryon Oscillation Spectroscopic Survey clustering data \citep[BOSS data release 12,][]{alam/etal:2017}. After unblinding, we updated the cosmological model  in our covariance calculation to use the results from the combined KiDS-1000 and galaxy clustering analysis of \cite{heymans/etal:2020} and repeated the inference process on the real data. This iterative approach for the covariance is advocated in J20. As the best-fitting parameter values in \cite{troster/etal:2020a}, \cite{heymans/etal:2020}, and our cosmic shear analysis are all very close, we only perform a single iteration that is then used for both the cosmic shear only and combined probe analysis of the KiDS-1000 data. This iteration has a negligible effect on our results. While our fiducial results and the consistency test with \textit{Planck} are based on the most accurate and updated covariance model, the internal consistency tests and the nuisance parameter sensitivity analyses, which we completed before unblinding employ the original covariance matrix (see \app\ref{app:unblinding} for details). 

\subsection{Fiducial results}
\label{sec:fid}

In \figs\ref{fig:COSEBIs}, \ref{fig:bp} and \ref{fig:xipm} we show the data vectors and their corresponding predictions by the best-fitting model\footnote{We note that in all cases the data is reported at discrete points, as a result of binning for 2PCFs and band powers, or by definition in the case of COSEBIs. Hence, the theory values are also discrete, although connected to each other for visual guidance.} 
for COSEBIs, band powers and shear correlation functions, respectively.
Each panel is labelled according to the pair of tomographic redshift bins used to measure the data. 
The red curves show the best-fitting predictions for each statistic
which are the sums of the gravitational lensing-only signal and
the intrinsic alignment terms (see \Eqt\ref{eq:c_ee}). 
The signal without the intrinsic alignments is presented by the blue dashed curves (GG).
The top sections in \figs\ref{fig:COSEBIs} and \ref{fig:bp} show the E-modes, while the bottom ones display the B-modes. 
In \fig\ref{fig:xipm} the top and bottom triangles show $\xi_\pm$ and the data points in the shaded regions are excluded from the cosmological analysis, due to their increased sensitivity to smaller physical scales (see \fig\ref{fig:compare} and section 5.1 of \citealt{hildebrandt/etal:2019}).

In all three figures we see that the intrinsic alignments of galaxies have the largest effect on the combinations of high- and low-redshift bins, most prominently z-15. The intrinsic alignment signal is dominated by the gravitational-intrinsic (GI) correlations, especially for pairs of tomographic bins where overlap in redshift is minimal, which produces anti-correlations for positive values of $A_{\rm IA}$. The intrinsic-intrinsic correlations (II) are mostly sub-dominant. The best-fitting value for $A_{\rm IA}$ is in all cases positive (see \tab\ref{tab:bestfitall}), resulting in a combined signal that is lower than the pure gravitational lensing term. 

In \fig\ref{fig:bp} we show the theoretical prediction for the band power B-modes, although these data points are not used in the analysis. The E/B-mode mixing in the band powers is small; nevertheless, it becomes visible at low angular frequencies in the higher-redshift bin combinations, where the E-mode signal is more significant (see \Eqt\ref{eq:bp}).  We find that the B-modes are consistent with zero ($p$-value = $0.4$).

We used the first five COSEBI E-modes for our cosmological analysis and therefore only display them in \fig\ref{fig:COSEBIs} \citep[adding more modes has a negligible impact on the constraints, e.g. see][]{asgari/etal:2020a}. G20, however, used both the first 5 and 20 COSEBIs B-modes to test the level of residual systematics in the data, which they found to be consistent with zero in both cases ($p$-value = $0.04$ and $0.38$, respectively). As adjacent COSEBI modes are highly correlated (see for example \fig\ref{fig:crosscov}), we caution the reader against a visual inspection of the goodness-of-fit of the model to the data. 

In \tab\ref{tab:bestfit} we report the goodness-of-fit of our best-fitting models (corresponding to the maximum of the full posterior), along with point estimates for the best-fitting values of $S_8$. We estimate the degrees of freedom for our data using the effective number of model parameters, $N_{\Theta}=4.5$ (see section 6.3 of J20). This value was obtained for a mock cosmic shear analysis very similar to ours by fitting a $\chi^2$ distribution to a histogram of minimum $\chi^2$ values from best fits to 500 mock data vectors. The number of varied parameters (12 for COSEBIs and band powers, 13 for 2PCFs, see \tab\ref{tab:priors}) is substantially larger than $N_{\Theta}$, which can have a significant effect on the goodness-of-fit estimates of the model, especially when the data vector is small.
Despite the differences between these two-point statistics, we expect them to have a similar sensitivity to cosmological parameters and therefore employ the same $N_{\Theta}$ for all of them.
We find acceptable goodness-of-fit for all three summary statistics with $p$-values (probability to exceed the given $\chi^2$) ranging from $0.16$ (COSEBIs) to $0.01$ (band powers). 

In the last column of \tab\ref{tab:bestfit} we show the peak of the marginal distribution of $S_8$ and its credible region derived from the highest posterior density of the marginal distribution. 
As shown in J20, section 6.4, this estimate can be shifted with regards to the true value of the cosmological parameters. It was therefore proposed to additionally report the maximum a posteriori (MAP) estimate and an associated credible interval using the projected joint highest posterior density, PJ-HPD, which ensures that the MAP value is within the credible region and in the case of a one-dimensional posterior reduces to the marginal credible region. We show the MAP and PJ-HPD in the fifth column of \tab\ref{tab:bestfit}. 
The best fit values for all parameters are shown in \tab\ref{tab:bestfitall}. The maximum marginal values are almost identical to the MAP in the case of $S_8$, but can in principle differ more substantially for other parameters.
The $p$-values for band powers and 2PCFs are considerably lower than for COSEBIs; however, since their best-fitting values are very similar, we conclude that this is a result of the noise realisation or low-level systematics that affect 2PCFs and band powers, but do not mimic a cosmological signal.

\begin{figure*}
   \begin{center}
     \begin{tabular}{c}
      \includegraphics[height=0.47\hsize]{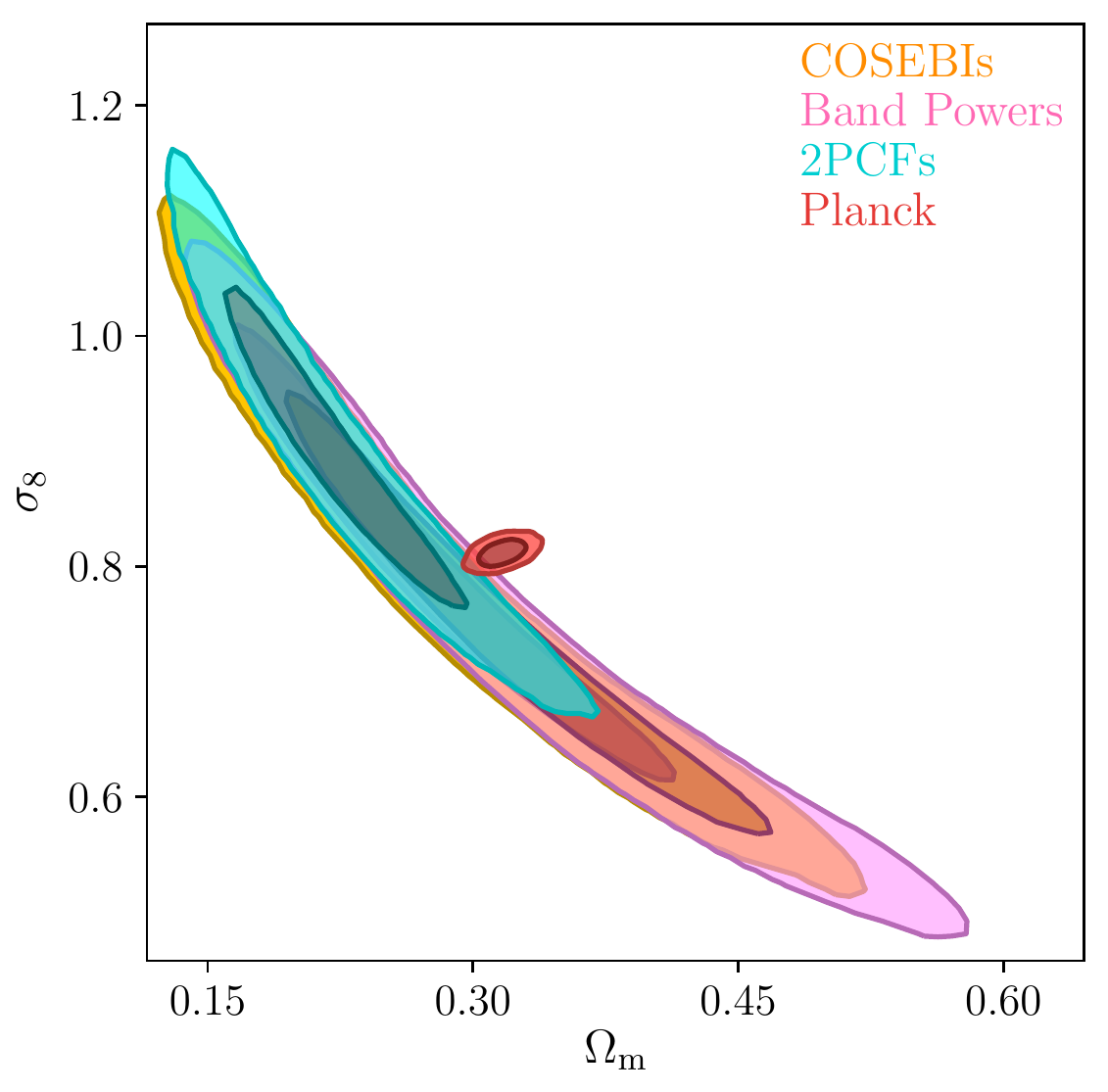}
      \includegraphics[height=0.47\hsize]{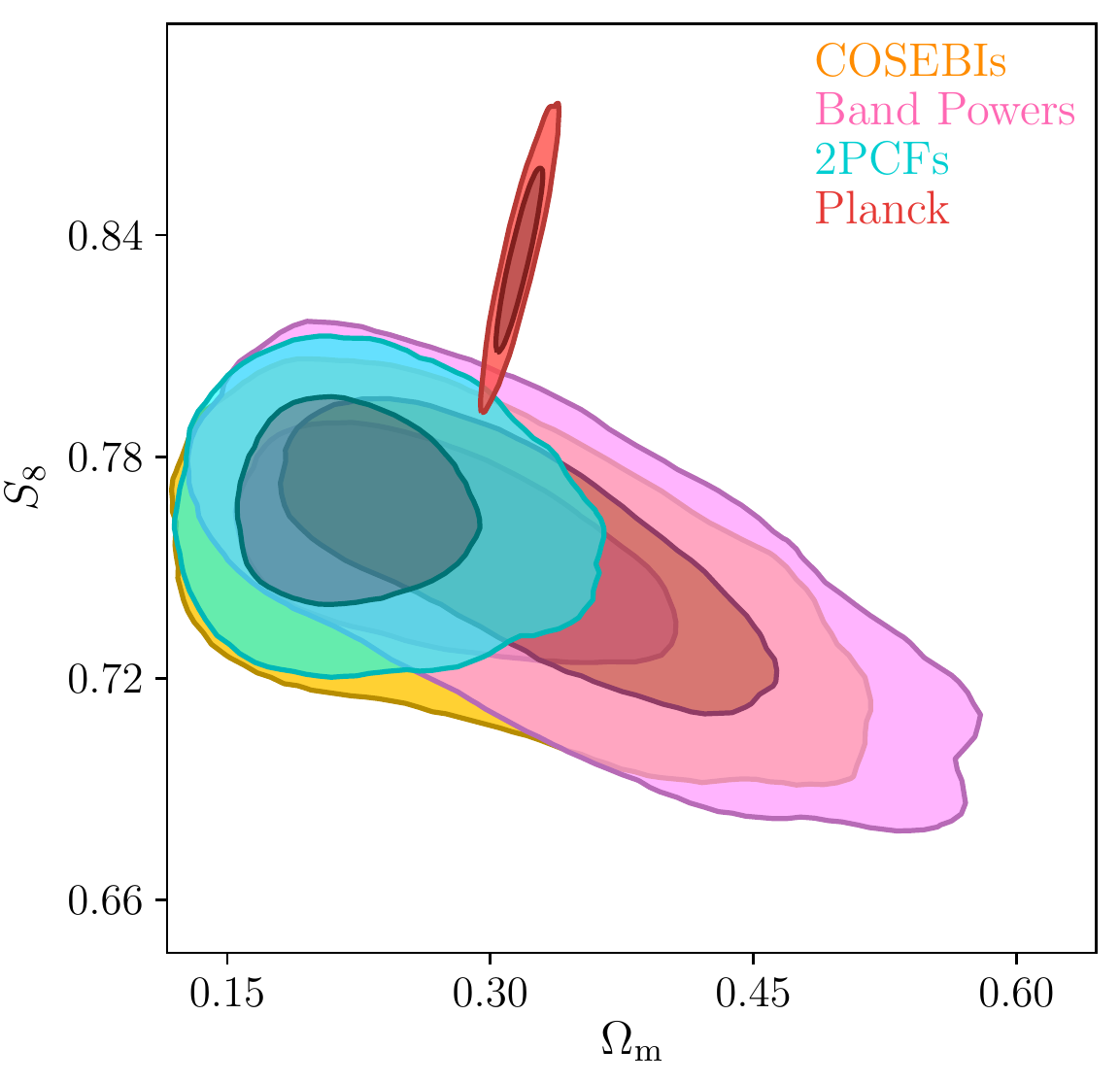}
     \end{tabular}
   \end{center}
     \caption{Marginalised constraints for the joint distributions of $\sigma_8$ and $\Om$ (\textit{left}), as well as $S_8$ and $\Om$ (\textit{right}). The $68\%$ and $95\%$ credible regions are shown for COSEBIs (orange), band powers (pink) and the 2PCFs (cyan). \textit{Planck} (2018, TT,TE,EE+lowE) results are shown in red.}
     \label{fig:sig8Om}
 \end{figure*}

Cosmic shear results are usually shown in terms of $\sigma_8$ and $\Om$, or $S_8$ and $\Om$. In \fig\ref{fig:sig8Om} we show our results for these parameters and compare them to the \textit{Planck} results. 
In the left panel we see that the constraints from these three statistics move along the degeneracy direction of $\sigma_8$ and $\Om$; however, they show good agreement in the value of $S_8$ as we saw in \tab\ref{tab:bestfit}. This movement is expected and will depend on the noise realisation in conjunction with the weighting of the data. In \fig\ref{fig:compare} we saw that our three sets of statistics show varying sensitivities to different angular scales. Hence, we can obtain different parameter constraints given the same noise realisation. We discuss this further and show mock data results in \app\ref{app:stagetests}. The left panel of \fig\ref{fig:sig8Om} shows that the extent of the $\xi_\pm$ contours appears smaller  than that of the other statistics. This is because the posterior is truncated at low $\Om$ by the prior. We also see in \tab\ref{tab:bestfit} that the constraints from $\xi_\pm$ for $S_8$ are tighter than those for both COSEBIs and band powers, whereas we would have expected similar constraining power for these three statistics. The right-hand panel of \fig\ref{fig:sig8Om} illustrates that the $\xi_\pm$ contours are horizontal  in $\Om$ and $S_8$, while the marginal posterior for COSEBIs and especially for band powers is tilted, showing that $S_8$ is not perpendicular to the degeneracy between $\sigma_8$ and $\Om$ for the latter two statistics.

The current established definition for $S_8$ is $\sigma_8(\Om/0.3)^\alpha$, with $\alpha=0.5$. Previously \citep[see for example][]{kilbinger/etal:2013}, the value of $\alpha$ was fitted to the contours, to find the tightest constraints from the data. As \fig\ref{fig:sig8Om} clearly shows, $\alpha=0.5$ does not provide an optimal description for the $\sigma_8$-$\Om$ degeneracy of either COSEBIs or band powers. In general, the value of $\alpha$ depends on the weighting of the angular scales entering the analysis, which probe different physical scales for different redshifts. In order to avoid confusion, we keep the established definition of $S_8$ with $\alpha=0.5$, but also include results for
\begin{equation}
\label{eq:Sigma8def}
\Sigma_8 := \sigma_8\, (\Om/0.3)^\alpha\,,
\end{equation}
where $\alpha$ is fitted to the contours. In \app\ref{app:extra} we describe our fitting method and show contours for $\Sigma_8$ and $\Om$ (see \fig\ref{fig:Sig8}).

In \tab\ref{tab:bestfitSig8} we present best-fitting values for $\alpha$ and constraints for its corresponding $\Sigma_8$. As expected, $\alpha \approx 0.5$ for the 2PCFs, i.e. $S_8$ remains a good summary parameter for this composition of the data vector. For COSEBIs and band powers we find $\alpha=0.54$ and $\alpha=0.58$, respectively, showing that they have a significantly different degeneracy to what is captured with $S_8$\footnote{The fit error for $\alpha$ is about $10^{-3}$.}.
Here we see that the sizes of the $\Sigma_8$ credible intervals for the different statistics are much closer to each other compared to the $S_8$ constraints in \tab\ref{tab:bestfit}. The constraints from $\xi_\pm$ are still slightly tighter. We expect this to occur when the noise realisation pushes the contours closer to the edges of the prior region, especially since the halo model used for predicting the matter power spectrum is not calibrated for very high and low values of $\sigma_8$ and $\Om$ and therefore becomes less likely to match the data. 
The standard deviation of the best-fitting $\Sigma_8$ for COSEBIs is $0.019$, for band powers it is $0.020$ and for 2PCFs it is $0.018$. We note that their central values cannot be directly compared, unless $\Om$ is fixed to $0.3$.

\begin{table*}
\centering
\caption{Goodness of fit and $S_8$ constraints.}
\label{tab:bestfit}
\begin{tabular}{c c c c c c}
\hline\hline\\[-2.2ex]
           & $\chi^2$ &       DoF & $p$-value &  $S_8$, best fit + PJ-HPD &     $S_8$, max + Marginal \\
\hline\\[-2ex]
   COSEBIs &     82.2 &  $75-4.5$ &     0.160 & $0.759^{+0.024}_{-0.021}$ & $0.758^{+0.017}_{-0.026}$ \\[0.8ex]
Band Power &    152.1 & $120-4.5$ &     0.013 & $0.760^{+0.016}_{-0.038}$ & $0.761^{+0.021}_{-0.033}$ \\[0.8ex]
     2PCFs &    260.3 & $225-4.5$ &     0.034 & $0.764^{+0.018}_{-0.017}$ & $0.765^{+0.019}_{-0.017}$ \\[0.4ex]
\hline
\end{tabular}
\tablefoot{$\chi^2$ and $p$-values (probability to exceed the given $\chi^2$ value) for the best-fitting parameters, given the effective number of degrees of freedom (DoF). The effective number of parameters is estimated using a $\chi^2$ fitted to results of mock data analysis. The first column shows which statistic is used.  In the fifth column we show the multivariate maximum posterior (MAP) for $S_8=\sigma_8(\Om/0.3)^{0.5}$ and its $68\%$ credible interval (CI) calculated using its projected joint highest posterior density (PJ-HPD). In the rightmost column we show the peak of the marginal distribution of $S_8$ and its associated $68\%$ credible interval.}
\end{table*}

\begin{table*}
\centering
\caption{Best-fit $\Sigma_8$ and $\Om$--$\sigma_8$ degeneracy line.}
\label{tab:bestfitSig8}
\begin{tabular}{c c c c}
\hline\hline\\[-2.2ex]
           & fitted $\alpha$ & $\Sigma_8$, best fit + PJ-HPD & $\Sigma_8$, max + Marginal \\
\hline\\[-2ex]
   COSEBIs &            0.54 &     $0.753^{+0.026}_{-0.016}$ &  $0.752^{+0.017}_{-0.021}$ \\[0.8ex]
Band Power &            0.58 &     $0.765^{+0.018}_{-0.024}$ &  $0.756^{+0.020}_{-0.020}$ \\[0.8ex]
     2PCFs &            0.51 &     $0.762^{+0.018}_{-0.017}$ &  $0.763^{+0.019}_{-0.017}$ \\[0.4ex]
\hline
\end{tabular}
\tablefoot{$\Sigma_8=\sigma_8(\Om/0.3)^\alpha$ values with fitted $\alpha$ to the $\sigma_8$ and $\Om$ posterior samples for each set of statistics. The second column shows the best-fitting $\alpha$, the third shows the best-fitting $\Sigma_8$ for that $\alpha$ and its credible interval PJ-HPD. The last column shows the maximum and 1$\sigma$ region around it for the marginal distribution of $\Sigma_8$. We note that the values of $\Sigma_8$ between different statistics cannot be directly compared with each other, since they correspond to different values of $\alpha$. }
\end{table*}

With our cosmic shear data we can put a tight constraint on the $\Sigma_8$ parameter, but with the exception of the intrinsic alignment amplitude $A_{\rm IA}$, we are largely prior-dominated for the remainder of the sampled parameters (see \tab\ref{tab:priors}). This is also reflected in the effective number of parameters that we record in \tab\ref{tab:bestfit}. Nevertheless, we show results for other parameter combinations in \app\ref{app:extra}.

\subsection{Impact of nuisance parameters and data divisions}
\label{sec:nuisance}

\begin{figure*}
   \begin{center}
     \begin{tabular}{c}
      \includegraphics[width=\hsize]{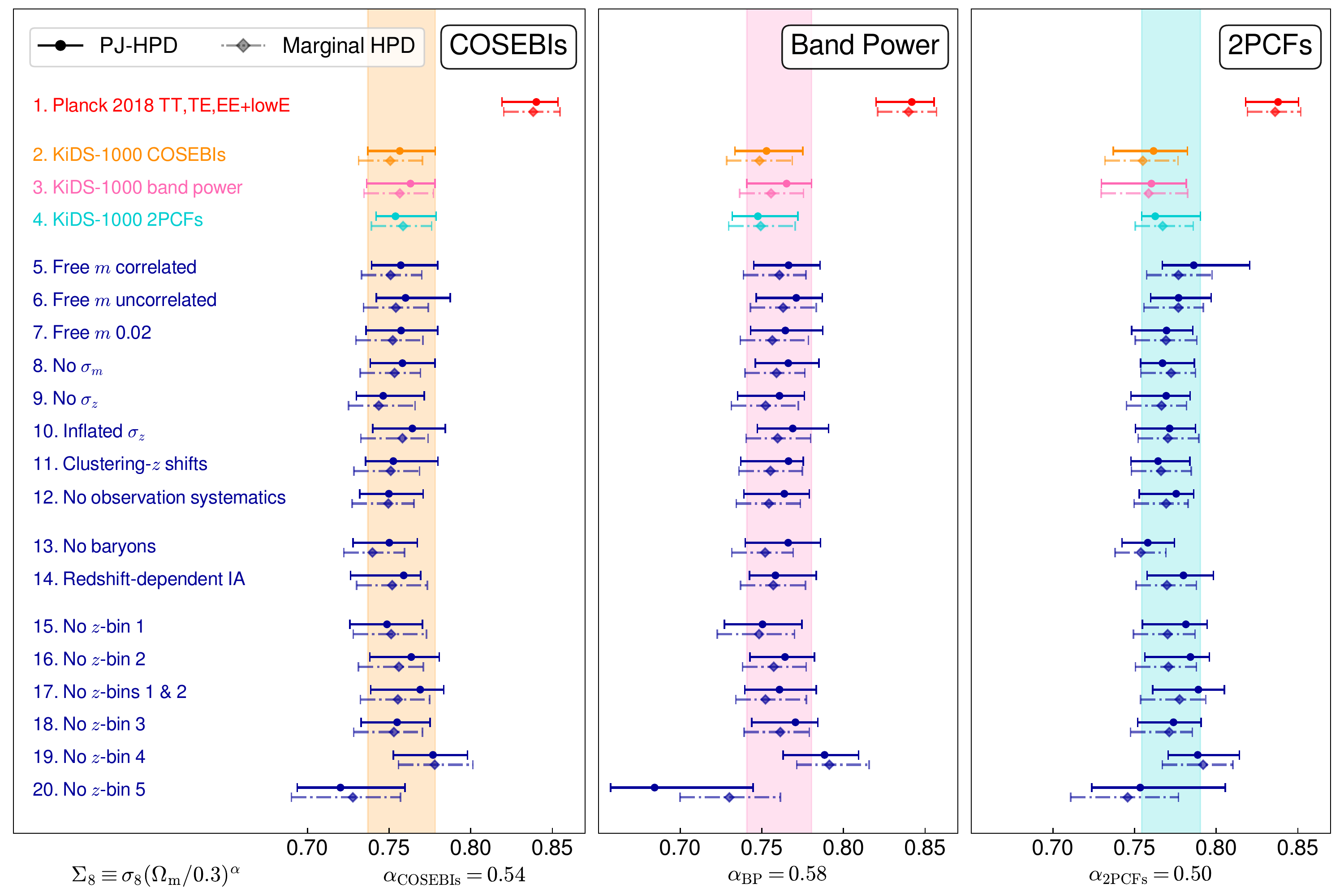}
     \end{tabular}
   \end{center}
     \caption{Impact of nuisance parameter treatment and tomographic bin exclusion on $\Sigma_8$ constraints. Results are shown for COSEBIs (\textit{left}), band powers (\textit{centre}) and 2PCFs (\textit{right}), with fiducial constraints in orange, pink, and cyan, respectively. We use the best-fitting value of $\alpha$ for the fiducial chain of each set of statistics to define $\Sigma_8$ (\Eqt\ref{eq:Sigma8def}) using the covariance matrix generated from the \cite{troster/etal:2020a} values instead of the iterative covariance used in \sect\ref{sec:fid}. The value of $\alpha$ for each panel is given underneath. Two sets of credible regions are shown for each case: the multivariate maximum posterior (MAP, circle) with PJ-HPD (solid) credible interval and the maximum of the $\Sigma_8$ marginal posterior (diamond) with its highest density credible interval (dot-dashed). The shaded regions follow the fiducial PJ-HPD results of the corresponding statistics. We show \textit{Planck} results (red), as well as the fiducial results of the other two statistics for the given $\alpha$ of each panel for comparison. Cases 5 to 12 show the impact of different observational systematics, while cases 13 and 14 show results for the impact of astrophysical systematics. The last six cases present the effect of removing redshift bins and their cross-correlations from the analysis. }
     \label{fig:summary_nuisance}
 \end{figure*}

\begin{figure*}
   \begin{center}
     \begin{tabular}{c}
      \includegraphics[height=11cm]{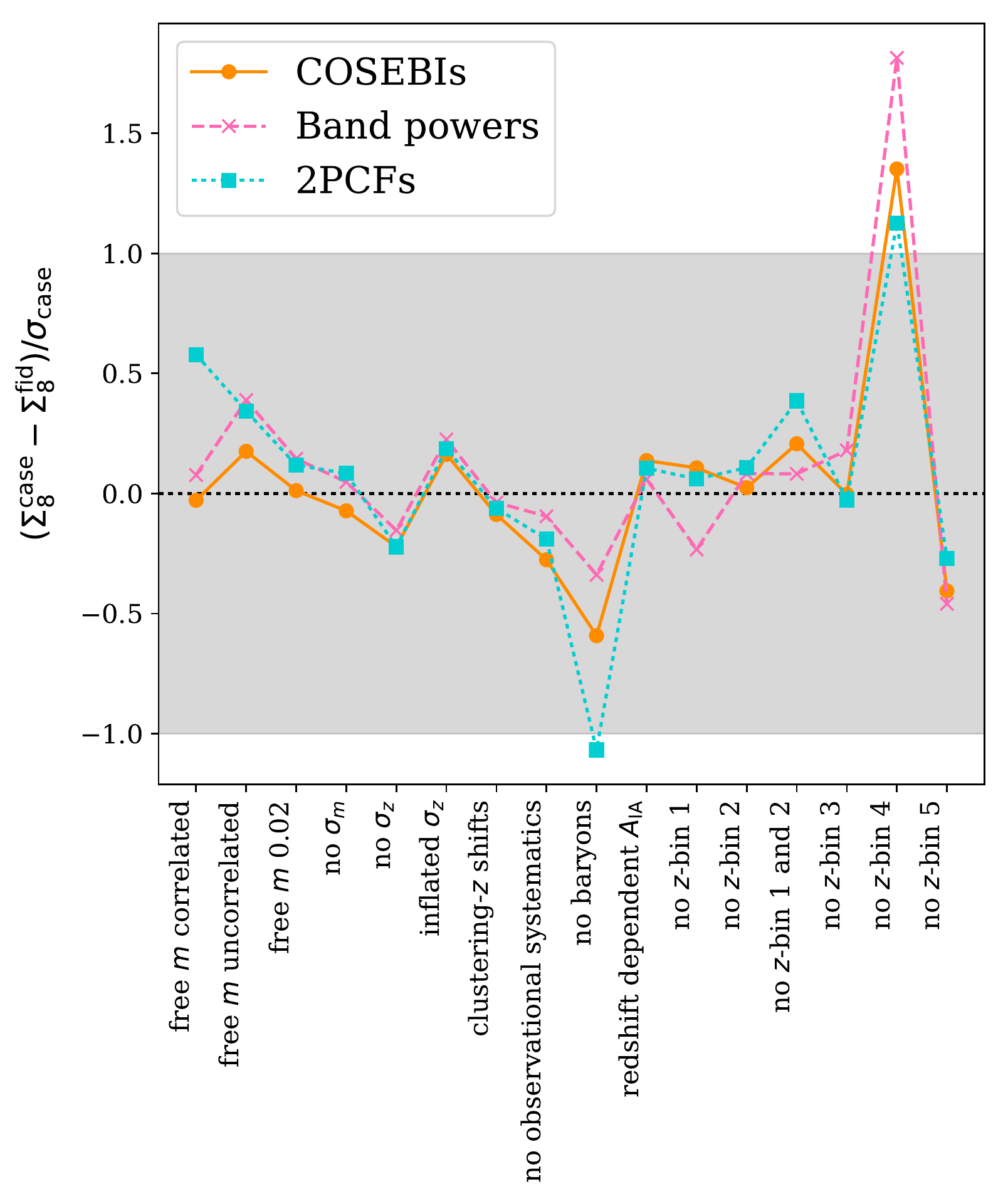}
       \includegraphics[height=11cm]{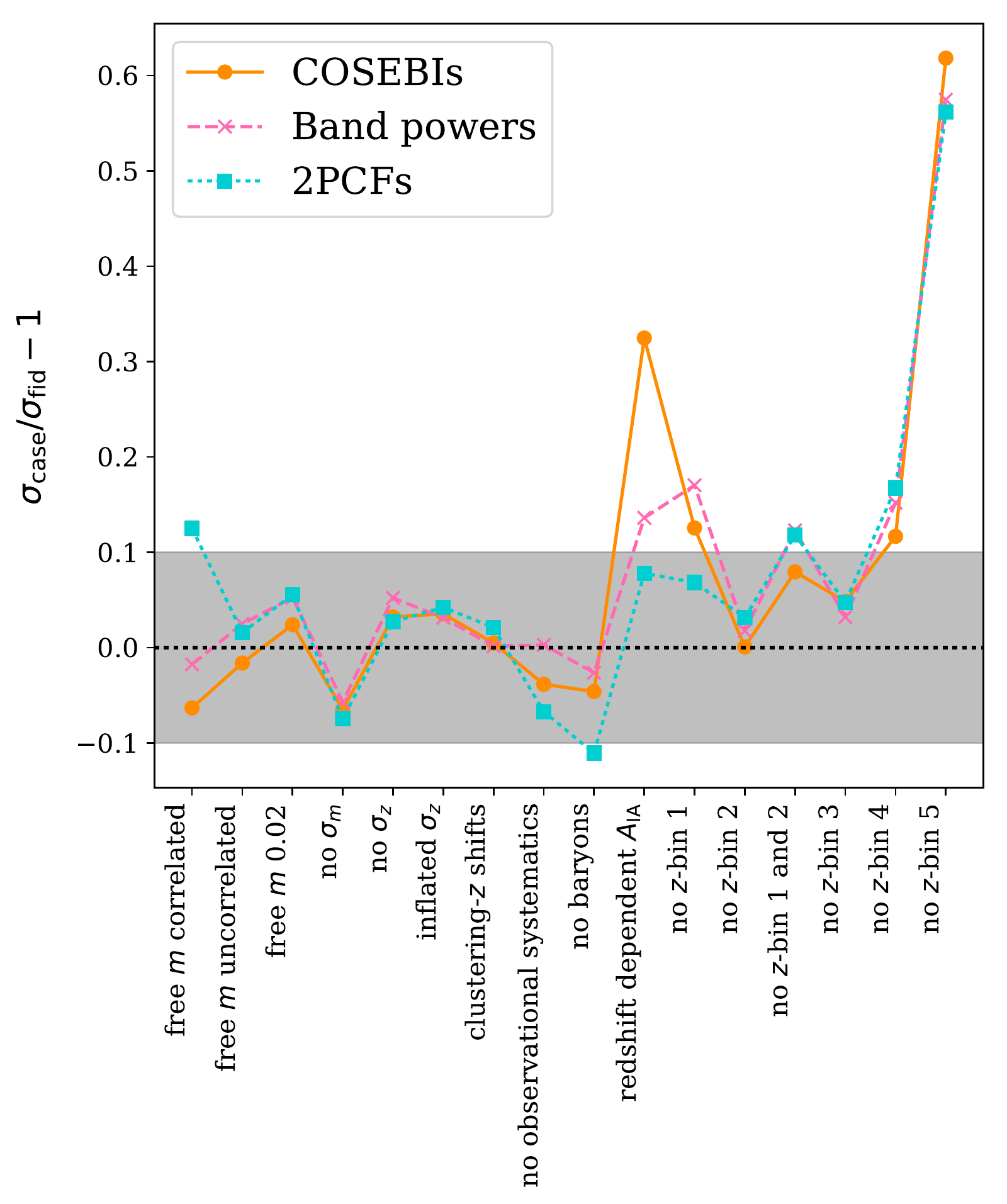}
     \end{tabular}
   \end{center}
     \caption{Relative impact of nuisance parameters and the removal of redshift bins. Each of the cases explored in \fig\ref{fig:summary_nuisance} is compared to their corresponding fiducial results. COSEBIs are shown as orange circles, band powers as pink crosses and 2PCFs as cyan squares. {\it Left:} The difference between the upper edge of the marginal $\Sigma_8$ posterior for each case and its fiducial chain, normalised by half of the length of the marginal credible interval of the case. The grey shaded area indicates the region in which systematic shifts remain below the $1\sigma$ statistical error. {\it Right:} Comparison of constraining power between the fiducial and the other cases. Here $\alpha$ is fitted to each chain separately to find the tightest $\Sigma_8=\sigma_8(\Om/0.3)^\alpha$ constraint for each case. We show the fractional difference between the standard deviations of the case and the fiducial one.}
     \label{fig:summary_of_summary}
 \end{figure*}

In our analysis we have a number of astrophysical and nuisance parameters which are marginalised over. Here we test the sensitivity of our data to the choice of these parameters and their priors. Furthermore, we investigate the impact of removing individual redshift bins from the analysis, as well as the lowest two redshift bins jointly. In the following we first introduce \figs\ref{fig:summary_nuisance} and \ref{fig:summary_of_summary} and then provide the details of each case.

The results of these tests are summarised in \fig\ref{fig:summary_nuisance}. Here we use $\Sigma_8$ with $\alpha$ fitted to the fiducial chain for each of the statistics to assess the impact of the nuisance parameters and the exclusion of redshift bins. We show two sets of point estimates and associated error bars for each case, the MAP and PJ-HPD credible interval, as well as the marginal mode and highest-posterior density credible interval. We note that PJ-HPD intervals are expected to have an error of about $10\%$ in their boundaries (see section 6.4 of J20). 

Each panel shows results for one of the two-point statistics, COSEBIs, band powers and 2PCFs; however,
in the first section of each panel we also show the fiducial results for the other two cosmic shear statistics (using the same $\alpha$) and \textit{Planck} for comparison. 
The shaded regions correspond to the PJ-HPD credible interval of the fiducial chain for the relevant statistics of each panel. The second section of the figure shows results for the impact of observational systematics. In the third section we explore the effect of astrophysical systematics. The fourth section allows for an inspection of the significance of the data in each redshift bin.

We also test the impact of removing the largest two $\theta$-bins from the analysis of $\xi_+$ and find its impact to be negligible. The mean of $S_8$ is lowered by $0.1\sigma$ compared to our fiducial case and its standard deviation is increased by $4\%$. This final test assesses the Gaussian likelihood approximation since the distribution of $\xi_+$ is significantly non-Gaussian for these bins (see figure 17 of J20). 

To quantify the impact of the different setups shown in \fig\ref{fig:summary_nuisance}, we extract two key properties of each test analysis, relative to the fiducial case. In the left-hand panel of \fig\ref{fig:summary_of_summary} we plot the difference between the upper edge of the marginal credible interval shown in \fig\ref{fig:summary_nuisance} for the fiducial setup, $\Sigma_{8}^{\rm fid}$, and the cases named on the abscissa, $\Sigma_{\rm 8}^{\rm case}$. We normalise $\Delta\Sigma_8 := \Sigma_{8}^{\rm case}-\Sigma_{8}^{\rm fid}$ by half of the length of the marginal credible interval that we found for each case, $\sigma_{\rm case}$. We chose the upper edge since we are primarily interested in a comparison with the {\it Planck} inferred value for $\Sigma_8$ which is larger than our measurements.
We show results for all three statistics, COSEBIs (orange), band powers (pink) and 2PCFs (cyan). 

The right-hand panel of \fig\ref{fig:summary_of_summary} compares the size of the constraints on $\Sigma_8$ between different cases and the fiducial case. The $\Sigma_8$ for each case is defined with its own corresponding best-fit $\alpha$. As the width of the $\Om$--$\sigma_8$ degeneracy is the main parameter that we constrain, this definition allows us to do an approximate figure-of-merit comparison between the different test cases and identify the ones that have a larger impact on our constraining power. For this plot we use the standard deviation of the marginal distributions as they are not affected by smoothing which affects the marginal credible intervals, or by the small number of samples that produce the PJ-HPD. 
J20  argued for a $0.1\sigma$ error on our constraints, coming from smoothing and sampling of the likelihood surfaces to set their requirements on the modelling and data systematics. Here we show the $0.1\sigma$ region in grey.

\subsubsection{Shear calibration uncertainty}

The first nuisance parameter that we consider is the error on the multiplicative shear calibration, $m$, that is applied to the ellipticity measurements, $\sigma_{\rm m}$. The value of $m$ is estimated using image simulations (see \sect\ref{sec:data} and \citealt{kannawadi/etal:2019}). The assumptions made when producing the image simulations can affect the value of this calibration parameter. In our fiducial chains we absorb this uncertainty into the covariance matrix; however, we could instead allow $m$ to vary as a free model parameter, one per redshift bin. In the covariance matrix estimation we use different values of $\sigma_{\rm m}$ for each redshift bin (see \tab\ref{tab:dataprops}) and assume that they are fully correlated. To produce the priors for the $m$ parameters, we can take the same approach or instead assume that we do not know the extent of this correlation and use larger uncorrelated priors that encompass any expected correlations between the redshift bins \citep[see for example][]{hoyle/etal:2018}. To do so, we multiply each of the $\sigma_{\rm m}$ values by the square root of the total number of redshift bins, $\sqrt{5}$. This way we produce two setups with free $m$, labelled \enquote{free $m$ correlated} and \enquote{free $m$ uncorrelated}. 

These setups cover all possible scenarios for the error on $m$. The $m$ calibration in the simulations is determined per tomographic bin, so that the estimates are independent.
However, the surface brightness profiles are modelled as Sersic profiles, and any model bias arising from mismatches with the true morphologies will be shared across the bins. Hence assuming that the $m$-values are fully correlated, as we have done in the fiducial analysis is an extreme scenario, whereas the scenario where $m$ is uncorrelated represents the other extreme. A more consistent estimate requires multi-band image simulations to capture the correlation between photometric redshift determination and shear estimation. 

For the cosmic shear analysis of KV450 a more conservative route was taken, where a $\sigma_{\rm m}=0.02$ was employed for all bins, equal to the largest value of $\sigma_{\rm m}$ that we use. Similar to our fiducial analysis, these studies included $\sigma_{\rm m}$ in the covariance matrix, assuming full correlation. Here we also test the effect of this assumption, but with free, correlated $m$ parameters (\enquote{free $m$ 0.02}). We then compare all of these setups with a zero $\sigma_{\rm m}$ case (\enquote{no $\sigma_{\rm m}$}) to fully capture the impact of this nuisance parameter\footnote{For all the cases where $\sigma_{\rm m}$ is not included in the covariance matrix, the fiducial $m$ correction is applied to the theory rather than the data vectors.}.

Comparing the $\Sigma_8$ values for these different choices, we see an at most $0.5\sigma$ shift corresponding to the \enquote{free $m$ correlated} results of the 2PCFs. With the \enquote{no $\sigma_{\rm m}$} and \enquote{free $m$ 0.02} cases we do not see a significant change in $\Sigma_8$. The impact of the uncertainty on $m$ on the standard deviations of the marginal distributions of $\Sigma_8$ is at most $10\%$.

\subsubsection{Photometric redshift uncertainty}

Another component of the data that is calibrated using simulations is the mean of the SOM redshift distribution of galaxies in each tomographic bin. In the fiducial chains we allow for a free $\delta_z$ parameter per redshift bin, but with correlated informative priors, through the covariance matrix between the $\delta_z$ values estimated from the MICE2 simulations (see H20b). 
To assess the impact of this freedom in the analysis, we fix the $\delta_z$ to their fiducial values (\enquote{no $\sigma_{\rm z}$}). Another case that we consider is the impact of inflating the priors taken from MICE2 by a factor of 3 instead of a factor of 2 that we used in the fiducial case (\enquote{inflated $\sigma_{\rm z}$}). H20b investigated cross-correlations with spectroscopic reference samples as a complementary, independent method for calibrating the redshift distributions. We use their quoted $\delta_z^{\rm CZ}$ shifts (see table 3 of their paper) in combination with their estimated covariance to create the \enquote{Clustering-$z$ shifts} case.
The $\delta_z$ uncertainty and mean values that we consider here have a negligible impact on our analysis. This is true for both the impact on the marginal value of $\Sigma_8$ and its constraints, as can be seen in \fig\thinspace\ref{fig:summary_of_summary}.

\subsubsection{Impact of all observational systematics}
To evaluate the joint impact of observational systematics, we re-analyse the data by setting $m$ and $\delta_z$ errors to zero. For the 2PCFs chains, we additionally fix the value of $\delta_{\rm c}$. We call this setup \enquote{no observational systematics}. From \fig\thinspace\ref{fig:summary_of_summary} we deduce that the impact of our observational systematics is small, whether we consider them separately or jointly. We remind the reader that variations of order $0.1\sigma$ are expected to occur between different instances of the sampling of the same posterior surface.

\subsubsection{Sensitivity to astrophysical modelling choices}

Our astrophysical nuisance parameters are the baryon feedback parameter, $A_{\rm bary}$, and the amplitude of the intrinsic alignments of galaxies, $A_{\rm IA}$. We test the impact of $A_{\rm bary}$ by assuming a no-feedback case with $A_{\rm bary}$ fixed to $3.13$ (\enquote{no baryons}). 
As illustrated by \fig\ref{fig:summary_of_summary} the no-baryons case has a significantly larger effect on $\xi_\pm$, which is expected since the 2PCFs are more sensitive to small physical scales as we saw in \fig\ref{fig:compare}. Contrary to expectations, COSEBIs appear to be more sensitive to baryon feedback compared to the band powers. 
This is not caused by the scale sensitivity, but is rather a result of this particular noise realisation. In \fig\ref{fig:triangle} we can see that the constraints on $A_{\rm bary}$ for band powers are skewed towards larger values, indicating that they prefer a model with weaker baryon feedback (see also \tab\ref{tab:bestfitall}). 
Therefore, the difference between band powers analysed with and without baryon feedback is smaller than for COSEBIs, which have a rather uniform $A_{\rm bary}$ marginal distribution. For the 2PCFs, however, we find a similarly uniform distribution. The increased sensitivity of the 2PCFs to baryon feedback is thus a result of the small scales that impact their modelling. This is true for both the upper edge of the marginal credible region and to a lesser extent the width of the constraints for $\Sigma_8$. In \app\ref{app:stagetests} we discuss that the marginal distributions of poorly constrained parameters, such as  $A_{\rm bary}$, can be skewed due to noise in the data.

In our fiducial analysis we assume that the amplitude of the intrinsic alignment model, which describes the response of projected galaxy ellipticities to the local quadrupole of the dark matter distribution, is independent of redshift (see section 2.4 of J20). 
However, this model can be modified empirically to include a redshift dependence (see equation 16 of J20), by multiplying its three-dimensional power spectra with factors of
\begin{equation}
  \left(\frac{1+z}{1+z_{\rm pivot}}\right)^{\eta_{\rm IA}}\;.
\end{equation}
As a test case we allow $\eta_{\rm IA}$ to vary uniformly in $[-5,5]$ and set $z_{\rm pivot}=0.3$ for a more straightforward comparison with previous KiDS and intrinsic alignment analyses \citep[e.g.][]{joachimi/etal:2011}. We call this case \enquote{redshift-dependent IA}. 

In \fig\ref{fig:summary_of_summary} we see that the redshift dependence of $A_{\rm IA}$ has little impact on the upper edge of the marginal credible region of $\Sigma_8$, however it can result in wider constraints. 
This redshift-dependence for the COSEBIs analysis produces a bimodal likelihood distribution, which results in a larger standard deviation. This is not seen with the other two statistics, which we therefore conclude is an effect of the cross-talk between the noise realisation and this extra freedom in the analysis. This has been seen in other analyses, when the additional redshift of the intrinsic alignment model is allowed to vary within broad priors \citep[for example][]{joudaki/etal:2017,joudaki/etal:2020,asgari/etal:2020a}. The inclusion of this freedom in the analysis does not impact the goodness-of-fit in a significant way.

\subsubsection{Removing tomographic redshift bins}
\label{sec:no_zbin}
Aside from the effect of nuisance parameters, we determine the impact of each tomographic redshift bin by removing them and their cross-correlations in turn from the data vector. These results are labelled as \enquote{no $z$-bin $i$}, with $i$ denoting the removed redshift bin. The first two redshift bins have a lower signal-to-noise and are mostly sensitive to the intrinsic alignments of galaxies. To capture the impact of an unconstrained intrinsic alignment model, we also run chains where both redshift bins 1 and 2 are removed from the analysis (\enquote{no $z$-bins 1 and 2}). 

Of these setups the no $z$-bin 4 case has the largest impact on $\Sigma_8$ marginal values (left panel of \fig\ref{fig:summary_of_summary}).
For this case, depending on the statistics used, we obtain between $1.1\sigma_{\rm case}$ to $1.8\sigma_{\rm case}$ differences in $\Sigma_8$. The significance of these shifts however depends on which values from the distributions are compared with each other. For example, for the  no $z$-bin 5 case we find larger deviations if we consider the maximum of the marginal distribution or the MAP values. In \app\ref{app:internal_consistency} we perform a series of internal consistency tests which do not flag the differences between these redshift bins as statistically significant.

When removing redshift bins we see that the constraining power does not change by more than $0.15\sigma$ unless the fifth bin is removed (right panel of \fig\ref{fig:summary_of_summary}). Without this bin our errorbars inflate by $60\%$. This shows that the inclusion of higher-redshift bins is crucial for increasing the statistical power of a cosmic shear analysis. 

\subsection{Internal consistency} 
\label{sec:consist}

In this section we summarise our internal consistency results. For details see \app\ref{app:stagetests} and \app\ref{app:internal_consistency}. 

Our cosmological analysis has been performed independently, using three sets of two-point statistics. We do not expect to find the exact same constraints from these statistics, since they place different weights on a given angular scale. That said, the statistics are measured within the same survey volume and using the same galaxies, so that it is reasonable to assume some level of redundancy between these measurements. Given these two competing factors, it is not immediately clear what level of variation is expected. In other words, are the results in \tab\ref{tab:bestfit} consistent? Or is the difference between $S_8$ constraints caused by systematic effects being picked up by one statistic but not another? 

To answer these questions, we apply a series of tests on mock data realisations, produced from multivariate Gaussian distributions. In our primary test we draw correlated noise realisations given the full covariance, including cross-correlations between 2PCFs, COSEBIs and band powers, estimated from the {\sc Salmo} simulations (see \fig\ref{fig:crosscov}). We choose a fiducial cosmology and create 100 realisations of the data vector, including all three sets of two-point statistics. We analyse each set and realisation separately with a similar setup to our fiducial analysis explained in \sect\ref{sec:data} and derive parameter constraints.  We compare the maximum of the marginal distributions for $S_8$ between the two-point statistics for each realisation and find that the distribution of $\Delta S_8 := S_8^{\rm stat1} - S_8^{\rm stat2}$, where $S_8^{\rm stat1/2}$ are the maximum marginal values for one of the statistics, is only $20-30\%$ narrower than the width of the marginal distributions for $S_8$ per two-point statistic. Therefore, we conclude that differences of up to $0.7-0.8\sigma$ between the results of COSEBIs, 2PCFs and band powers are expected to occur frequently (for about $68\%$ of the realisations). For our KiDS-1000 analysis we find the maximum  $\Delta S_8$ for the marginal posterior modes of COSEBIs and 2PCFs, which is a difference of about $0.4\sigma$. 

Among the significantly constrained parameters in our data analysis, only $A_{\rm IA}$ displays a notable difference, with the marginal posterior peaking roughly at double the value for band powers in comparison with correlation functions and COSEBIs. In our mock analysis we see differences of this level or higher in $A_{\rm IA}$ in $~5\%$ of the cases. Given the full consistency between the $S_8$ values we conclude that the results between the three sets of summary statistics are in agreement. 

While the two-point statistics have different scale sensitivities, we expect their response to biases in the redshift distributions to be similar, as that will mainly affect the relative amplitude of the data vectors. H20b conducted tests of the KiDS-1000 redshift distributions by comparing them with simulations as well as cross-correlations with clustering-redshifts as discussed in \sect\ref{sec:data} and \sect\ref{sec:nuisance}. However, we note that these tests are not very sensitive to discrepancies that may exist in the tails of the redshift distributions, beyond their impact on the mean redshift.

We also follow the methodology of \cite{kohlinger/etal:2019} and perform three tiers of Bayesian consistency tests, comparing the cosmological inference from all bin combinations involving a given redshift bin with that from the remainder of the data vector. We find consistent results between all redshift bins, except for the second tomographic bin which covers the range $0.3 < z_{\rm B} < 0.5$. 
Analyses using this bin and its cross-correlations, compared to using all other bins, produce results that conflict by up to $3\sigma$ in some parameters (for more details see \sect\ref{app:internal_consistency}). Also in \fig\ref{fig:deltaz} we see that the data favours a $\delta_{z,2}$ parameter that shift the redshift distribution of this bin to larger values.
While this inconsistency warrants further investigation in the future, we find that removing the second redshift bin, or indeed the first and second bin, from the analysis has a negligible impact on the cosmological parameter constraints (see \sect\ref{sec:no_zbin}). 


\subsection{Comparison with other surveys}
\label{sec:constst_external}

\begin{figure}
   \begin{center}
     \begin{tabular}{c}
      \includegraphics[width=\hsize]{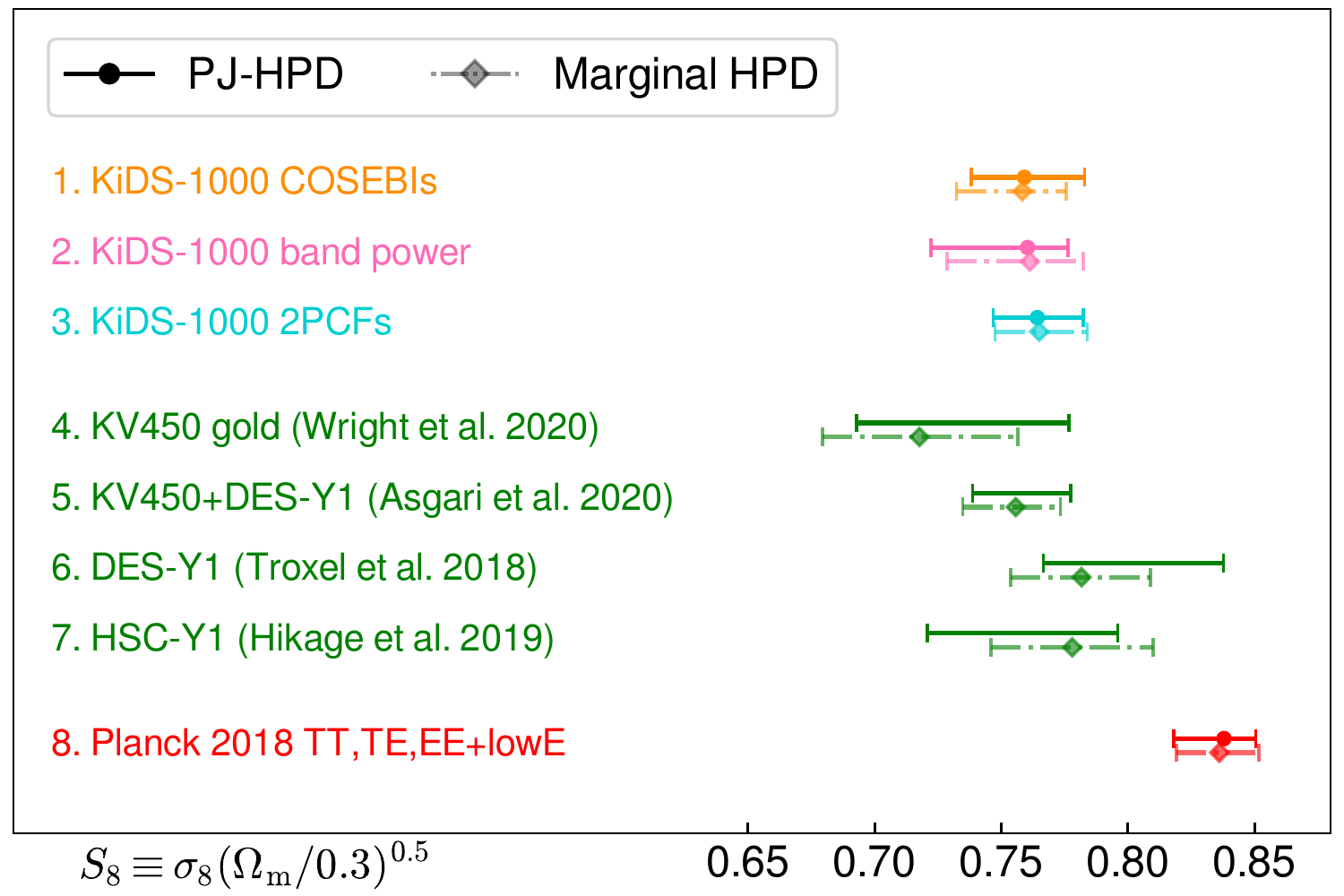}
     \end{tabular}
   \end{center}
     \caption{Comparison between $S_8$ values for different surveys. All results are shown for both multivariate maximum posterior (MAP) and PJ-HPD (upper solid bar), as well as the marginal mode and the marginal $S_8$ credible interval (lower dot-dashed bar). The top three points show our fiducial KiDS-1000 results. The next four show a selection of recent cosmic shear analyses from external data as well as previous KiDS data releases. We note that $S_8$ does not fully capture the degeneracy direction for all of the analysis above (see the discussion in \sect\ref{sec:fid} and \app\ref{app:extra}). For example for the HSC-Y1 contours $\alpha=0.45$ was found to be the best fitting power. The last entry shows the \textit{Planck} 2018 (TT,TE,EE+lowE) constraints. An extended version of this plot can be found in \app\ref{app:extra}.}
     \label{fig:summary_surveys}
 \end{figure} 
 
 \begin{figure*}
   \begin{center}
     \begin{tabular}{c}
      \includegraphics[width=0.5\hsize]{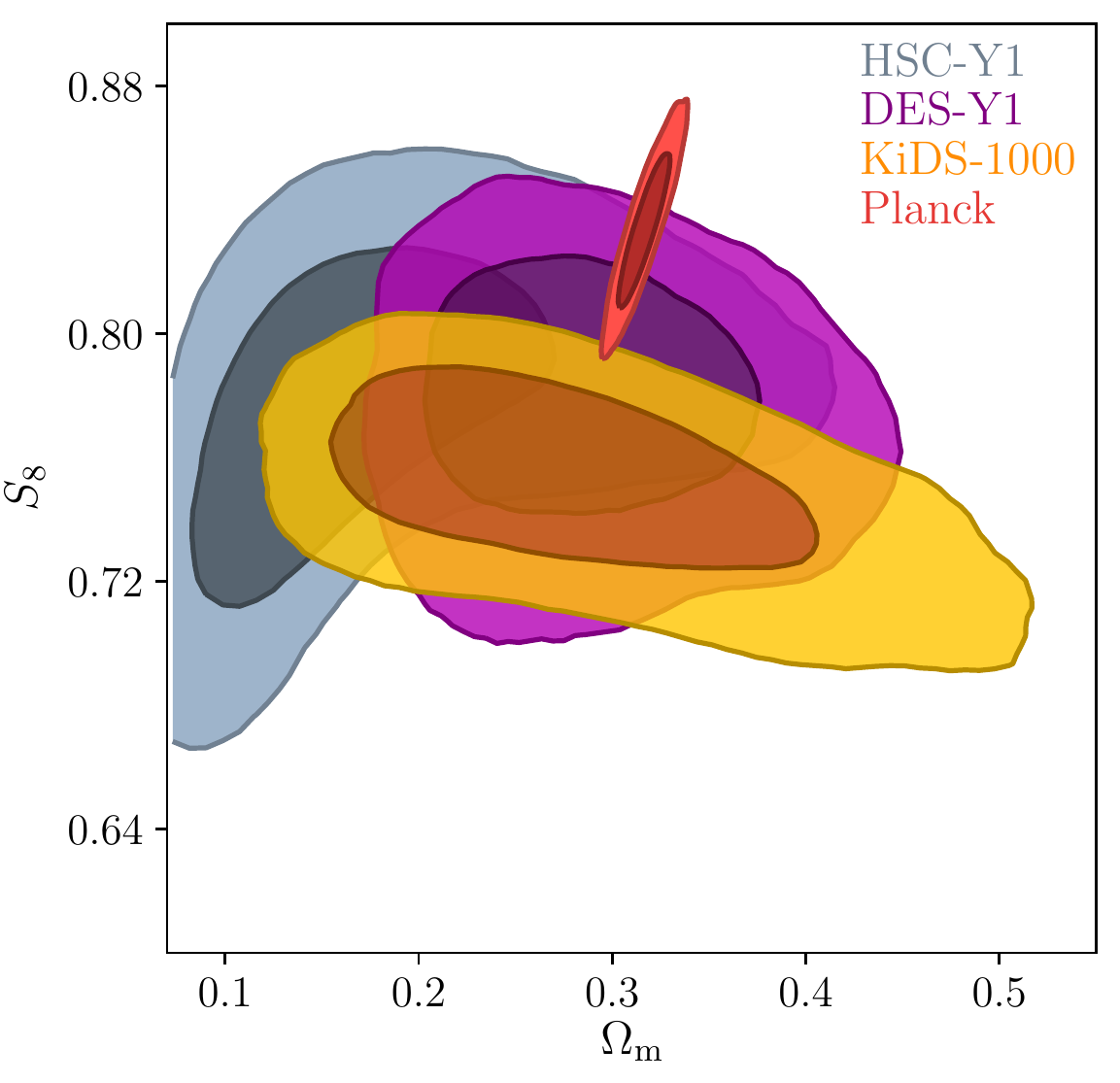} 
      \includegraphics[width=0.5\hsize]{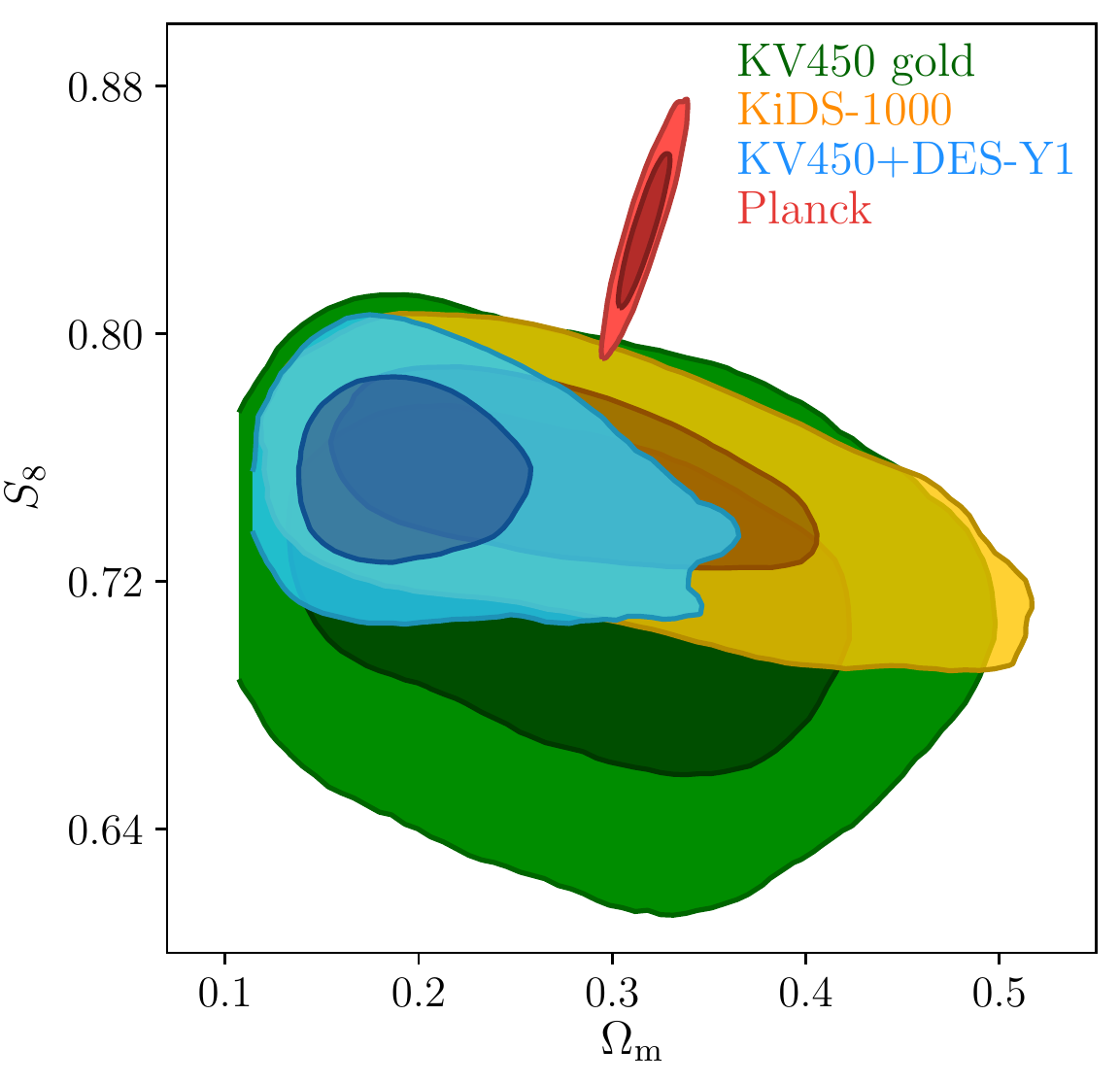}
     \end{tabular}
   \end{center}
     \caption{Comparison between KiDS-1000 and other surveys in the $S_8-\Om$ plane. The fiducial KiDS-1000 results which use COSEBIs (orange) and the {\it Planck} primary anisotropy constraints (red) are shown in both panels. The DES-Y1 results of \citet[purple]{troxel/etal:2018a} and HSC-Y1 results of \citet[grey]{Hikage18} are shown in the left panel, while the KV450 constraints of \citet[green]{wright/etal:2020b} and the joint KV450 and DES-Y1 results of \citet[blue]{asgari/etal:2020a} are shown in the right panel. A summary of these constraints in $S_8$ can be found in \fig\ref{fig:summary_surveys}.}
     \label{fig:2d_compare}
 \end{figure*}

In this section we compare our parameter constraints with previous results from cosmic shear surveys and \textit{Planck}.
Figure\thinspace\ref{fig:summary_surveys} contrasts our $S_8$ constraints with a selection of recent cosmic shear results shown in green (see \fig\ref{fig:summary_surveys_all} for an extended selection). 
The final entry shows the \textit{Planck} results. For each case we show two sets of error bars, corresponding to the marginal highest-posterior density region and the PJ-HPD. Since we do not have a good estimate of the MAP from the public chains, we do not show best-fitting values for the external cosmic shear results. 

The different cosmic shear analysis presented in \fig\ref{fig:summary_surveys} constrain slightly different degeneracy directions in the $\sigma_8-\Om$ plane and therefore $S_8$ does not necessarily capture their best constrained parameter combination. Hence we also compare the results of these surveys in the $S_8-\Om$ plane displayed in \fig\ref{fig:2d_compare}. We note that for all cosmic shear analyses presented here, the $\Om$ constraints are prior dominated.  Consequently, no meaningful conclusions can be drawn from the differences that can be seen in the figure when it comes to this parameter.

Of the external cosmic shear data, the \cite{wright/etal:2020b} result is the closest to our methodology in terms of the calibration of the redshift distributions. This KV450 analysis employed 2PCFs measured on less than half of the imaging area that we analyse ($777$ deg$^{2}$ versus $341$ deg$^2$). We find that our  results are in good agreement with \cite{wright/etal:2020b}, with the multivariate maximum posterior values of $S_8$ agreeing to within $0.003$ for the 2PCFs, the statistics used in both\footnote{The MAP value for \cite{wright/etal:2020b} is taken from the {\sc MultiNest} chain, $S_8 = 0.765$.}. Marginal errors decrease by more than a factor two, reflecting the increase in survey area, and the reduced impact of calibration uncertainties in our KiDS-1000 analysis. Our KiDS-1000 constraints are similar to the joint KV450 and DES-Y1 analysis of \cite{asgari/etal:2020a} both in their constraining power and value. We find that the DES-Y1 and HSC-Y1 results of \cite{troxel/etal:2018a} and \cite{Hikage18} are also both in agreement with our constraints. It is evident from this plot that all of these cosmic shear analyses measure a lower $S_8$ than the \textit{Planck} inferred value under a flat $\Lambda$CDM model, although with varying levels of significance. 

We use two complementary methods to estimate the level of tension between our results and \textit{Planck}. For this we choose the COSEBIs analysis, which has the best goodness of fit. The first method is to simply compare the results in $\Sigma_8$, the only parameter that we can set tight constraints on with cosmic shear that is also shared by \textit{Planck}. We use the conventional method, 
\begin{equation}
\label{eq:tention}
\tau = \frac{\overline{\phi}^{Planck}-\overline{\phi}^{\rm COSEBIs}}{\sqrt{{\rm Var}[\phi^{Planck}]+{\rm Var}[\phi^{\rm COSEBIs}]}}\;
\end{equation}
where $\phi$ is either $S_8$ or $\Sigma_8$, $\overline{\phi}$ is the mean of $\phi$ and ${\rm Var}[\phi]$ is its variance. With this definition we find that the \textit{Planck} predictions are $3.4\sigma$ larger than our measured $\Sigma_8$ value. The difference in $S_8$ is $3\sigma$, but note that this parameter does not fully capture the tension due to the residual correlation with $\Om$. We use a complementary method which takes the full shape of the marginal distributions into account, bypassing the Gaussian distribution assumption used in \Eqt\eqref{eq:tention}, and find slightly larger values of $3.2\sigma$ for $S_8$ and $3.5\sigma$ for $\Sigma_8$ (Hellinger; see appendix F.1 of \citealt{heymans/etal:2020} for details).
These methods of estimating differences between cosmological analyses ignore the possible complexities of the multi-dimensional parameter space. 
Other methods that test consistencies within the full posterior are generally less stable owing to the difficulty in estimating the statistical properties of this distribution to a sufficiently high accuracy. On the other if the tension is truly in one aspect of the model, summarised in a single parameter, then including extra dimensions to the tension metric will likely dilute the significance of the results. 

A Bayesian approach compares the full likelihood between an analysis of the two sets of data separately and their combined analysis. 
The Bayes factor can be used in conjunction with the Jeffreys' scale to assess the tension between the data sets. We find that the base 10 logarithm of the Bayes factor is $0.54-1.15$, with a preference for two separate cosmologies, corresponding to a substantial to strong evidence for disagreement (the two values are estimated via the importance nested sampling and the traditional methods, see \app\ref{app:internal_consistency} for more details). 
This result is in qualitative agreement with the simple marginal distribution comparison above. \cite{suspect:2019} suggested using a different measure called suspiciousness, $S$, which is less sensitive to the choice of priors compared to the raw evidence comparison of the Bayes factor. 
We measure this quantity but are unable to cast it into a meaningful scale of disagreement. 
To do so we need to have a robust measure of the degrees of freedom for \textit{Planck}, KiDS-1000 and their joint analysis. In J20 we saw that the dimensionality method that is currently used in conjunction with suspiciousness produces biased estimates of the effective number of parameters (see \app\ref{app:external_consistency} for more details). The alternative methods proposed there require analysing many mock realisations of the data with computationally expensive posterior sampling. Future work is required to develop a robust way to derive the sampling distribution for suspiciousness.

\section{Summary and conclusions}

\label{sec:conclusions}

We have presented a cosmic shear analysis of the fourth Data Release of the Kilo-Degree Survey \citep[KiDS-1000,][]{kuijken/etal:2019}, making use of circa $1000\,{\rm deg}^2$ of deep nine-band optical-to-infrared photometry with exquisite image quality in the $r$-band for gravitational shear estimates. In addition to more than doubling the survey area with respect to earlier KiDS analyses \citep{hildebrandt/etal:2019}, this work incorporated the following major updates:
\begin{itemize}
\setlength\itemsep{0.5em}
\item[$\bullet$] The galaxies entering our five tomographic redshift bins are selected to have good representation by objects with spectroscopic redshifts (the \enquote{gold} sample), which are subsequently re-weighted via an unsupervised machine learning approach to provide accurate redshift distributions \citep{wright/etal:2020a,hildebrandt/etal:2020}.

\item[$\bullet$] The multiplicative shear calibration is based on image simulations containing COSMOS-emulated galaxies \citep{kannawadi/etal:2019}. This analysis was repeated for the new sample selection, with a revised determination of the residual calibration uncertainties that is now derived per tomographic bin from the spread in a number of conservative settings implemented in the simulations. 

\item[$\bullet$] The accuracy of the covariance models, likelihood, and inference pipeline has been validated on an extensive suite of KiDS-1000 mock catalogues. The key cosmological quantity constrained by cosmic shear, the parameter $S_8=\sigma_8 (\Omega_{\rm m}/0.3)^{0.5}$, is now used as a sampling parameter in evaluating the posterior, enabling us to impose a wide top-hat prior that is more conservative than previous analyses relying on the primordial power spectrum amplitude, $A_{\rm s}$, or a function thereof.

\item[$\bullet$] The analysis was conducted independently with three cosmic shear two-point statistics: the angular shear correlation functions $\xi_\pm$, Complete Orthogonal Sets of E/B-Integrals (COSEBIs), and angular band powers. The latter two are constructed as linear combinations of $\xi_\pm$ that offer a clean separation into cosmological E-modes and systematics-driven B-modes (exact for COSEBIs and approximate for band powers), as well as additional data compression (the COSEBIs and band powers data vectors are $66\,\%$ and $46\,\%$ smaller than the 2PCFs data vector). Both derived statistics inherit the beneficial lack of sensitivity to the survey mask and galaxy ellipticity noise from the correlation functions, but avoid the very broad responses of $\xi_\pm$ to Fourier modes, which lead to increased non-Gaussianity in the likelihood due to small $\ell$-modes and increased sensitivity to small-scale features in the modelling (large $\ell$-modes), such as baryon feedback. 

\end{itemize}

These additions have increased the constraining power of KiDS with little change in our best-fitting value for $S_8$. Comparing the similar setups of our correlation function analysis with the results from \citet{wright/etal:2020b} who worked with KiDS Data Release 3, we find a decrease in the marginal $S_8$ errors by $54\,\%$. The marginal posterior mode of $S_8$ has increased by 0.05 in KiDS-1000; however, the multivariate maximum posterior agrees to within $3 \times 10^{-3}$ for the two analyses, so the shift in the marginal distribution is solely due to the different shape of the posterior distribution. 
Our results are in good agreement with those of the DES and HSC surveys, reducing marginal $S_8$ errors by $14\,\%$ with respect to \citet{troxel/etal:2018a} and by $32\,\%$ with respect to \citet{Hikage18}.

From a theoretical point of view we conclude that there is a strong case for favouring COSEBIs and/or band power statistics over the standard shear correlation functions in the likelihood analysis, with COSEBIs providing the cleanest and most compact data vector, and band powers offering intuition through directly tracing the angular power spectra predicted from theory. Both of these methods allow for an E and B-mode decomposition, which are mixed with each other in the case of the correlation functions. This will be of particular importance for analysis of future data with improved constraining power. 

Despite these differences, we find the KiDS-1000 $S_8$ constraints derived from the three statistics to be in excellent agreement. Due to the different scales probed, the analyses trace different sections of the $\Omega_{\rm m}$--$\sigma_8$ degeneracy line, which causes $S_8$ to not fully capture the constraining power transverse to the degeneracy in all cases. Fitting the parameter $\Sigma_8=\sigma_8 (\Omega_{\rm m}/0.3)^\alpha$ to the posterior, we find a best fit of $\alpha=0.51$ for $\xi_\pm$, i.e. $S_8$ is very close to the optimal summary parameter as found in previous KiDS analyses. For COSEBIs and band powers, $\alpha=0.54$ and 0.58, respectively. The constraining power on the optimal $\Sigma_8$ is then nearly identical between the three statistics. 

Constraining a spatially flat $\Lambda$CDM model, we obtain $S_8= 0.758^{+0.017}_{-0.026} \;(68\%\;{\rm CI})$ for our fiducial setup using COSEBIs. The quoted values are extracted from the mode and highest posterior density of the marginal $S_8$ posterior (denoted by M-HPD). Since the analysis of mock data shows that the marginal posterior mode or mean can be shifted significantly from the global best fit, due to a high-dimensional posterior with complex shape, we additionally provide the multivariate posterior maximum with an associated projected credible interval (PJ-HPD), $S_8=0.759^{+0.024}_{-0.021}$. For KiDS-1000 cosmic shear the two credible intervals are in very good agreement though, with nearly identical point estimates for $S_8$ and credible interval sizes differing by less than $5\,\%$ (this is also true for \textit{Planck} CMB constraints). The goodness of fit is acceptable, ranging from a $p$-value of 0.16 for COSEBIs to 0.03 for 2PCFs and 0.01 for band powers. Since the latter two preferentially extract information from higher angular frequencies relative to COSEBIs, this could indicate an as yet insignificant limitation in our non-linear modelling, e.g. in the intrinsic alignment of galaxies. On the other hand given the consistency between the values of $S_8$ for COSEBIs, 2PCFs, and band powers, this could be a result of an unfortunate noise realisation that affects the higher $\ell$-modes. 

Due to the tighter constraints of KiDS-1000, the tension in $\Sigma_8$ with \citet{Planck2018} has increased to $3.4\sigma$, i.e. a 7 in 10\,000 chance of a mere statistical fluctuation between the low and high-redshift probes assuming Gaussian distributions ($3\sigma$ in the less constrained $S_8$). Whether this discrepancy is mitigated by extensions to our cosmological model will be further investigated by \cite{troster/etal:2020b}, but the most obvious routes are unlikely to provide a satisfactory solution. For instance, KiDS and \textit{Planck} would be reconciled in significantly open cosmologies \citep{joudaki/etal:2017}, but \textit{Planck} prefers a positive curvature whose significance is still under debate (see \citealp{efstathiou20} and references therein). We argue that the tension with the CMB indeed manifests in the parameter $S_8$ (or $\Sigma_8$ if $S_8$ retains significant correlations with $\Omega_{\rm m}$), as was also observed in \citet{troster/etal:2020a}. Bayesian tension measures that act on the full shared parameter space between KiDS and \textit{Planck} are also provided, showing a substantial to strong evidence for disagreement.

We demonstrate that our constraints are robust to changes in the calibration procedures of multiplicative calibration in gravitational shear estimates, as well as of the redshift distributions. The $S_8$ credible intervals are not significantly affected by these changes either, which indicates that the KiDS-1000 constraints are statistics dominated. We also find no unexpected shifts in the inferred $S_8$ value when removing baryon feedback from the matter power spectrum model, when introducing additional flexibility to the intrinsic alignment model, or when removing all tomographic bin combinations involving a certain bin from the data vector. A Bayesian internal consistency analysis of tomographic bin splits reveals significant tension (up to $3\sigma$) when isolating all bin combinations involving the second bin, whose signals have higher amplitude than expected for its mean redshift. This will be a priority to investigate further in forthcoming work. 
However, excluding all elements of the KiDS-1000 data vector dependent on the second bin does not affect our cosmological constraints, which we therefore consider robust to this effect.

Looking ahead to the Legacy analysis of the complete KiDS survey, the statistical power of cosmic shear measurements is going to further improve thanks to a $35\,\%$ increase in sky area and a second pass in the $i$-band over the full survey. New, dedicated VST observations in spectroscopic survey fields will consolidate the redshift calibration and yield gains especially at redshifts beyond unity, unlocking the potential for very high signal-to-noise cosmic shear signals beyond our current highest-redshift bin. An upgrade to full multi-band image simulations will improve both the precision and accuracy of the shear calibration. Together with the innovation and cross-comparison opportunities provided by the contemporaneous DES and HSC cosmic shear measurements, we can therefore be optimistic that decisive new insights into the structure-growth tension will be delivered even before the next generation of powerful weak lensing surveys will begin to take data.

\begin{acknowledgements}

The chains are plotted with {\sc ChainConsumer} \citep{Hinton2016}: samreay.github.io/ChainConsumer. We are grateful to Eric Tittley, especially for saving our data. We thank Matthias Bartelmann, out external blinder, for keeping the key to our blinded data, which he revealed to us on the 9th of July. We also acknowledge Joe Zuntz for his help over the years with {\sc CosmoSIS}. We thank Joachim Harnois-Déraps, Shahab Joudaki,  Mohammadjavad Vakili and Ziang Yan for useful discussions. We are also thankful to the anonymous referee for their constructive comments.

This project has received funding from the European Union's Horizon 2020 research and innovation programme: We acknowledge support from the European Research Council under grant agreement No.~647112 (CH, MA, CL, BG and TT) and 770935 (HHi, AHW, AD and JLvdB).
CL is grateful for the working environment kindly provided by WPC Systems Ltd. during the pandemic.
TT acknowledges support under the Marie Sk\l{}odowska-Curie grant agreement No.~797794. 
CH acknowledges support from the Max Planck Society and the Alexander von Humboldt Foundation in the framework of the Max Planck-Humboldt Research Award endowed by the Federal Ministry of Education and Research. 
HHi is supported by a Heisenberg grant of the Deutsche Forschungsgemeinschaft (Hi 1495/5-1). 
HHo acknowledges support from Vici grant 639.043.512, financed by the Netherlands Organisation for Scientific Research (NWO).
KK acknowledges support by the Alexander von Humboldt Foundation. 
This work was partially enabled by funding from the UCL Cosmoparticle Initiative (BS).
MB is supported by the Polish Ministry of Science and Higher Education through grant DIR/WK/2018/12, and by the Polish National Science Center through grants no. 2018/30/E/ST9/00698 and 2018/31/G/ST9/03388.
JTAdJ is supported by the Netherlands Organisation for Scientific Research (NWO) through grant 621.016.402.
LM acknowledges support from STFC grant ST/N000919/1.
HYS acknowledges the support from NSFC of China under grant 11973070, the Shanghai Committee of Science and Technology grant No.19ZR1466600 and Key Research Program of Frontier Sciences, CAS, Grant No. ZDBS-LY-7013.\\

The KiDS-1000 results in this paper are based on data products from observations made with ESO Telescopes at the La Silla Paranal Observatory under programme IDs 177.A-3016, 177.A-3017 and 177.A-3018, and on data products produced by Target/OmegaCEN, INAF-OACN, INAF-OAPD and the KiDS production team, on behalf of the KiDS consortium. \\

{ {\it Author contributions:}  All authors contributed to the development and writing of this paper.  The authorship list is given in three groups:  the lead authors (MA, CL, BJ) followed by two alphabetical groups.  The first alphabetical group includes those who are key contributors to both the scientific analysis and the data products.  The second group covers those who have either made a significant contribution to the data products, or to the scientific analysis.}
\end{acknowledgements}


\bibliographystyle{aa}
\bibliography{references} 
 

\appendix
\section{Constraints on all parameters, additional tables and figures}
\label{app:extra}

\begin{figure}
   \begin{center}
     \begin{tabular}{c}
      \includegraphics[width=\hsize]{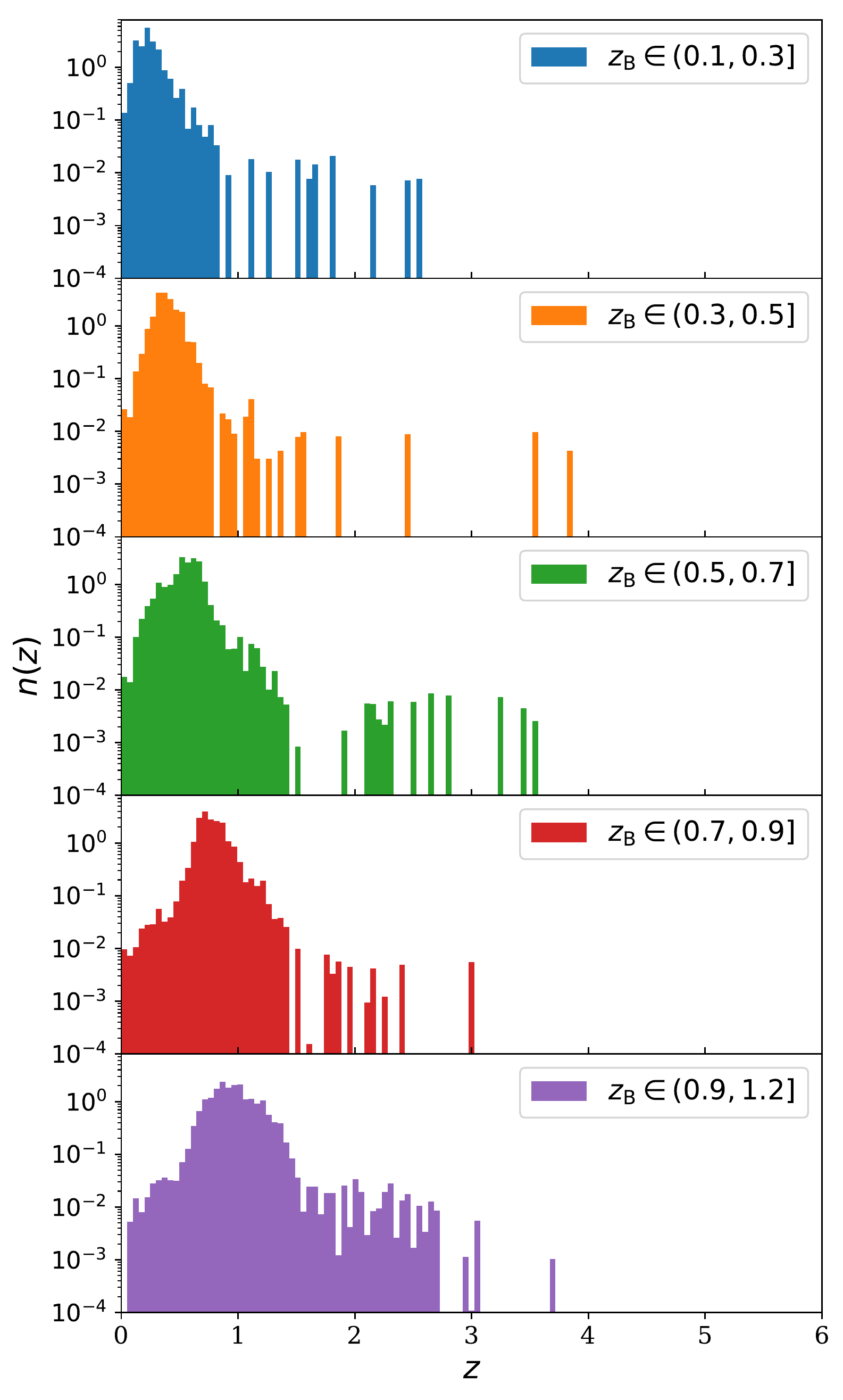} 
     \end{tabular}
   \end{center}
     \caption{Redshift distribution of sources in log-space. The distributions for each bin is shown for the full range of redshifts used in this analysis. Compare with \fig\ref{fig:photoz}.  }
     \label{fig:pofz_log}
 \end{figure}

\begin{table*}
\centering
\caption{Statistical properties of the redshift distribution of galaxies in each tomographic bin.}
\label{tab:nofz_stats}
\begin{tabular}{ c  c c c c c }
\hline
\hline \\[-2.2ex]
                               & bin 1                                 & bin 2                                    & bin 3                                  & bin 4                                   & bin 5 \\ \\[-2.2ex] \hline  \\[-2.2ex]
 $z_{\rm B}$ range & $0.1 < z_{\rm B} \leq 0.3$ &  $0.3 < z_{\rm B} \leq 0.5$ & $0.5 < z_{\rm B} \leq 0.7$  & $0.7 < z_{\rm B} \leq 0.9$  & $0.9 < z_{\rm B} \leq 1.2$\\ \\[-2.2ex]
mean                      &   0.26                               & 0.40                                     &  0.56                                    & 0.79                                    & 0.98                                \\ \\[-2.2ex]
std                          &   0.16                               &  0.16                                    &  0.20                                    & 0.16                                   & 0.25                                 \\ \\[-2.2ex]
Fraction with $z>2$  &   $0.1\%$                         & $0.1\%$                                &  $0.3\%$                             & $0.1\%$                             &  $1\%$                             \\ \\[-2.2ex]
\hline
\end{tabular}
\tablefoot{The first row lists the number associated with the tomographic bins. In the second row we show the  range of best-fitting photometric redshifts, $z_{\rm B}$, for each bin. The third and forth rows present the mean and standard deviation of the redshift distributions and finally the last row shows the fraction of galaxies with redshifts larger than 2.}
\end{table*}

\begin{figure*}
   \begin{center}
     \begin{tabular}{c}
      \includegraphics[height=0.46\hsize]{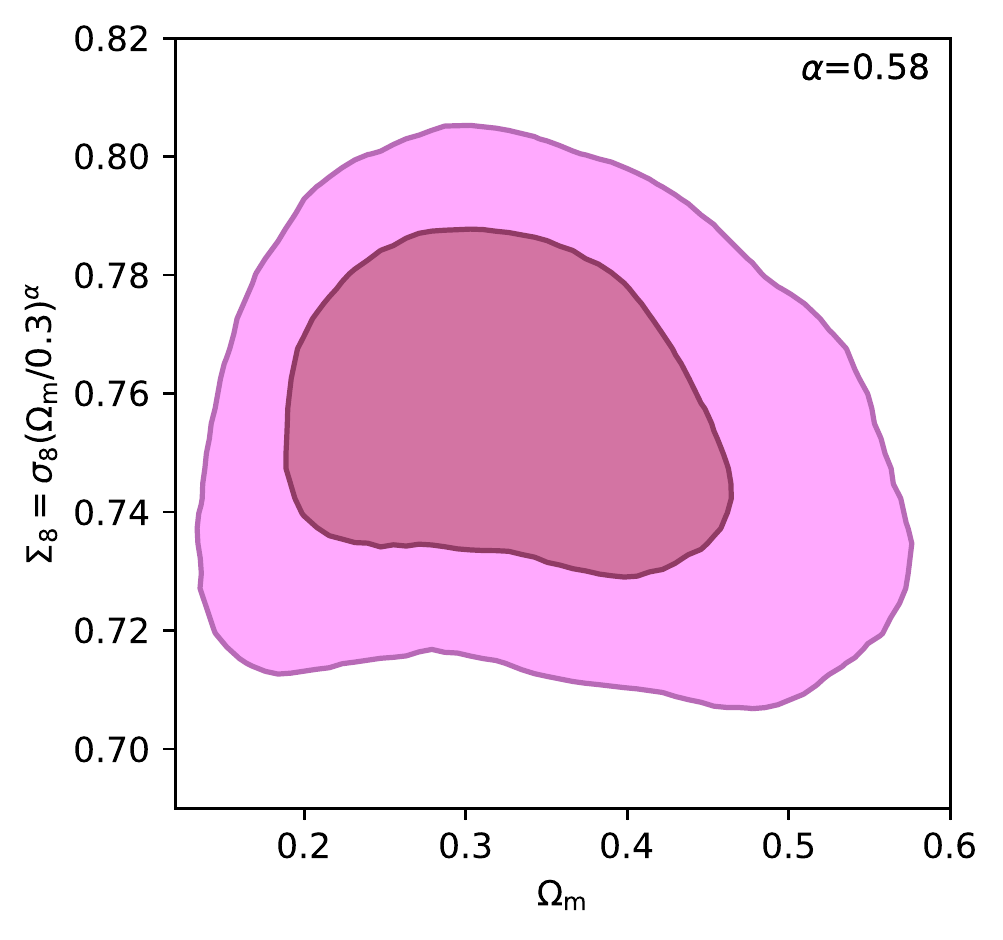} 
      \includegraphics[height=0.46\hsize]{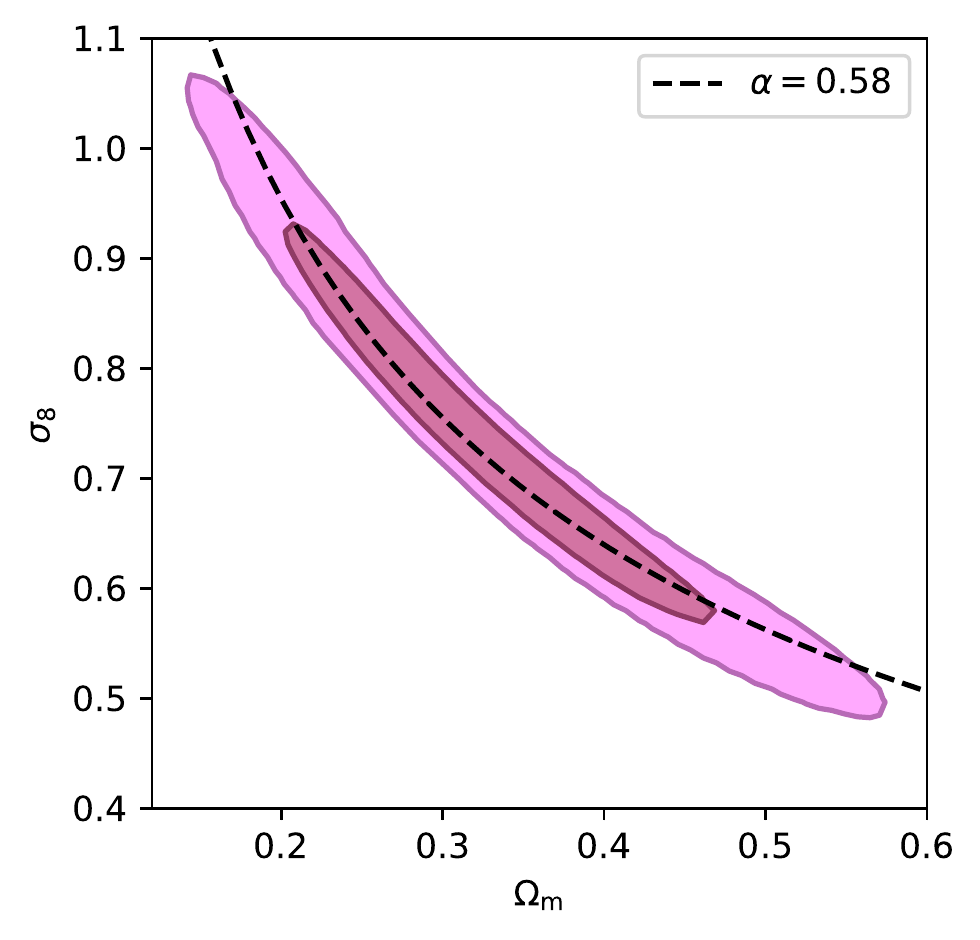} 
     \end{tabular}
   \end{center}
     \caption{The best-fitting curve of the form $\sigma_8=\Sigma_8(\Om/0.3)^{-\alpha}$ and its resulting $\Sigma_8$. Here we demonstrate the fitting method using band powers. The dashed curve in the right-hand panel shows the best-fitting function to all samples in the $\sigma_8$ and $\Om$ plane for which we find $\alpha=0.58$. The left-hand panel shows the resulting marginal $\Sigma_8$ posterior against $\Om$.}
     \label{fig:Sig8}
 \end{figure*}

\begin{figure*}
   \begin{center}
     \begin{tabular}{c}
      \includegraphics[width=\hsize]{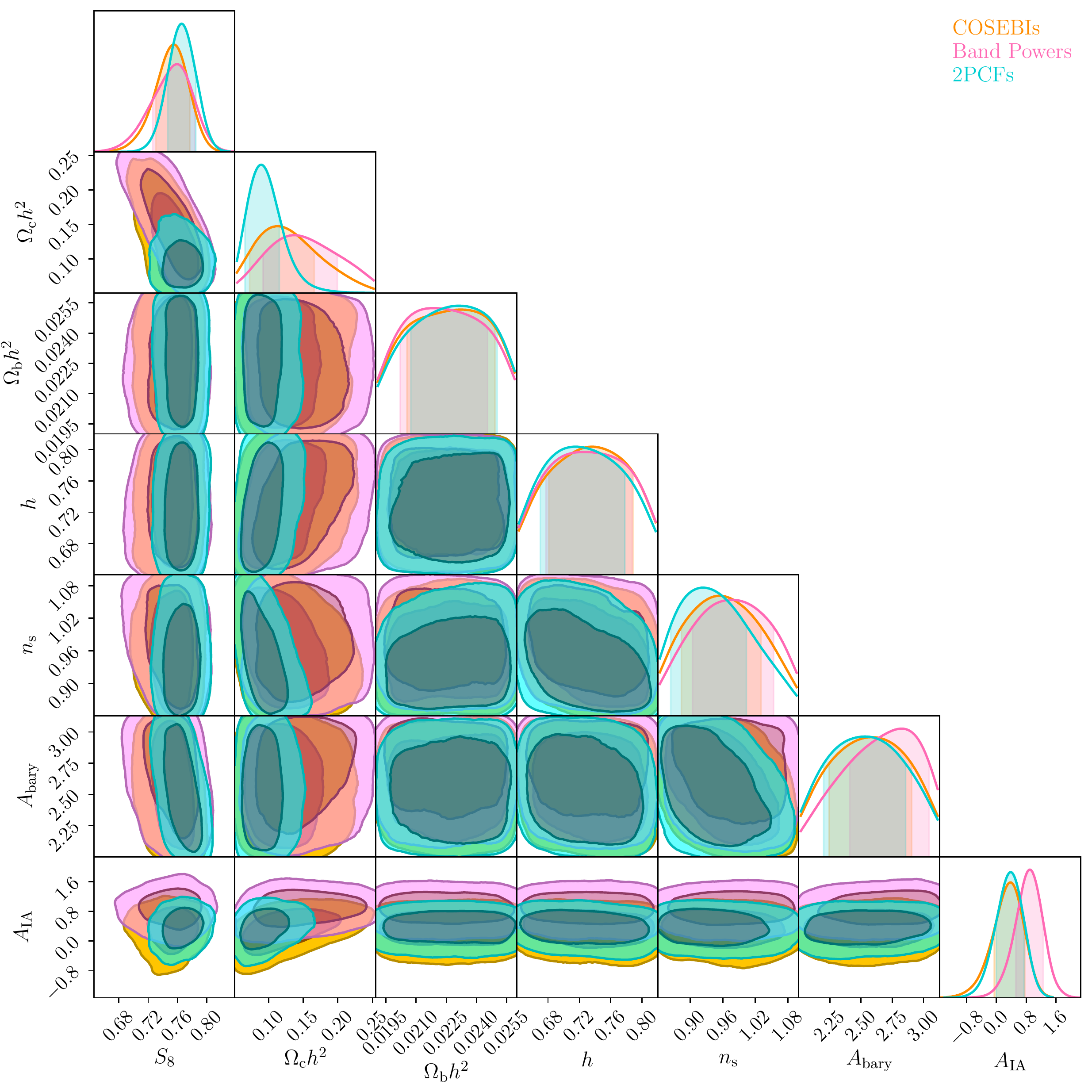}
     \end{tabular}
   \end{center}
     \caption{Constraints on sampled cosmological and astrophysical parameters. Results are shown for COSEBIs (orange), band powers (pink) and the 2PCFs (cyan). We use kernel density estimation to smooth the distributions, which in the case of poorly constrained parameters can produce artificial constraints near the prior boundaries (for example constraints on $h$ or $\Omega_{\rm b}h^2$).}
     \label{fig:triangle}
 \end{figure*}

\begin{figure*}
   \begin{center}
     \begin{tabular}{c}
      \includegraphics[width=\hsize]{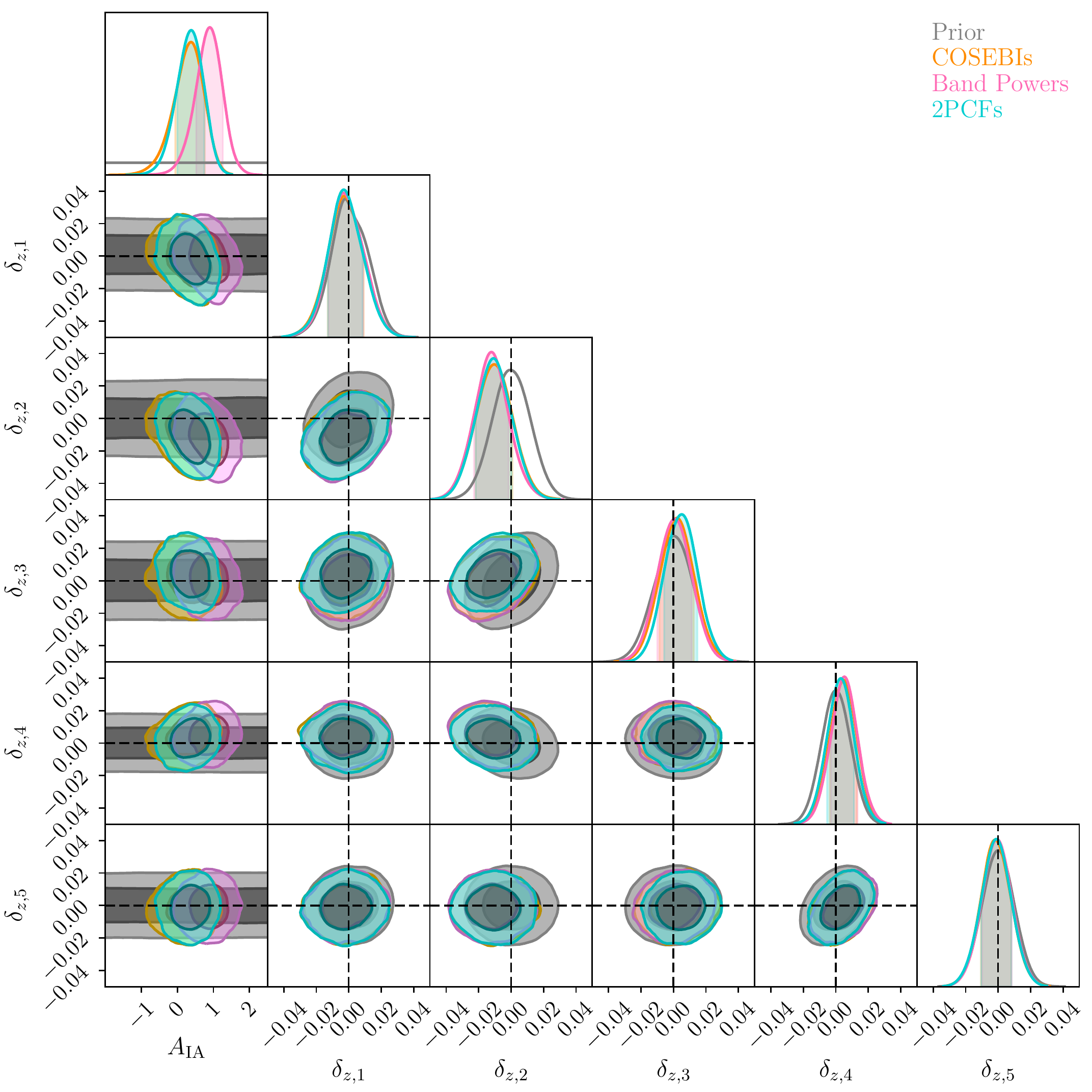}
     \end{tabular}
   \end{center}
     \caption{Constraints on $\delta_z$ and the intrinsic alignment amplitude $A_{\rm IA}$. The $\delta_z$ nuisance parameters represent our uncertainty in the mean of the redshift distributions.
     The input prior region is shown in grey. The prior for $A_{\rm IA}$ is flat within its boundaries (the full range is between $-6$ and $6$), while correlated Gaussian priors are used for the $\delta_z$ nuisance parameters (the $\delta_z$ priors are shifted to have a zero mean). Results are shown for COSEBIs (orange), band powers (pink) and the 2PCFs (cyan). }
     \label{fig:deltaz}
 \end{figure*}

\begin{figure}
   \begin{center}
     \begin{tabular}{c}
      \includegraphics[width=\hsize]{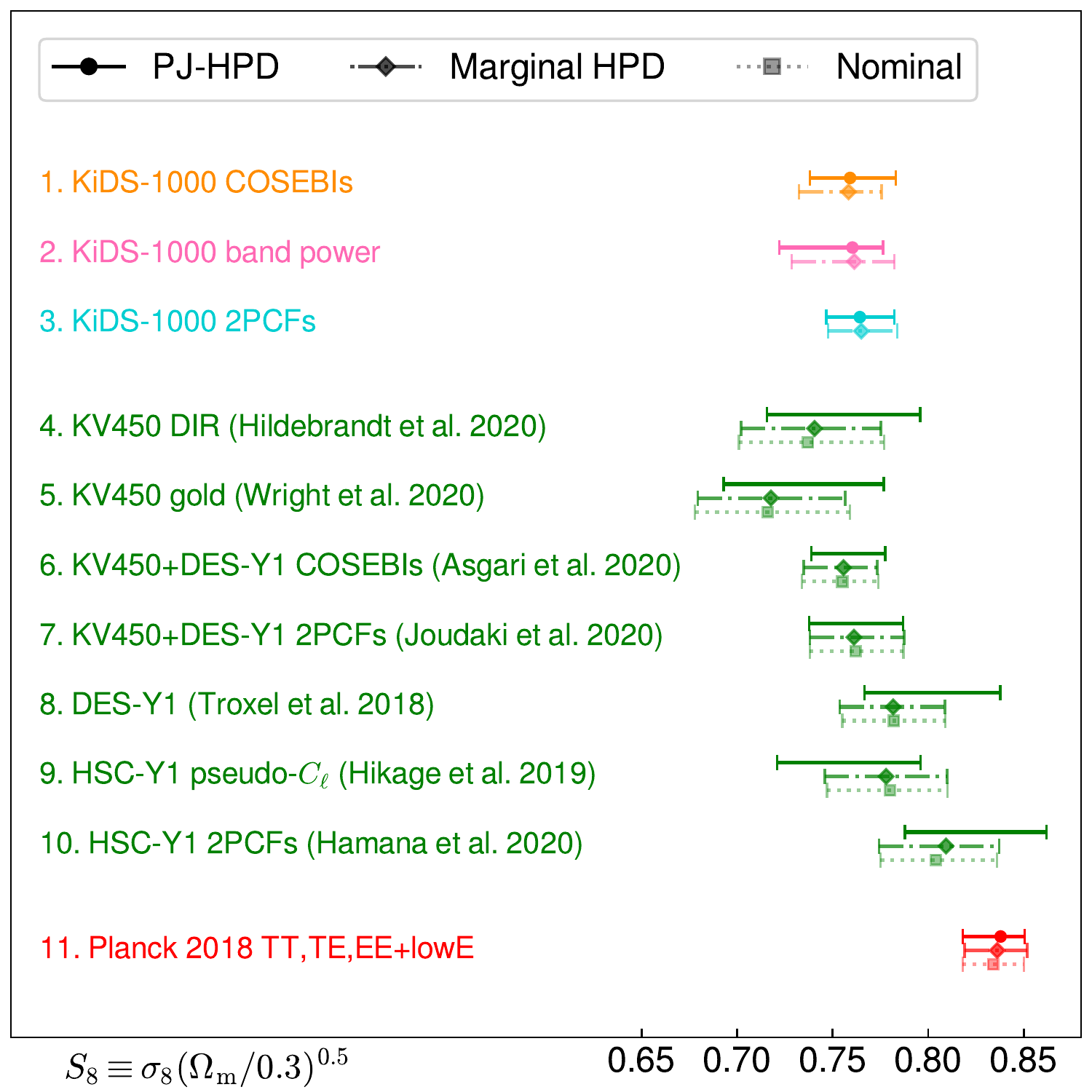}
     \end{tabular}
   \end{center}
     \caption{Comparison between $S_8$ constraints of different surveys (extended version of \fig\ref{fig:summary_surveys}). The top three group of bars show our KiDS-1000 results, for COSEBIs, band powers and 2PCFs. The green bars show the constraints from other cosmic shear surveys and the red ones refer to {\it Planck} 2018 results. The solid bar in each set shows the projected joint highest posterior density (PJ-HPD) credible region encompassing $68.3\%$ of all sampled points (with the multivariate maximum posterior where determined).
The dot-dashed bar displays the $1\sigma$ credible region around the maximum of the marginal distribution of $S_8$ (Marginal HPD). For the external results we plot a third bar (dotted) showing their nominal reported values. }
     \label{fig:summary_surveys_all}
 \end{figure}

In this appendix we provide additional material that complement the findings presented in the main body of the paper. 
We compared the redshift distributions of the five tomographic bins in \fig\ref{fig:photoz}. Figure\thinspace\ref{fig:pofz_log} shows these distributions for the full range of redshifts that we consider in this analysis, $0 \leq z \leq 6$. We use a logarithmic scale for the vertical axis in this figure, to show the level of suppression at the tails of the distributions. Table\thinspace\ref{tab:nofz_stats} presents the mean, standard deviation and the fraction of galaxies with redshift beyond $z>2$. 

In \sect\ref{sec:fid} we showed our main results, focusing mainly on $S_8$ and $\Sigma_8$ constraints, where we argued that, in general, $S_8$ does not capture the best-constrained direction perpendicular to the $\Om$--$\sigma_8$ degeneracy. Figure\thinspace\ref{fig:Sig8} demonstrates our fitting method on band powers, which we use to find an appropriate $\alpha$ that captures the direction perpendicular to the degeneracy line. 
In the right-hand panel we show the best-fitting $\sigma_8=\Sigma_8(\Om/0.3)^{-\alpha}$ (dashed curve) to the sampled $\sigma_8$ and $\Om$ posterior points for the band powers. 
We find $0.58$ to be the best-fit value for $\alpha$. 
In the left-hand panel we show the resulting $\Sigma_8$ and $\Om$ for $\alpha$ fixed to $0.58$. Here we calculate $\Sigma_8$ for each point in the samples separately. Comparing the left-hand side of this figure with \fig\ref{fig:sig8Om} we see that $\Sigma_8$ has smaller correlation with $\Om$ than $S_8$.

While we showed best-fit values and credible regions for $S_8$ and $\Sigma_8$ in \sect\ref{sec:fid}, here we provide credible regions for all constrained parameters in \tab\ref{tab:bestfitall}. We note that our constraints for most parameters are prior-dominated and therefore in the case of flat priors the credible regions are affected by smoothing and the sampler reaching the edge of the prior range. In the table we report constrained parameters in bold.  To assess which parameters are constrained, we consider the relative height of the $2\sigma$ levels and the maximum of a one dimensional Gaussian,
\begin{equation}
\frac{{\rm Pr}(\theta=\mu\pm2\sigma)}{{\rm Pr}(\theta=\mu)} \approx 0.135\;,
\end{equation}
where ${\rm Pr}(\theta)$ is a Gaussian distribution with mean $\mu$ and variance $\sigma^2$. If the relative amplitude of the marginal distribution for a parameter between its two extremes and its maximum is smaller than $0.135$ we deduce that this parameter is constrained (this is done with the binned distributions without any extra smoothing). With this criterion we see that $S_8$ and $A_{\rm IA}$ are the only two physical parameters that are constrained. We do not use this criterion for the second group of parameters, which are derived from the first group, since their prior is non-flat. 
 We note that the $\Om$ constraints are also prior dominated as demonstrated in J20.
The last group of parameters in \tab\ref{tab:bestfitall} have Gaussian priors; therefore they, by definition, pass the criterion described above. Their constraints, however, are very similar to the size of the input Gaussian priors, hence we do not show them in bold.

Figure\thinspace\ref{fig:triangle} shows marginalised credible regions for all the sampled cosmological and astrophysical parameters. We show results for COSEBIs (orange), band powers (pink) and 2PCFs (cyan). We apply a kernel density estimation (KDE) method to smooth the distributions. For parameters with poor constraints, e.g. $\Omega_{\rm b}h^2$, KDE smoothing creates artificial constraints by smoothing the edges of the distribution where it hits the limits of the flat prior range. In \fig\ref{fig:deltaz} we show our constraints for $\delta_{z,i}=z_i^{\rm est}-z_i^{\rm true}$ (see \sect\ref{sec:data}) and $A_{\rm IA}$. Here we have shifted the contours by the $\Delta z$ values in \tab\ref{tab:dataprops} to centre the prior, shown in grey, on zero (dashed lines).  Any shift from zero for the $\delta_z$ parameters is indicative of a self-calibration by the cosmic shear data. We see that the $\delta_z$ contours mostly recover the input prior and that $\delta_z$ values are consistent with zero within their $1\sigma$ marginal region. The largest deviation is found for the second tomographic bin, where we see an almost $1\sigma$ shift towards negative values, indicating a preference for a redshift distribution with a larger mean. This suggests that the shifted SOM redshift distributions have underestimated the mean of the true redshift of the galaxies in bin 2. We have seen other indications in the data for an anomaly in the distribution of the second bin. 
Our internal consistency tests (see \app\ref{app:consistency} for more details) also flag the second bin as an anomaly. Nevertheless, in \fig\ref{fig:summary_nuisance} we showed that excluding redshift bin 2 has a negligible effect on our final results. 

In \fig\ref{fig:triangle} we see a mild correlation between the $\delta_z$ parameters and the $A_{\rm IA}$. This correlation decreases for higher redshift bins where the signal is less affected by intrinsic alignments of galaxies.  Band powers show a preference for a higher $A_{\rm IA}$ compared to COSEBIs and 2PCFs, with a maximum marginal value that is $0.53$ larger. In our mock analysis, described in \app\ref{app:stagetests}, we find that a $\Delta A_{\rm IA}\geq 0.53$ occurs in about $5\%$ of the noise realisations. We conclude that this difference is a result of the particular noise realisation in our data, given that all three summary statistics show consistent constraint for $S_8$. 

In \fig\ref{fig:summary_surveys} we compared the KiDS-1000 constraints to a selection of recent cosmic shear and the {\it Planck} results. In \fig\ref{fig:summary_surveys_all} we show results for a larger selection of cosmic shear surveys and also include the reported nominal $S_8$ constraints by each external analysis (using various estimates of central values and credible intervals). We see that the nominal results are, in all cases, very close to our estimated marginal highest density credible region. Our cosmic shear results are consistent with all the results shown here, which all report $S_8$ values that are smaller than the inferred value from {\it Planck}.

\begin{table*}
\centering
\caption{Marginal constraints on all model parameters.}
\label{tab:bestfitall}
\resizebox{2\columnwidth}{!}{
\begin{tabular}{ccccccc}
\hline\hline\\[-2.2ex]
                         & \multicolumn{2}{c}{COSEBIs} & \multicolumn{2}{c}{Band Power} & \multicolumn{2}{c}{2PCFs} \\
                         &          Best fit + PJ-HPD &             Max + marginal &          Best fit + PJ-HPD &             Max + marginal &          Best fit + PJ-HPD &             Max + marginal \\
\hline\\[-2ex]
                   $S_8$ & $\mathbf{ 0.759}^{\mathbf{+0.024}}_{\mathbf{-0.021}}$ & $\mathbf{ 0.758}^{\mathbf{+0.017}}_{\mathbf{-0.026}}$ & $\mathbf{ 0.760}^{\mathbf{+0.016}}_{\mathbf{-0.038}}$ & $\mathbf{ 0.761}^{\mathbf{+0.021}}_{\mathbf{-0.033}}$ & $\mathbf{ 0.764}^{\mathbf{+0.018}}_{\mathbf{-0.017}}$ & $\mathbf{ 0.765}^{\mathbf{+0.019}}_{\mathbf{-0.017}}$ \\[0.8ex]
  $\Omega_\mathrm{c}h^2$ & $ 0.118^{+0.034}_{\mathbf{-0.054}}$ & $ 0.105^{+0.056}_{\mathbf{-0.033}}$ & $ 0.107^{\mathbf{+0.078}}_{-0.027}$ & $ 0.132^{\mathbf{+0.063}}_{-0.040}$ & $ 0.079^{+0.032}_{\mathbf{-0.012}}$ & $ 0.088^{+0.024}_{\mathbf{-0.021}}$ \\[0.8ex]
  $\Omega_\mathrm{b}h^2$ & $ 0.026^{+0.000}_{-0.005}$ & $ 0.023^{+0.002}_{-0.003}$ & $ 0.026^{+0.000}_{-0.005}$ & $ 0.022^{+0.002}_{-0.002}$ & $ 0.019^{+0.005}_{-0.000}$ & $ 0.023^{+0.002}_{-0.002}$ \\[0.8ex]
                     $h$ & $ 0.767^{+0.047}_{-0.065}$ & $ 0.727^{+0.065}_{-0.045}$ & $ 0.640^{+0.124}_{-0.000}$ & $ 0.704^{+0.087}_{-0.025}$ & $ 0.666^{+0.110}_{-0.000}$ & $ 0.711^{+0.066}_{-0.042}$ \\[0.8ex]
          $n_\mathrm{s}$ & $ 0.901^{+0.100}_{-0.055}$ & $ 0.949^{+0.082}_{-0.065}$ & $ 1.001^{+0.043}_{-0.108}$ & $ 0.999^{+0.059}_{-0.091}$ & $ 0.927^{+0.093}_{-0.049}$ & $ 0.928^{+0.068}_{-0.068}$ \\[0.8ex]
         $A_\mathrm{IA}$ & $\mathbf{ 0.264}^{\mathbf{+0.424}}_{\mathbf{-0.337}}$ & $\mathbf{ 0.389}^{\mathbf{+0.354}}_{\mathbf{-0.413}}$ & $\mathbf{ 0.973}^{\mathbf{+0.292}}_{\mathbf{-0.383}}$ & $\mathbf{ 0.917}^{\mathbf{+0.332}}_{\mathbf{-0.357}}$ & $\mathbf{ 0.387}^{\mathbf{+0.321}}_{\mathbf{-0.374}}$ & $\mathbf{ 0.370}^{\mathbf{+0.364}}_{\mathbf{-0.339}}$ \\[0.8ex]
       $A_\mathrm{bary}$ & $ 2.859^{+0.199}_{-0.497}$ & $ 2.558^{+0.352}_{-0.316}$ & $ 3.130^{+0.000}_{-0.623}$ & $ 2.842^{+0.231}_{-0.402}$ & $ 2.816^{+0.046}_{-0.611}$ & $ 2.583^{+0.275}_{-0.388}$ \\[0.4ex]
\hline\\[-2ex]
              $\sigma_8$ & $ 0.838^{+0.140}_{-0.141}$ & $ 0.772^{+0.146}_{-0.123}$ & $ 0.730^{+0.116}_{-0.134}$ & $ 0.723^{+0.124}_{-0.130}$ & $ 0.887^{+0.084}_{-0.107}$ & $ 0.895^{+0.095}_{-0.095}$ \\[0.8ex]
     $\Omega_\mathrm{m}$ & $ 0.246^{+0.101}_{-0.060}$ & $ 0.253^{+0.088}_{-0.074}$ & $ 0.326^{+0.115}_{-0.077}$ & $ 0.313^{+0.088}_{-0.094}$ & $ 0.223^{+0.065}_{-0.033}$ & $ 0.211^{+0.051}_{-0.038}$ \\[0.8ex]
          $A_\mathrm{s}$ & $ 2.422^{+5.379}_{-1.238}$ & $ 1.134^{+1.907}_{-0.816}$ & $ 2.095^{+2.011}_{-1.354}$ & $ 0.822^{+1.237}_{-0.597}$ & $ 4.300^{+2.445}_{-1.903}$ & $ 2.866^{+2.217}_{-1.305}$ \\[0.4ex]
\hline\\[-2ex]
            $\delta z_1$ & $ 0.002^{+0.009}_{-0.011}$ & $ 0.003^{+0.009}_{-0.012}$ & $ 0.002^{+0.008}_{-0.012}$ & $ 0.003^{+0.008}_{-0.012}$ & $ 0.002^{+0.009}_{-0.010}$ & $ 0.003^{+0.009}_{-0.011}$ \\[0.8ex]
            $\delta z_2$ & $ 0.009^{+0.011}_{-0.010}$ & $ 0.009^{+0.010}_{-0.011}$ & $ 0.011^{+0.014}_{-0.008}$ & $ 0.009^{+0.011}_{-0.009}$ & $ 0.009^{+0.012}_{-0.009}$ & $ 0.009^{+0.010}_{-0.011}$ \\[0.8ex]
            $\delta z_3$ & $-0.014^{+0.010}_{-0.009}$ & $-0.016^{+0.011}_{-0.009}$ & $-0.013^{+0.011}_{-0.009}$ & $-0.015^{+0.011}_{-0.009}$ & $-0.018^{+0.008}_{-0.011}$ & $-0.018^{+0.010}_{-0.009}$ \\[0.8ex]
            $\delta z_4$ & $-0.016^{+0.007}_{-0.008}$ & $-0.016^{+0.008}_{-0.007}$ & $-0.016^{+0.007}_{-0.009}$ & $-0.016^{+0.007}_{-0.008}$ & $-0.014^{+0.008}_{-0.008}$ & $-0.014^{+0.008}_{-0.008}$ \\[0.8ex]
            $\delta z_5$ & $ 0.007^{+0.010}_{-0.007}$ & $ 0.008^{+0.007}_{-0.010}$ & $ 0.007^{+0.008}_{-0.009}$ & $ 0.007^{+0.008}_{-0.009}$ & $ 0.007^{+0.008}_{-0.009}$ & $ 0.007^{+0.009}_{-0.008}$ \\[0.8ex]
         $10^4 \delta_c$ & - & - & - & - & $-0.006^{+1.936}_{-2.371}$ & $-0.205^{+2.386}_{-1.935}$ \\[0.4ex]
\hline
\end{tabular}
}
\tablefoot{We show two sets of estimates for each parameter and the three two-point statistics employed in this work. The estimates are the maximum posterior of the full multivariate distribution (MAP) together with the projected joint highest posterior density (PJ-HPD) interval, and the maximum of the one-dimensional marginal distributions with the marginal highest density credible interval (CI). In the first block of parameters we show estimates in bold, if they are constrained on both sides given the criterion in the text. For parameters that are only constrained on one side we show only the corresponding error in bold.}
\end{table*}

\section{Consistency tests}

\label{app:consistency}

We perform a number of internal consistency tests on the KiDS-1000 data at the level of parameter estimates and posteriors. In \app\ref{app:stagetests} we detail tests of consistency between the constraints from the three different two-point statistics. We follow the methodology of \cite{kohlinger/etal:2019} to quantify the internal consistency between different divisions of the data based on tomographic bins (\app\ref{app:internal_consistency}). The details of the consistency test with respect to the primordial {\it Planck} results are shown in \app\ref{app:external_consistency}. A summary of this appendix can be found in Sects.\thinspace\ref{sec:consist} and \ref{sec:constst_external}.

\subsection{Consistency between statistics}
\label{app:stagetests}

\begin{figure}
   \begin{center}
     \begin{tabular}{c}
      \includegraphics[width=\columnwidth]{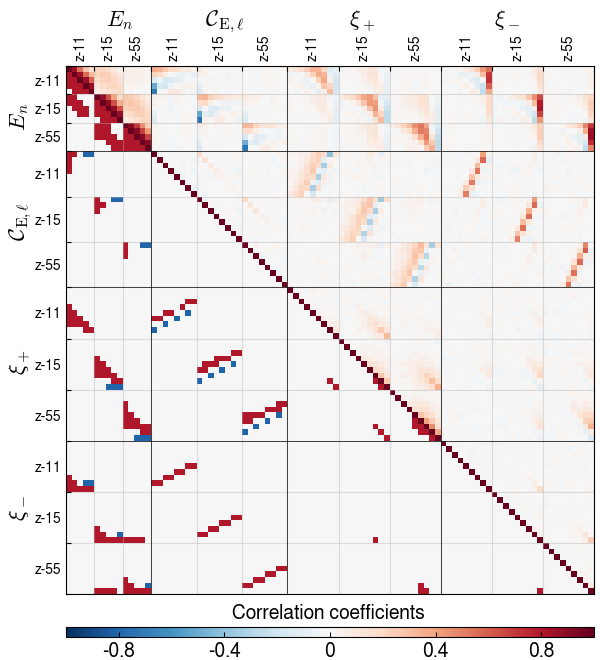} 
     \end{tabular}
   \end{center}
     \caption{Cross-correlation matrix between COSEBIs ($E_n$), band powers (${\mathcal C}_{{\rm E},l} $) and 2PCFs ($\xi_\pm$) from {\sc Salmo} mocks. The top triangle shows the cross-correlation values corresponding to the colour-bar. The bottom triangle highlights the entries with more than $20\%$ (red) or less than $-20\%$ (blue) correlation. We show results for tomographic bin combinations of the lowest and highest redshift bins only, i.e. three blocks per statistic containing the bin combinations 1-1, 1-5, and 5-5. }
     \label{fig:crosscov}
 \end{figure}

\begin{figure*}
   \begin{center}
     \begin{tabular}{c}
      \includegraphics[height=\columnwidth]{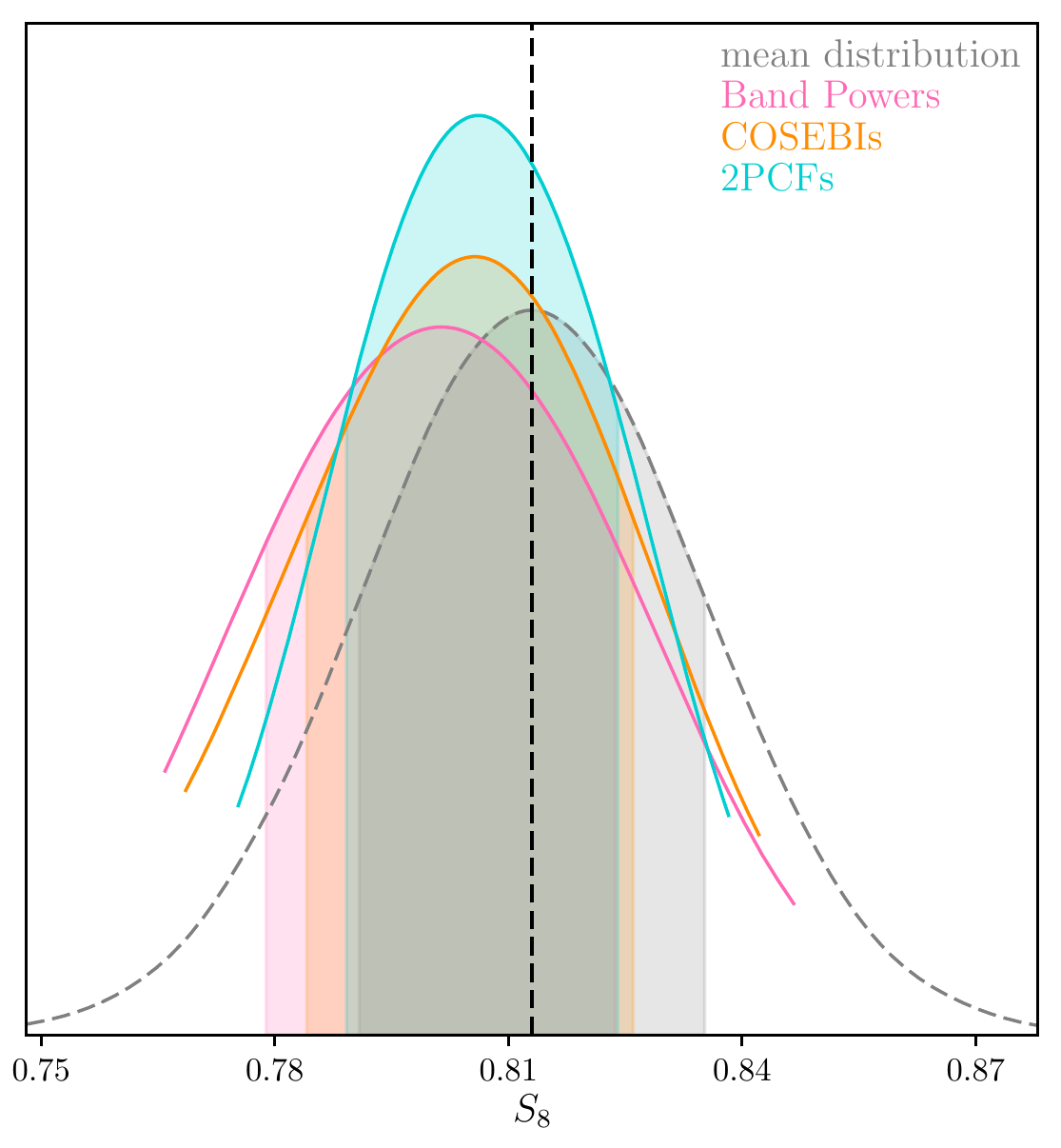} 
      \includegraphics[height=\columnwidth]{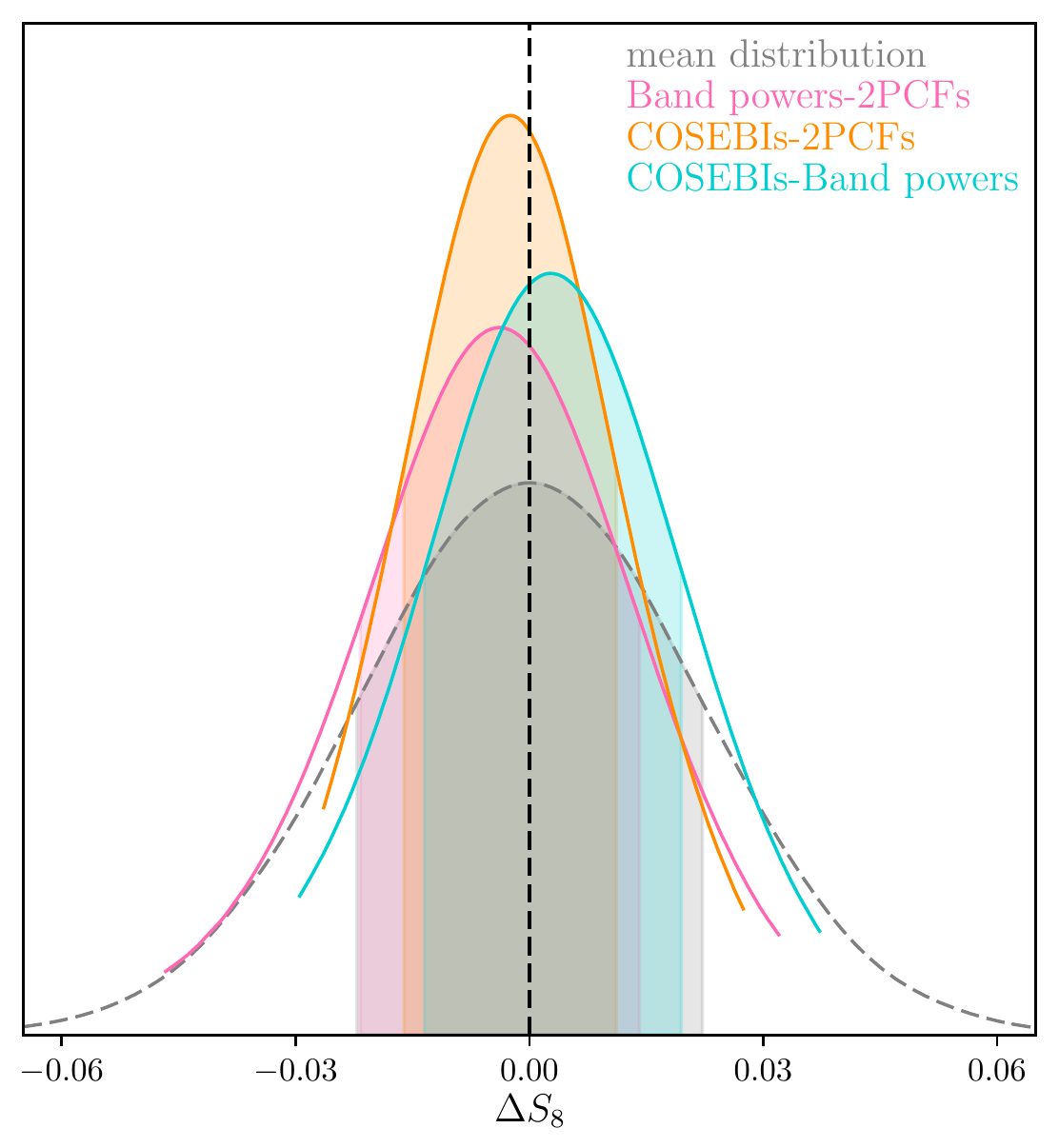}
     \end{tabular}
   \end{center}
     \caption{ Distribution of inferred $S_8$ values from 100 realisations of the data vector sampled from the covariance matrix. {\it Left:} The distribution of the maximum of the marginal distribution for $S_8$. Results are shown for COSEBIs (orange), band powers (pink) and 2PCFs (cyan). For comparison we show a Gaussian distribution centred at the input value of $S_8$ and a standard deviation equal to the mean of the individual standard deviations for each realisation and set of two-point statistics (grey dashed curve). {\it Right:} The difference between the $S_8$ posterior modes of pairs of two-point statistics (as indicated in the legend) given the same noise realisation. The same reference Gaussian distribution is shown in grey (dashed curve) but centred on zero. }
     \label{fig:stage3}
 \end{figure*}

\begin{figure}
   \begin{center}
     \begin{tabular}{c}
      \includegraphics[width=\columnwidth]{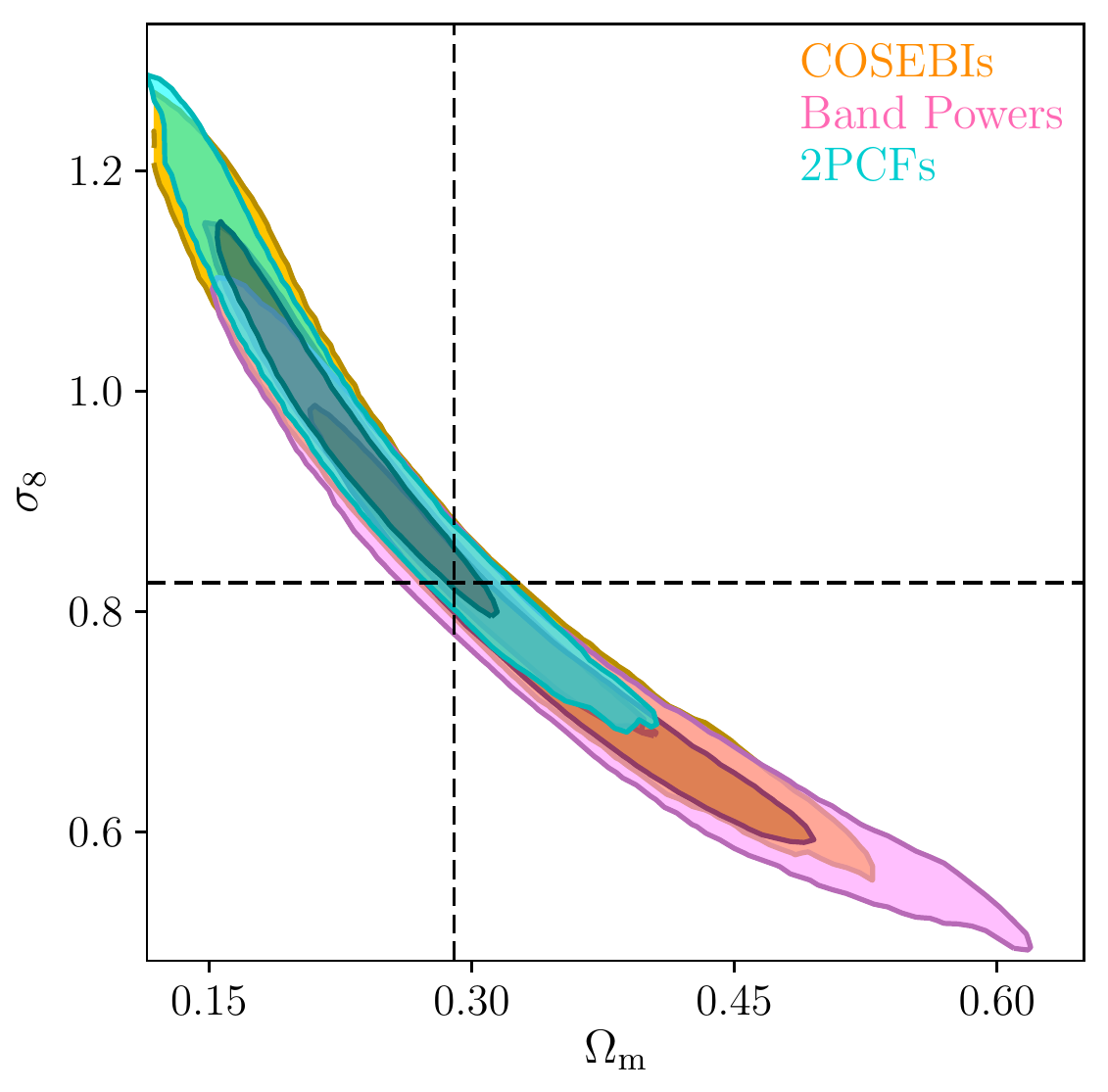} 
     \end{tabular}
   \end{center}
     \caption{Marginal posterior from mock data displaying a similar degeneracy to the real KiDS-1000 data (compare with \fig\ref{fig:sig8Om}). The input values for $\sigma_8$ and $\Om$ are shown with the dashed lines. These are results for one of the 100 mock realisations that we analysed. In the same set of realisations we find a number of similar results, with shortened contours for one or more of the statistics. }
     \label{fig:sig8Om_mock}
 \end{figure}

The two-point statistics that we consider have differing sensitivities to $\ell$-scales as shown in \fig\ref{fig:compare}. Therefore, despite being measured from the exact same data set, we do not expect them to find the same constraints on cosmological parameters. 
Previously, seemingly incompatible results from analysis of the same data with different two-point statistics has been seen. For example, the quadratic power spectrum estimator developed by \citet{koehlinger17} yielded a lower value of $S_8$ compared to the 2PCF analysis on the same KiDS data set (using Data Release 3). In addition, the HSC analysis of \cite{hamana/etal:2020} using 2PCFs found a higher value of $S_8$ compared to the pseudo-CL analysis of \cite{Hikage18}. Unlike these previous analyses, here we quantify the level of difference that we expect for constriants from our summary statistics.

To quantify the expected difference between 2PCFs, band powers, and COSEBIs we analyse mock data. We produce mock data by adding noise to a theoretical data vector. We draw the noise realisations from multivariate Gaussian distributions based on a cross-covariance between the different statistics. To estimate this cross-covariance we use the {\sc Salmo} simulations described in section 4 of J20.

In \fig\ref{fig:crosscov} we show the cross-correlation matrix between the three two-point statistics showing combinations with redshift bins 1 and 5. We can see sub-matrices for the auto-correlations of each of the statistics, as labelled in the figure. The top triangle entries show the level of cross-correlations, while the bottom triangle shows all the values that exceed $\pm 20\%$ (red for positive and blue for negative values). We see that our two-point statistics have non-negligible cross-correlations, with highest values belonging to correlations between $\xi_-$ and band powers or COSEBIs. The figure also shows negative elements presenting anti-correlations. These are most pronounced in the case of COSEBIs and band powers.
In addition, we see that many of the elements of the cross-covariance are small, showing a lack of correlation. For example, the small-scale $\xi_\pm$ is not used by the other statistics, and also COSEBIs are uncorrelated with the high-$\ell$ modes of the band powers.

With this cross-covariance we produce 100 realisations of a data vector containing COSEBIs, band powers and 2PCFs. We then divide the data vector and covariance matrix based on each set of statistics that we used in our fiducial analysis. We apply the same setup and pipeline as described in \sect\ref{sec:data} to these mock data and find parameter constrains. Figure\thinspace\ref{fig:stage3} shows the resulting distribution of the maximum of the marginals for $S_8$  (left panel). For comparison we also show a Gaussian distribution centred on the input $S_8$ with the averaged standard deviation of all the chains. The right-hand panel shows the distribution of the difference between the $S_8$ posterior modes shown in the left-hand panel for each pair of two-point statistics given the same noise realisation.

From the left-hand side of \fig\ref{fig:stage3} we can immediately see that the distribution for band powers is wider than for COSEBIs which in turn is wider than for 2PCFs. 
This results from the choice of $\alpha$ for $S_8$, which is  not perpendicular to the $\sigma_8$ and $\Om$ degeneracy for COSEBIs and band powers. As expected, we find the maximum of the marginal distribution to be biased with respect to the input $S_8$ (also see section 6.4 of J20). 2PCFs show the smallest bias, however they also possess the tightest distribution, resulting in a similar relative bias compared to their width (see the discussion on MAP versus maximum marginal values in \sect\ref{sec:fid}). On the right-hand side we see that the $\Delta S_8$ between two statistics has a comparable size to the mean distribution shown in grey. To assess the level of difference that we expect for $\Delta S_8$, we compare the width of each distribution with the mean distribution using $\sigma$ values coming from the two statistics that are compared. We find that $\Delta S_8$ is only $20-30\%$ tighter than its corresponding mean values, comparing any two of the statistics. 
This means that we do not expect to find perfect agreement between the results of different two-point statistics. In the KiDS-1000 analysis we find the largest $S_8$ difference to be between COSEBIs and 2PCFs. Based on the analysis here we conclude that this difference of $0.4\sigma$ is expected. 

To assess the fidelity of this result, we estimated a theoretical covariance between COSEBIs and 2PCFs for a non-tomographic analysis and repeated the analysis with mock data produced with the theoretical cross-covariance. We find consistent results between this test and the previous one.

Our parameter constraints for $\sigma_8$ and $\Om$ in \fig\ref{fig:sig8Om} show that the $\xi_\pm$ results are shifted along the degeneracy line towards high $\sigma_8$ and low $\Om$, such that they touch the edge of our prior range. This seemingly large effect is fully consistent with a noise fluctuation, and among the aforementioned 100 mock realisations we saw many examples with similar trends. Figure\thinspace\ref{fig:sig8Om_mock} shows one such realisation. In some of the other realisations COSEBI or band power contours are shifted high along the degeneracy direction. In general, we find that the contours for poorly constrained parameters can move towards the edge of their prior range producing one-sided constraints, while shortening the marginalised posterior distributions. Given the hard cut at the prior edge, this will appear as a tighter constraint on a parameter, although it is fully dependent on the noise realisation. We see another example of this effect in the KiDS-1000 data in \fig\ref{fig:triangle} where the constraints on $A_{\rm bary}$ with band powers appear tighter than the results of COSEBIs or 2PCFs (also see the $n_{\rm s}$ constraints). 

\subsection{Internal consistency of KiDS data}
\label{app:internal_consistency}

\begin{figure*}
   \begin{center}
     \begin{tabular}{c}
      \includegraphics[width=1.7\columnwidth]{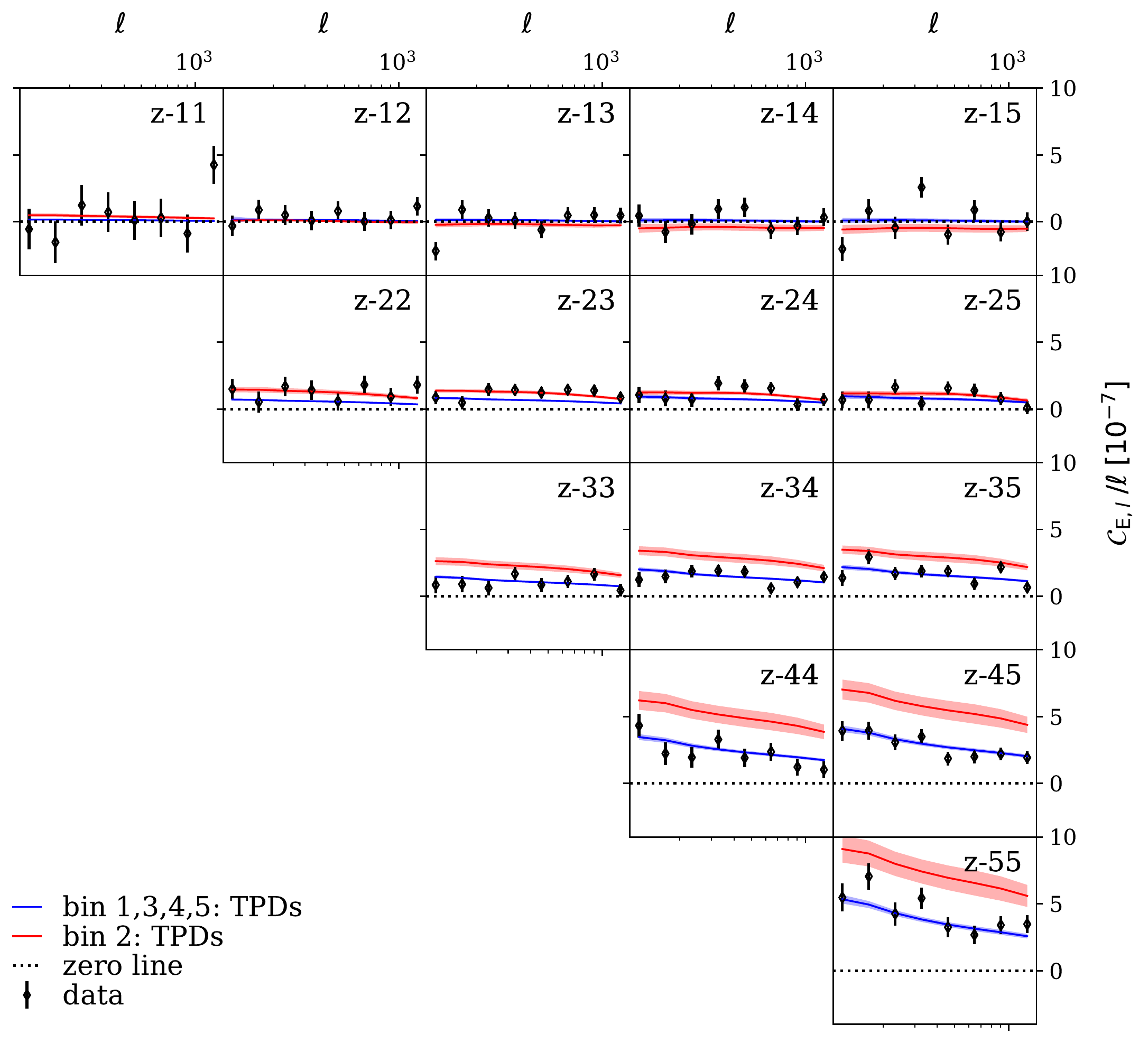} 
     \end{tabular}
   \end{center}
     \caption{Band power data compared to the best-fit model from the internal consistency test that isolates all bin combinations involving the second tomographic bin. The red curves show the translated posterior distributions (TPDs) resulting from the second bin and its cross-correlations. The blue curves are the TPDs derived from the remainder of the tomographic bins and their combinations.  The shaded bands around the curves show their standard deviations. }
     \label{fig:bp_bin2vsAll}
 \end{figure*} 
 
 \begin{figure}
   \begin{center}
     \begin{tabular}{c}
      \includegraphics[width=\columnwidth]{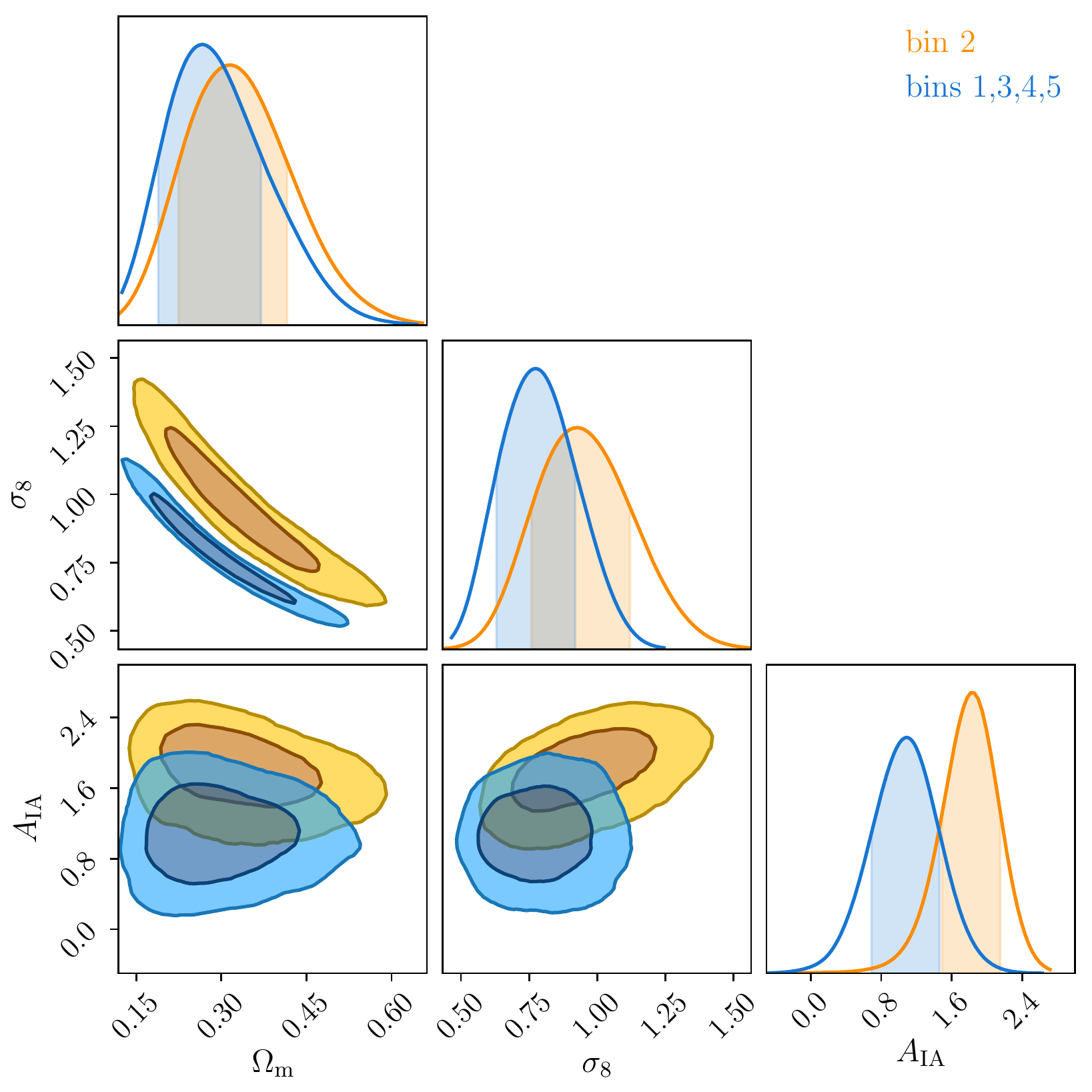} 
     \end{tabular}
   \end{center}
     \caption{Marginal posteriors using band powers in the internal consistency test that isolates all bin combinations involving the second tomographic bin. The test duplicates the sampling parameters (with fixed $\delta_z$) and assigns them to the two parts of the data vector. The orange contours refer to the split including the second bin and all its cross-correlations, while the blue ones present the constraints from the the remainder of the redshift bins (and their cross-correlations). The cross-covariance between the two parts of the data are included via the data covariance matrix. Other divisions of the data show much more consistent results. }
 \label{fig:contours_bin2_all}
 \end{figure}

\begin{table*}
\caption{Tier 1 ({\it left}) and tier 2 ({\it right}) test results for COSEBIs, band powers (BP) and 2PCFs. }
\label{tab:tier12_cosebis}
\let\center\empty
\let\endcenter\relax
\centering
\resizebox{1.7\columnwidth}{!}{\begin{tabular}{c| l c c c | c c c }
 & data split &     $\log_{10}R ({\rm trad.}) $ & $\text{log}_{10}R ({\rm import.}) $ &        $\ln S$ & $\Delta(S_8,A_{\rm IA})$ & $\Delta(S_8)$ & $\Delta(A_{\rm IA})$    \\ \hline 
 \parbox[t]{2mm}{\multirow{5}{*}{\rotatebox[origin=c]{90}{COSEBIs}}} 
& $z$-bin 1 vs. all others &    0.57  &  1.12    & 0.22  & $0.7\sigma$ &  $1.4\sigma$ &          $0.1\sigma$ \\
& $z$-bin 2 vs. all others &   $-1.89$  &  $-1.56$   & $-8.82$  &  $2.2\sigma$ &  $2.7\sigma$ &         $2.1\sigma$ \\
& $z$-bin 3 vs. all others &    1.69 &  2.47    & 0.14    & $0.1\sigma$ &  $0.1\sigma$ &         $0.5\sigma$ \\
& $z$-bin 4 vs. all others &    0.95  & 1.82   & $-2.06$    &  $1.2\sigma$ &  $1.1\sigma$ &         $1.4\sigma$ \\
& $z$-bin 5 vs. all others &    0.82  & 1.47   & $-2.77$    &  $1.3\sigma$ &  $1.3\sigma$ &         $1.2\sigma$ \\ \hline
 \parbox[t]{2mm}{\multirow{5}{*}{\rotatebox[origin=c]{90}{BP}}} 
 & $z$-bin 1 vs. all others &       0.05  &   0.65  &    $-0.25$ & $1.6\sigma$ &  $2.0.\sigma$ &         $0.4\sigma$ \\
& $z$-bin 2 vs. all others &      $-2.46$  &   $-1.76$  &    $-9.61$ & $2.8\sigma$ &  $3.0\sigma$ &         $1.5\sigma$ \\
& $z$-bin 3 vs. all others &     1.56  &     2.26  &     0.23  &  $0.1\sigma$ &  $0.2\sigma$ &         $0.1\sigma$ \\
& $z$-bin 4 vs. all others &    0.05  &      0.75  &    $-3.64$  &  $1.5\sigma$ &  $1.6\sigma$ &         $1.5\sigma$ \\
& $z$-bin 5 vs. all others &    1.25  &     2.00  &    $-0.75$  &  $1.0\sigma$ &  $0.8\sigma$ &         $1.4\sigma$ \\
  \hline
\parbox[t]{2mm}{\multirow{5}{*}{\rotatebox[origin=c]{90}{2PCFs}}} 
& $z$-bin 1 vs. all others &     1.20 &       2.00  &    0.99  & $0.3\sigma$ &  $1.1\sigma$ &         $0.1\sigma$ \\
& $z$-bin 2 vs. all others &     $-2.07$  &    $-1.23$  &   $-9.92$  &  $2.2\sigma$ &  $2.4\sigma$ &         $2.1\sigma$ \\
& $z$-bin 3 vs. all others &      2.35  &     3.13  &    0.42  &  $0.7\sigma$ &   $1.2\sigma$ &         $0.5\sigma$ \\
& $z$-bin 4 vs. all others &      0.69 &      1.14  &  $-4.65$  &  $0.9\sigma$ &   $1.1\sigma$ &         $1.2\sigma$ \\
& $z$-bin 5 vs. all others &     0.78  &      1.61  &  $-3.93$  &  $1.0\sigma$ &  $0.7\sigma$ &         $1.5\sigma$ \\
\end{tabular}
}
\tablefoot{The data split is done by separating a tomographic bin and all its cross-correlations from the rest of the data. The sampled parameters are duplicated and each part of the data is allowed to constrain one set, while the cross-correlations within the data are taken into account through their covariance matrix. The $\delta_z$ parameters are fixed to the mean of their prior for both parts of the chain (see the discussion in the text). 
The error on the estimated Bayes factor $\log_{10} R$ and the suspiciousness $\ln S$ is about $0.05$ for all cases. We show $\log_{10} R$ for both the traditional (trad.) and importance nested sampling (import.) methods. The tier 2 results measure the significance of the differences between the marginalised distributions of the duplicated parameters indicated in the table heading. }
\end{table*}

Following the methodology of \cite{kohlinger/etal:2019}, we perform three tiers of tests on divisions of the data based on splitting according to tomographic bins and all their cross-correlations. With the tier 1 test we compare the Bayesian evidence, 
\begin{equation}
Z \equiv {\rm Pr}({\rm data} | {\tens M}) = \int {\rm d} \vec{p}\; {\rm Pr}({\rm data} | \vec{p},{\tens M})\,  {\rm Pr}(\vec{p} | {\tens M}) \;, 
\end{equation}
where $\tens M$ is the model under consideration, with parameters $\vec{p}$. The evidence is calculated for two cases: the fiducial run (1-cosmo henceforth) and an analogous run where the parameters are duplicated for each split of the data (2-cosmo henceforth). In the 2-cosmo run each part of the data has its own set of parameters to constrain, but the correlations within the data are taken into account via the data covariance matrix. We compare the evidences using the Bayes factor,
\begin{equation}
\label{eq:bayes}
R = \frac{{\rm Pr}({\rm data} | {\tens M}_1)}{{\rm Pr}({\rm data} | {\tens M}_2)} = \frac{{\rm Pr}( {\tens M}_1|{\rm data} )}{{\rm Pr}({\tens M}_2|{\rm data}) } \frac{{\rm Pr}({\tens M}_2)}{{\rm Pr}({\tens M}_1)}
\;.
\end{equation}
We assume that the a-priori probabilities of the two models are equal, ${\rm Pr}({\tens M}_2)={\rm Pr}({\tens M}_1)$. 
With this assumption, the Bayes factor compares the probability of the models given the data. If $R<1$ then ${\tens M}_2$ is preferred by the data and vice versa. 
For our internal consistency test $ {\tens M}_1$ is the 1-cosmo model, where all the parameters are shared between the two parts of the data and $ {\tens M}_2$ is the 2-cosmo case.

We use the {\sc MontePython} package \citep{audren/etal:2013,brinckmann_lesgourgues:2018} where our internal consistency tests are developed. We find very good consistency between our {\sc MontePython} and {\sc KCAP} likelihood codes (better than our $0.1\sigma$ threshold).  
Currently, {\sc MontePython} does not allow for sampling over parameters with non-flat priors. To circumvent this issue it is common practice to include the prior in the likelihood values. 
This can result in biased estimates of $Z$ for non-flat priors. In our fiducial chains the $\delta_z$ shifts have a Gaussian prior (as does $\delta_c$ for the 2PCFs). 
In \sect\ref{sec:nuisance} we showed that fixing these to their fiducial value has little impact on the constraints (the no $\sigma_z$ case). 
Given this limitation in {\sc MontePython} and the negligible impact of the $\delta_z$ shifts, we fix these parameters (and  $\delta_c$ for the 2PCFs), for both 1 and 2-cosmo runs. 

Given a high-dimensional parameter space, it is difficult to estimate the evidence accurately. 
Alternative methods to {\sc MultiNest} have been proposed which aim to provide a more reliable value for $Z$ \citep[e.g. {\sc Polychord}, ][]{handly/etal:2015}. 
These alternatives are however several times slower than the {\sc MultiNest} runs; thus we estimate the evidence from the {\sc MultiNest} output using two methods: the standard approach employing the posterior sample (trad.) and an importance nested sampling (import.) version generated automatically by  {\sc MultiNest}. 
To estimate the traditional method we use the {\sc anesthetic} processing tool \citep{anesthetic}. 
We find that in general the differences between these estimates of $Z$ are larger than their associated errors, while it is not clear which one is closer to the truth. 
To assess this, we run one  {\sc Polychord} chain for a case where we found the largest difference between the traditional and importance sampling values. We find that the {\sc Polychord} estimate of $\log_{\rm 10} Z$ is in-between these two values. 
Therefore, we report the Bayes factor for both of these estimates. 

\cite{lemos/etal:2020} suggested using suspiciousness, $S$, instead of the Bayes factor for the tier 1 test, as it has much reduced sensitivity to the volume of the prior. This is particularly useful for the tier 1 test as the 2-cosmo model is inherently penalised due to the doubling of parameter space. We show $\ln S$ values for all cases, but refrain from translating them into the popular $\tau\sigma$ measure. To do so, we need a robust estimate of the effective number of parameters, $N_{\Theta}$, for both the 1-cosmo and 2-cosmo runs. 
In section 6.3 of J20 we see that the dimensionality measure which has so far been used in conjunction with suspiciousness is in general a biased estimator of $N_{\Theta}$. 
The other methods suggested in J20 involve running multiple computationally expensive chains. 
Therefore, here we only report the values for $\ln S$ and leave their further interpretation to future work. 

With the tier 2 test we consider the posterior of the difference between the two instances of the same parameters that result from the 2-cosmo analysis. 
We count the fraction of samples in this distribution with lower density than the posterior density at the origin\footnote{The posterior at the origin is estimated from a fit to the sampled posterior, using a kernel density estimation with a Gaussian.}, where the results for both sections of the data are perfectly matched. The smaller this fraction the less likely it is to have agreement between the two parts of the data. 
This fraction is then cast into an $\tau\sigma$ value based on the fractional differences between the peak and the tails of a one dimensional Gaussian distribution \citep[more details in section 2.2 of][]{kohlinger/etal:2019}. 

The only fully constrained parameters with KiDS-1000 data are $S_8$ (or $\Sigma_8$) and $A_{\rm IA}$, as discussed in \app\ref{app:extra}. Hence, for the tier 2 tests we only consider the marginal distributions for these two parameters and their combinations.

\tab\ref{tab:tier12_cosebis} lists the tier 1 results in the left columns and the tier 2 results on the right. We report values for all three two-point statistics. 
Similar trends can be seen for the results of COSEBIs, band powers and 2PCFs. In all cases redshift bin 2 stands out, whereas the remaining tests return values consistent with noise. 
In this case the tier 1 test shows a negative $\log_{10} R$ indicating a preference for the 2-cosmo model. 
We use Jeffreys' scale to interpret the significance of the measured $\log_{10} R$, and find it to show strong to decisive evidence for the 2-cosmo model, depending on the statistics and the method used to estimate the evidences. 
The tier 2 test corroborates this result. We see that the parameter differences for this separation of the data are larger than all the other cases, with up to\footnote{We have seen even more significant differences when including some of the poorly constrained parameters. For example, when comparing the parameter estimates for $\Delta(S_8,A_{\rm IA}, \Om)$ we find differences of up to $3.6\sigma$, however as $\Om$ is not significantly constrained by our data, this more significant value may just be a result of additional fluctuations in the noise.} $3\sigma$. 

To better understand the origin of the inconsistency between the second tomographic bin and all others, we compare the translated posterior distributions \citep[TPD,][section 2.3]{kohlinger/etal:2019} that are produced from the 2-cosmo chains. We make predictions for all bins using the TPDs and compare them with the data. 
Figure\thinspace\ref{fig:bp_bin2vsAll} shows results for band powers. We choose band powers here as their data points are considerably less correlated compared to COSEBIs or 2PCFs, facilitating a visual inspection. 
The TPDs of bin 2 and its cross-correlations are shown in red, while the TPDs of all other bins are presented in blue. 
The width of the curves show the standard deviation of the TPDs. 
We see that the first bin and its cross-correlations, owing to their very low signal-to-noise, cannot distinguish between the two sets of TPDs, whereas for other pairs of redshift bins the two TPDs are clearly separated. 
The inconsistency of the bin 2 results with all other bins is clear here, with the former having larger signals than expected for their redshift distributions.

A reasonable explanation for this discrepancy is that a small but high-redshift population of galaxies has contaminated the second bin.
We expect a higher signal for higher-redshift galaxies, as their light passes more structures before reaching us producing stronger correlations between their observed shapes. 
Since here we have no freedom to change the redshift distributions,  
the model is forced to increase the amplitude of the power spectra to compensate for the higher amplitude in correlations with bin 2. 
This is done via varying both $A_{\rm IA}$ and $S_8$ as can be seen in \fig\ref{fig:contours_bin2_all}. This figure illustrates the tier 2 results, comparing the constraints for the parameters obtained from bin 2 and its cross-correlations (orange) with the rest of the data (blue). We show marginal distributions for the subset of parameters, $\sigma_8$, $\Om$ and  $A_{\rm IA}$.  In \fig\ref{fig:deltaz} we saw that the largest $\delta_z$ shift belonged to this bin. Including these shift parameters can mitigate these inconsistencies to some extent. 

In \sect\ref{sec:nuisance} we evaluated the impact of removing the second bin from the analysis and found its effect on our final results to be negligible. 
Consequently, we do not exclude this bin from our fiducial analysis (also see the discussion in \sect\ref{sec:consist}). Regardless, due to the excess signal in bin 2, including it in the analysis can only serves to increase the value of $S_8$ and decrease the tension with {\it Planck}.

\subsection{Quantifying tension with {\it Planck}}
\label{app:external_consistency}

In \sect\ref{sec:constst_external} we reported the tension in the marginal distributions of $S_8$ and $\Sigma_8$ for COSEBIs. Here we use a similar methodology to the tier 1 test in \app\ref{app:internal_consistency} to quantify the inconsistency between KiDS-1000 and {\it Planck}, also using COSEBIs, which are chosen owing to their better goodness-of-fit to the model. 

We compare the evidence for a single set of cosmological parameters for both KiDS-1000 and {\it Planck} by running a joint chain (1-cosmo) with the evidences found for their separate analysis (2-cosmo). The only difference here is that the two data sets are independent, allowing us to use the respective fiducial chains for the 2-cosmo runs. The Bayes factor can  now be written as,
\begin{equation}
R = \frac{{\rm Pr}(\text{KiDS-1000\; and}\; Planck | {\tens M}_1)}{{\rm Pr}(\text{KiDS-1000} | {\tens M}_{2,{\rm K}})\; {\rm Pr}(Planck | {\tens M}_{2,{\rm P}})}\;,
\end{equation}
where ${\tens M}_1$ is the model with shared parameters between KiDS-1000 and {\it Planck}, while ${\tens M}_2$ is the model with separate parameters for KiDS (${\tens M}_{2,{\rm K}}$) and {\it Planck} ($ {\tens M}_{2,{\rm P}}$). We find $\log_{10} R = -1.15$ (strong) using evidences from importance nested sampling and $\log_{10} R = -0.54$ (substantial) with the standard nested sampling method, both showing a preference for ${\tens M}_2$, i.e. tension between KiDS-1000 and \textit{Planck}. 
We also report the suspiciousness value, $\ln S = -2.94$ but find a negative value for the difference between the dimensionality of the 1-cosmo and 2-cosmo runs, although we expect a positive value. As discussed in \app\ref{app:internal_consistency}, the estimated values of dimensionality are generally biased with regards to the effective number of degrees of freedom as read off from the sampling distribution of the minimum $\chi^2$. As a result we are unable to cast this result into the more intuitive $\tau\sigma$ measure.

\section{Impact of survey pixel size on the size of constraints}

\label{app:pixelsize}


To calculate covariance matrices, we need to estimate the effective area, $A_{\rm eff}$, of the observed images; see \citet{joachimi/etal:2020}, appendix~E, for the details of the covariance model. 
The value of $A_{\rm eff}$ depends on the assumed pixel size, since we use a binary mask. The shape noise term is independent of $A_{\rm eff}$, since this term is estimated using the effective number of galaxy pairs. There are two other Gaussian terms in the covariance matrix which are impacted by the choice of $A_{\rm eff}$. For the cosmic variance (also known as sample variance) and the mixed terms the covariance scales approximately with the inverse of $A_{\rm eff}$. However, in the case of the mixed term, we include the effective number density of galaxies, $n_{\rm eff}$, which in turn depends on the effective area. We scale $n_{\rm eff}$ with respect to $A_{\rm eff}$ to keep the total number of galaxies in each tomographic bin constant and independent of the effective area. As a result, the only Gaussian term that is impacted by $A_{\rm eff}$ is the cosmic variance term. J20 argue for using the effective area of a survey with the same extent as KiDS-1000 but without the very small scale masks to calculate this term. One way to achieve this is by lowering the resolution of the mask, as we implement here.

\begin{figure}
   \begin{center}
     \begin{tabular}{c}
      \includegraphics[width=\hsize]{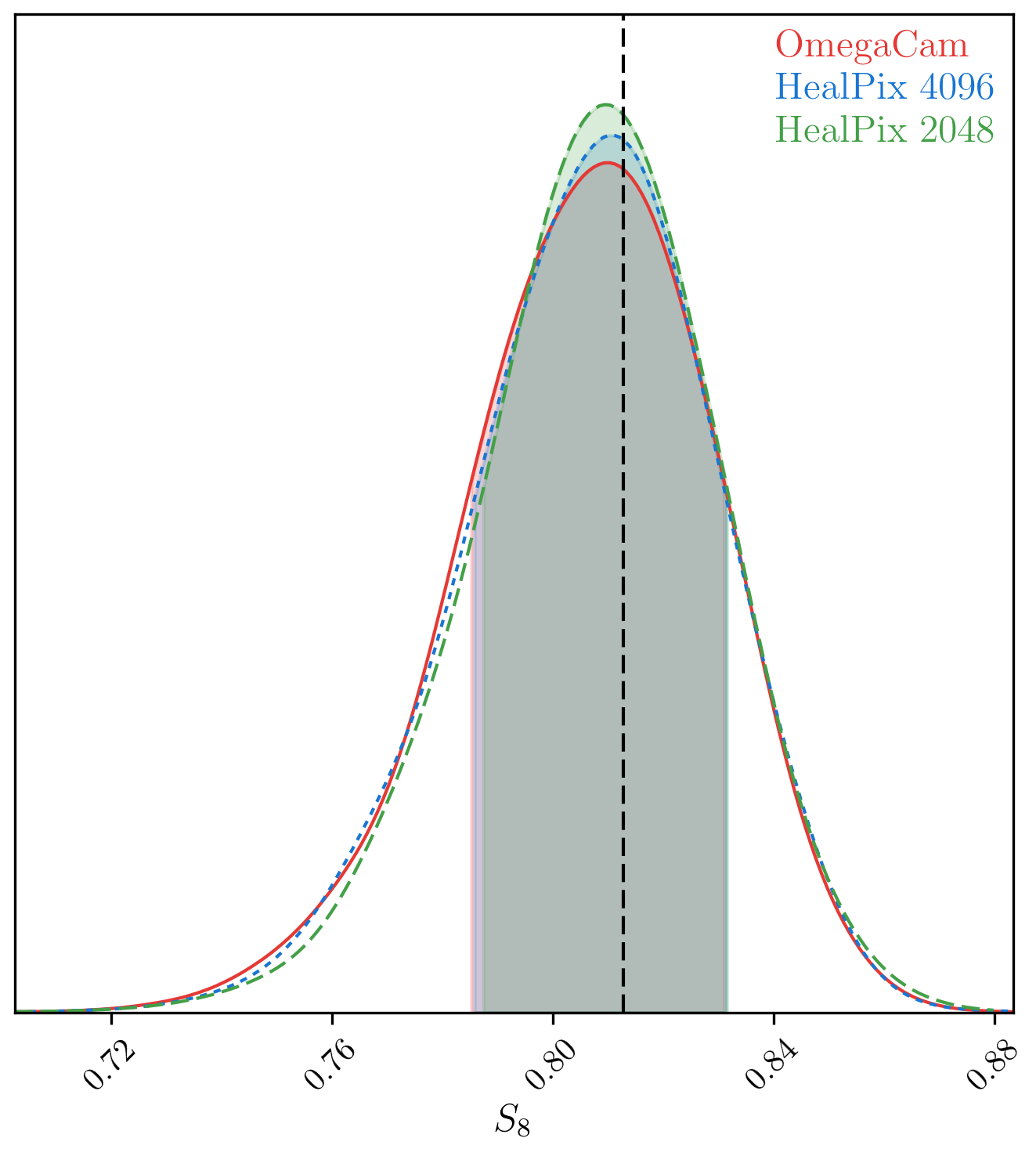}
     \end{tabular}
   \end{center}
     \caption{\small{ Impact of mask pixel size on $S_8$ constraints with mock data. The covariance matrices are calculated using the effective areas determined with the OmegaCam pixel size (red solid), {\sc HealPix} with $N_{\rm side}=4096$ (blue dotted) and {\sc HealPix} with $N_{\rm side}=2048$ (green dashed). The mock data is noise free and the dashed line shows the input $S_8$ value. }}
     \label{fig:covariance_area}
 \end{figure}

Figure\thinspace\ref{fig:covariance_area} shows marginal constraints for $S_8$ with the fiducial priors used in our analysis. Noise-free mock data is used to assess the impact of the pixel size. We consider three different resolutions of the survey mask: at the OmegaCam pixel size, resulting in $A_{\rm eff}=777.4\;{\rm deg}^2$; using {\sc HealPix} with $N_{\rm side}=4096$ ($A_{\rm eff}=867.0\;{\rm deg}^2$); and using {\sc HealPix} with $N_{\rm side}=2048$ ($A_{\rm eff}=904.2\;{\rm deg}^2$). The survey masks consider a pixel as observed, if some fraction of the sky area covered by it has unmasked imaging, which explains the increase in area as the resolution becomes coarser.
Here we  have kept the area for the sub-dominant non-Gaussian terms fixed to $A_{\rm eff}=867.0\;{\rm deg}^2$ and included the $m$ calibration covariance terms which are independent of the area. 

We see that the constraints are not significantly impacted by the effective area. 
Both the standard deviation of the sampled points and the peaks of the marginal distributions are unchanged well within our error margin of $0.1\sigma$. In figures 10 and 11 of J20 we see that the diagonal terms in the cosmic shear covariance matrix are dominated by the noise term, whereas the diagonals of the sub-matrices are dominated by the mixed term. 
Therefore, this result is expected.
In all three cases the maximum marginal value of $S_8$ is slightly biased towards smaller values, while the projection of the maximum posterior recovers the input.

\section{Modelling residual constant c-terms}

\label{app:cterm}

\begin{figure*}
   \begin{center}
     \begin{tabular}{c}
      \includegraphics[width=1.7\columnwidth]{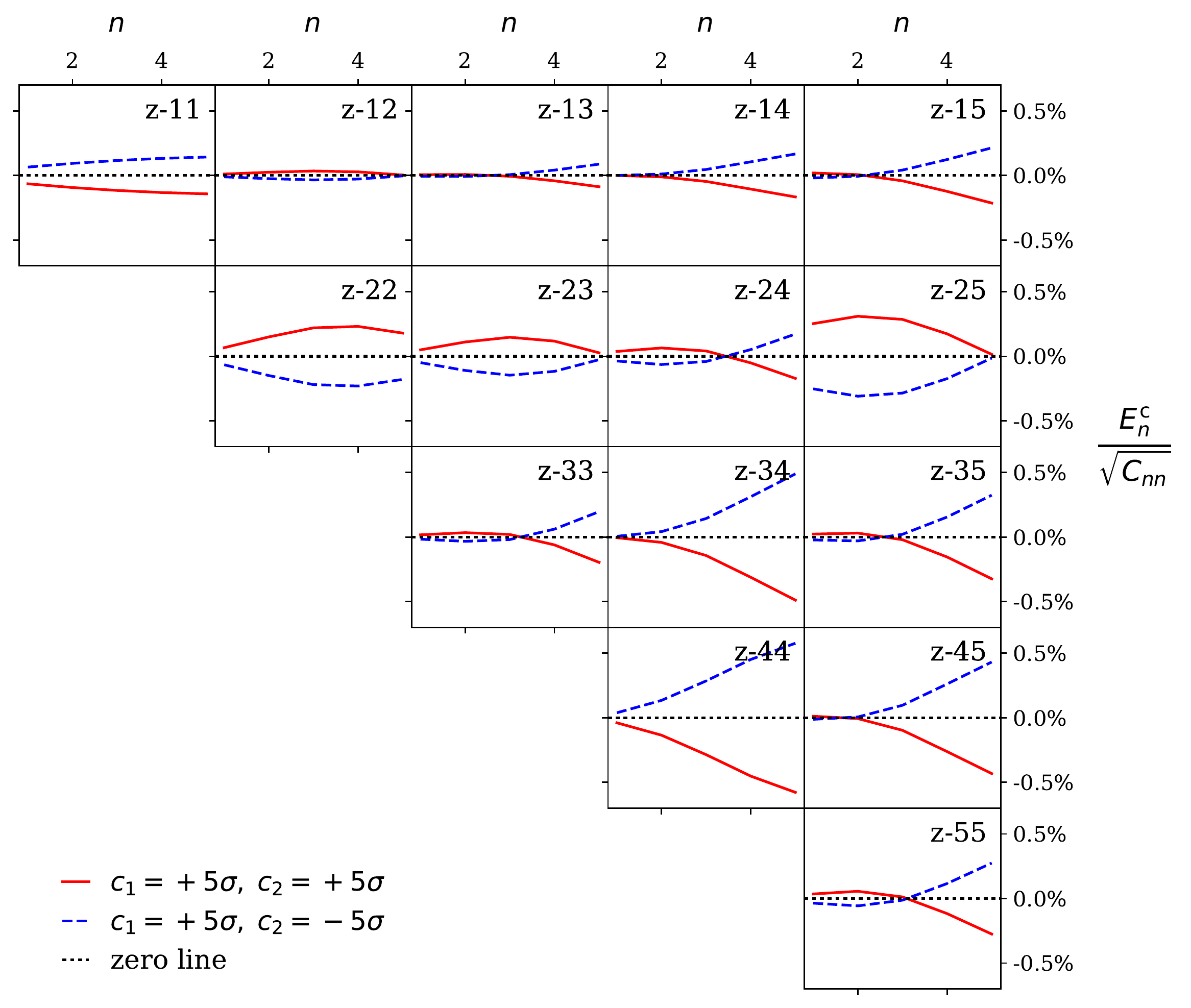} 
     \end{tabular}
   \end{center}
     \caption{Effect of a constant additive shear bias on COSEBIs. $E_n^{\rm c}$ is calculated for two extreme cases where $c_1=5\sigma$ and  $c_2=\pm 5\sigma$ of their allowed range for KiDS-1000 (\Eqt\ref{eq:Ectermintegral}). Here we use a KV450 footprint which results in a larger effect than KiDS-1000. }
     \label{fig:cterm}
 \end{figure*}

The measured ellipticities of galaxies can be biased by an additive term, usually dubbed the $c$-term. In our data we correct for a constant overall $c$-term for each of the ellipticity components (see G20 for more details).  There is an uncertainty on this parameter, such that there could be some residual signal from the term that remains in the data. We are able to marginalise over this uncertainty using additional free parameters. In our analysis we considered a single additive parameter $\delta_{\rm c}=\pm\sqrt{c_1^2+c_2^2}$ which only affects $\xi_+$. The other statistics that we consider are unaffected by this constant additive terms (to very good approximation in the case of band powers); however, they can still be affected by $c$-terms due to survey boundary effects. Here we first look at how $\xi_-$ is impacted by a constant $c_1$ and $c_2$, and then propagate through to COSEBIs and band powers. 

Under a flat-sky approximation we can write the correlation functions as,
\begin{align}
\label{eq:xipm_cart}
\hat{\xi}_+(\theta) & =\langle \epsao\epsbo+\epsat\epsbt \rangle(\theta)\;, \\ \nonumber
\hat{\xi}_-(\theta) & =\langle (\epsao \epsbo-\epsat \epsbt)\cos(4\phi)\\ \nonumber
&\quad+(\epsao \epsbt + \epsat \epsbo) \sin(4\phi)\rangle(\theta)\;, 
\end{align}
where $\epsao$ and $\epsat$ are the Cartesian ellipticity components of a galaxy and $\phi$ is the polar angle of the vector connecting the two galaxies, labelled as $a$ and $b$. The average is taken over all pairs of galaxies with separation angle within a defined $\theta$-bin. 

Let us assume that the observed ellipticity is only biased by the $c$-terms, $c_1$ and $c_2$, and write the observed ellipticity as,
\begin{equation}
\epsilon^{\rm obs}_i = \epsilon_i+c_i\;,
\end{equation}
where $i=1,2$. We can now find the observed $\xi_\pm$ by replacing $\epsilon_{i}$ with $\epsilon^{\rm obs}_i$ in \Eqt\eqref{eq:xipm_cart},
\begin{align}
\label{eq:xipm_c}
\hat{\xi}^{\rm obs}_+(\theta) & =\hat{\xi}_+(\theta) + c_1^2+c_2^2\;, \\ \nonumber
\hat{\xi}^{\rm obs}_-(\theta) & = \hat{\xi}_-(\theta) + (c_1^2-c_2^2)\langle\cos(4\phi)\rangle(\theta)+ 2c_1c_2\langle\sin(4\phi)\rangle(\theta)\;, 
\end{align}
 where we set $\langle \epsilon_i \rangle=0$. For a finite field $\langle\cos(4\phi)\rangle(\theta)$ and $\langle\sin(4\phi)\rangle(\theta)$ do not vanish and their values depend on $\theta$. Therefore, we expect to get a small contribution from the $c$-terms to $\xi_-$. We note that $\xi_\times$ is also similarly affected by the $c$-terms and an analogous equation can be written for this correlation.

Both COSEBIs and band powers are defined as integrals over $\xi_\pm$, therefore we can propagate the effect of the $c$-terms using \Eqt\eqref{eq:COSEBIsReal} and \eqref{eq:bp_xipm}. As a constant additive term is filtered out for these statistics, only the $\xi_-$ terms remain. 
First we define $\xi_-^{\rm c}(\theta) := \hat{\xi}^{\rm obs}_-(\theta) - \hat{\xi}_-(\theta)$. We can then write,
\begin{align}
\label{eq:Ecterm}
E^{\rm c}_n &\equiv E^{\rm obs}_n - E_n= B_n - B^{\rm obs}_n  \\ \nonumber
 & = \frac{1}{2} \int_{\theta_{\rm min}}^{\theta_{\rm max}}
 \d\theta\,\theta\: T_{-n}(\theta)\,\xi_-^{\rm c}(\theta)\;,
\end{align}
and
\begin{align}
\label{eq:BPcterm}
{\mathcal C}^{\rm c}_{{\rm E},l} &\equiv {\mathcal C}^{\rm obs}_{{\rm E},l}- {\mathcal C}_{{\rm E},l} = 
{\mathcal C}_{{\rm B},l}-{\mathcal C}^{\rm obs}_{{\rm B},l}  \\ \nonumber
&\approx\frac{\pi}{{\mathcal N}_l}\; \int_0^\infty \d \theta\, \theta\; T(\theta)\; \xi_-^{\rm c}(\theta)\; g_-^l(\theta)\;.
\end{align}
In practice, to marginalise over the effect of the $c$-terms on $\xi_-$, COSEBIs and band powers, we need to let both $c_1$ and $c_2$ vary independently. Here we have assumed that the $c$-terms are constant within the survey and as such they can be taken out of the integrals in \Eqt\eqref{eq:Ecterm} and \eqref{eq:BPcterm} to yield
\begin{align}
\label{eq:Ectermintegral}
E^{\rm c}_n  = & \frac{1}{2}(c_1^2-c_2^2)\; \int_{\theta_{\rm min}}^{\theta_{\rm max}}
 \d\theta\,\theta\: T_{-n}(\theta) \langle\cos(4\phi)\rangle(\theta)\\ \nonumber
&+ c_1c_2\; \int_{\theta_{\rm min}}^{\theta_{\rm max}}
 \d\theta\,\theta\: T_{-n}(\theta) \langle\sin(4\phi)\rangle(\theta)\;,
\end{align}
and 
\begin{align}
\label{eq:BPctermintegral}
{\mathcal C}^{\rm c}_{{\rm E},l}  \approx & (c_1^2-c_2^2)\;\frac{\pi}{{\mathcal N}_l}\; \int_0^\infty \d \theta\, \theta\; T(\theta) \; g_-^l(\theta) \langle\cos(4\phi)\rangle(\theta)\\ \nonumber
&+ 2c_1c_2\;\frac{\pi}{{\mathcal N}_l}\; \int_0^\infty \d \theta\, \theta\; T(\theta)\;  g_-^l(\theta) \langle\sin(4\phi)\rangle(\theta)\;.
\end{align}
To model the effect of the $c$-terms for $\xi_-$, we can use the position of galaxies in the data to measure the expectation value of $\cos(4\phi)$ and $\sin(4\phi)$, or the integrals containing them in \Eqt\eqref{eq:Ectermintegral} and \eqref{eq:BPctermintegral} in the case of COSEBIs and band powers. We can then use these values as inputs to model $\xi_-^{\rm c}$, $E_n^{\rm c}/B_n^{\rm c}$ and ${\mathcal C}^{\rm c}_{{\rm E/B},l}$. This can be done by running the same tree-code used to measure the 2PCFs with two separate runs where the ellipticities of galaxies are replaced by two sets of constant values. 

We estimate that this effect on COSEBIs, band powers and $\xi_-$ for the KiDS-1000 data is smaller than $1\%$ compared to the size of the error bars, where we used values for $c_1$ and $c_2$ taken from the $5\sigma$ limits of their estimated errors (see G20, section 3.5.1). Since $\langle\cos(4\phi)\rangle(\theta)$ and $\langle\sin(4\phi)\rangle(\theta)$ are non-zero due to survey boundaries and masks, the effect of the constant $c$-term is scale-dependent (increases with $\theta$). 
In \fig\ref{fig:cterm} we show this effect for COSEBIs, $E_n^{\rm c}$, with respect to the expected error on the measured COSEBIs through the covariance matrix $C_{nn}$, which we have used in our fiducial analysis. Here we have used a KV450 footprint and expect this effect to be even less significant if a KiDS-1000 footprint is employed. 

For larger surveys with contiguous coverage these terms should be small given angular scales that are well within the survey area. However, the measurement errors also decrease for these surveys. Therefore, their importance needs to be re-evaluated for future surveys.

\section{Distribution of the amplitude of COSEBIs in {\sc Salmo} simulations}

\label{app:dist}

\begin{figure}
   \begin{center}
     \begin{tabular}{c}
      \includegraphics[width=\hsize]{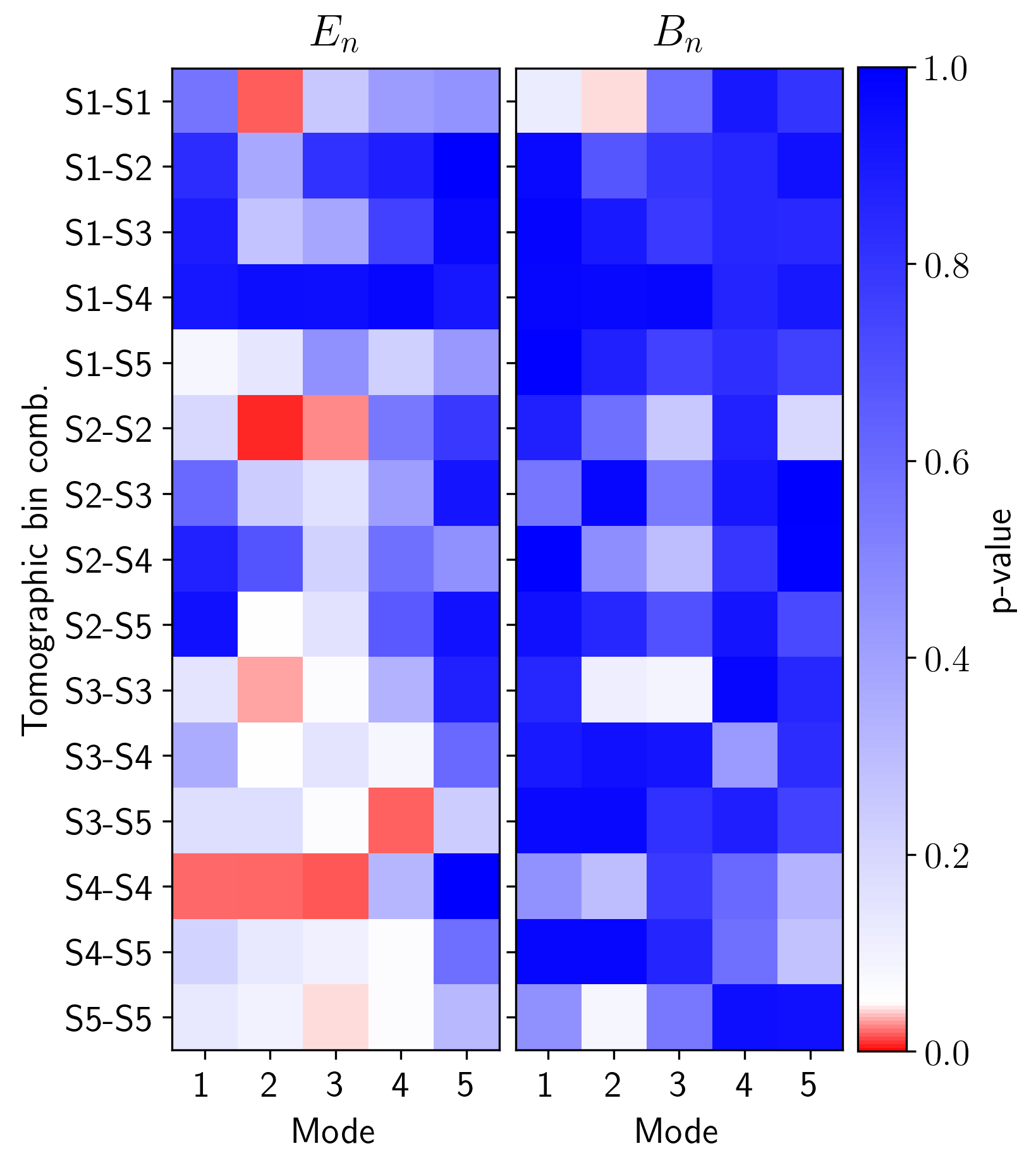}
     \end{tabular}
   \end{center}
     \caption{\small{ Distribution of COSEBI E- and B-modes in {\sc Salmo} simulations, as a function of tomographic bin combination and COSEBI mode. The plot shows the $p$-value of a Kolmogorov-Smirnov test of the sampling distribution from 1000 mocks compared to a Gaussian. The minimum $p$-value that we find is $0.01$, showing a marginally non-Gaussian distribution. }}
     \label{fig:dist}
 \end{figure}

We measure COSEBI E- and B-modes from the {\sc Salmo} simulations described in \citet[section 4]{joachimi/etal:2020} and compare their distribution to a Gaussian with the same mean and variance using a Kolmogorov-Smirnov (KS) test. \fig\ref{fig:dist} shows the $p$-values associated with this test. This figure can be contrasted with figure 17 in J20, where $\xi_+$ shows a low $p$-value for its largest two $\theta$-bins over all redshift bin combinations. The distributions of the COSEBI B-modes are consistent with Gaussian distributions, whereas there are a few smaller $p$-values (shown in hues of red) for the E-modes, with a minimum of 0.01. Considering \fig\ref{fig:compare}, we expect to get a similarly Gaussian distribution for COSEBIs as for $\xi_-(\theta_{\rm max})$. In the {\sc Salmo} simulations the distribution of $\xi_-$ is perfectly Gaussian. The $p$-values for $\xi_+(\theta_{\rm max})$ on the other hand go as low as $10^{-4}$, therefore the significance of their non-Gaussianity is much higher. 

Given this marginally non-Gaussian result for some of the COSEBIs modes for certain pairs of redshift bins, we test the distribution of their $\chi^2$ values in the simulations (comparing each mock $E_n$ with their mean value over all mocks) and find that to be consistent with a $\chi^2$ distribution with the correct degrees of freedom ($p$-value$ = 0.8$). Given these results, we conclude that the full distribution of COSEBIs is close enough to a Gaussian. For a likelihood analysis the $\chi^2$ is the quantity that is used and therefore the assumption that it is $\chi^2$-distributed is of more importance than the Gaussianity of the individual COSEBIs modes. 
With more simulations we can resolve whether or not the slightly low $p$-values persist.

\section{Changes after unblinding}

\label{app:unblinding}

Our blinding strategy is described in \cite{kuijken/etal:2015}. 
Prior to unblinding our data, we ran all the fiducial chains using a covariance matrix calculated with the fiducial set of model parameters used in J20. Since our blinding strategy allowed for comparing relative constraints between different setups, without major changes to the conclusions, we ran all of the systematics and internal consistency chains for one of the blinds only.

After unblinding we re-ran the systematics and internal consistency chains for the correct blind without changing the cosmological parameters used in the covariance matrix. After unblinding we changed the definition of $\delta_{\rm c}$ to take both positive and negative values and re-ran the 2PCFs chains. This update only impacted the results at a level consistent with variations between different chains. 

For our fiducial results we repeated the likelihood analysis with an updated covariance model based on the best-fit parameters of \cite{heymans/etal:2020}. These chains were run after the unblinding to test the effect of an iterated covariance model, which had a negligible impact (less than $0.1\sigma$) on our constraint of $S_8$. The combined  chain with {\it Planck} used in our external consistency test was also run after the unblinding, with the iterated covariance matrix.


\end{document}